\documentclass[amsmath,amssymb,english,aps,prb,twocolumn,floatfix,superscriptaddress,showpacs]{revtex4}
\usepackage[english]{babel}
\usepackage{verbatim}
\usepackage{graphicx}

\newcommand{\kkk}{\boldsymbol{k}}

\newcommand{\qqq}{\boldsymbol{q}}
\newcommand{\rrr}{\boldsymbol{r}}
\newcommand{\RRR}{\boldsymbol{R}}

\newcommand{\WW}{\mathbb{W}}
\newcommand{\CC}{\mathcal{C}}
\newcommand{\NN}{\mathcal{N}}
\newcommand{\dd}{$d$}
\newcommand{\pp}{$p$}
\newcommand{\sss}{$s$}

\newcommand{\ff}{$f$}

\newcommand{\JJ}{$J$}
\newcommand{\uk}{$\mathcal{U}$}
\newcommand{\vk}{$\mathcal{V}$}
\newcommand{\jk}{$\mathcal{J}$}
\newcommand{\wk}{$\mathcal{W}$}

\newcommand{\occ}{\textrm{occ}}
\newcommand{\unocc}{\textrm{unocc}}
\newcommand{\expo}{\textrm{e}}

\newcommand{\LDAU}{\mbox{LDA+U}}
\newcommand{\LDADMFT}{\mbox{LDA+DMFT}}

\newcommand{\crpa}{cRPA}
\newcommand{\clda}{cLDA}
\newcommand{\cldam}{\textrm{cLDA}}

\newcommand{\LAPWlo}{\mbox{(L)APW+lo}}

\def\eg{\ensuremath{e_{g}}}
\def\t2g{\ensuremath{t_{2g}}}
\def\dz2{\ensuremath{d_{3z^2-r^2}}}
\def\dx2y2{\ensuremath{d_{x^2-y^2}}}
\def\dxy{\ensuremath{d_{xy}}}
\def\dxz{\ensuremath{d_{xz}}}
\def\dyz{\ensuremath{d_{yz}}}

\def \srmo{SrMO$_{3}$}
\def \srmol{Sr$_{2}$MO$_{4}$}
\def \srvo{SrVO$_{3}$}
\def \srcro{SrCrO$_{3}$}
\def \srmno{SrMnO$_{3}$}
\def \srnbo{SrNbO$_{3}$}
\def \srtco{SrTcO$_{3}$}
\def \srmoo{SrMoO$_{3}$}
\def \srmool{Sr$_{2}$MoO$_{4}$}
\def \srtcol{Sr$_{2}$TcO$_{4}$}
\def \srruol{Sr${_2}$RuO$_{4}$}
\def \srrhol{Sr${_2}$RhO$_{4}$}
\def \srirol{Sr${_2}$IrO$_{4}$}

\begin{document}

\thispagestyle{empty}

\title{Hubbard $U$ and Hund's Exchange $J$ in Transition Metal Oxides: 
Screening vs. Localization Trends from Constrained Random Phase Approximation}
\author{Lo\"ig Vaugier} 
\affiliation{Centre de Physique Th\'eorique, Ecole Polytechnique, CNRS UMR 7644, 91128 Palaiseau, France}
\affiliation{Japan Science and Technology Agency, CREST, Kawaguchi 332-0012, Japan}
\author{Hong Jiang}
\affiliation{Beijing National Laboratory for Molecular Sciences, State Key Laboratory of Rare Earth Material Chemistry and Application, Institute of Theoretical and Computational Chemistry, College of Chemistry and Molecular Engineering, Peking University, 100871 Beijing, China}
\author{Silke Biermann}
\affiliation{Centre de Physique Th\'eorique, Ecole Polytechnique, CNRS UMR 7644, 91128 Palaiseau, France}
\affiliation{Japan Science and Technology Agency, CREST, Kawaguchi 332-0012, Japan}

\date{today}

\begin{abstract}
In this work, we address the question of calculating
the local effective Coulomb interaction matrix in materials 
with strong electronic Coulomb interactions from first
principles. To this purpose, we implement the constrained 
random phase approximation (cRPA) into a density functional 
code within the linearized augmented plane wave (LAPW) framework.

We apply our approach to the 3$d$ and 4$d$ early transition
metal oxides SrMO$_{3}$ (M=V, Cr, Mn) and (M=Nb, Mo, Tc)
in their paramagnetic phases. For these systems, we explicitly
assess the differences between two physically motivated 
low-energy Hamiltonians:
The first is the three-orbital model comprising the $t_{2g}$ states
only, that is often used for early transition metal oxides.
The second choice is a model where both, metal $d$- and oxygen $p$-states
are retained in the construction of Wannier functions, but the
Hubbard interactions are applied to the $d$-states only (``d-dp Hamiltonian'').
Interestingly, since -- for a given compound -- both $U$ and $J$
depend on the choice of the model, so do their trends within a family
of these compounds. In the 3$d$ perovskite series SrMO$_{3}$ 
the effective Coulomb interactions in the $t_{2g}$ Hamiltonian
decrease along the series, due to the more efficient screening.
The inverse -- generally expected -- trend, increasing interactions
with increasing atomic number, is however
recovered within the more localized ``d-dp Hamiltonian''.
Similar conclusions are established in the layered 4$d$ perovskites series
Sr$_{2}$MO$_{4}$ (M=Mo, Tc, Ru, Rh).
Compared to their isoelectronic and isostructural 3$d$ analogues, the 4$d$
perovskite oxides SrMO$_{3}$ (M=Nb, Mo, Tc) exhibit weaker
screening effects. Interestingly, this leads to an effectively larger
$U$ on 4$d$ than on 3$d$ shells when a $t_{2g}$ model is constructed.
\end{abstract}

\maketitle

\section{INTRODUCTION}
Naively, the calculation of the Coulomb repulsion between two charges 
may seem to be a simple textbook problem. 
However, calculating this repulsion for two charges in a solid 
is far from being trivial, as the electronic polarizability
screens the Coulomb potential leading to a renormalized repulsion strength. 
Obtaining a quantitative estimation
of the interactions, 
is particularly important in situations where the Coulomb interactions
dominate the behavior of the system \cite{review-imada}. 

The description of the excitations of such strongly correlated 
materials require a theoretical treatment that goes beyond the
one-particle picture of band theory.
Simply identifying the Kohn-Sham spectra of
density functional theory (DFT) in the local density approximation 
(LDA)~\cite{Kohn-nobel,hohenberg-kohn-1964,DFT-Kohn-Sham,DFT-review}
with the many-body spectra, for example, becomes then an inappropriate
oversimplification. 
Popular methods for introducing many-body corrections into this
description construct a multi-orbital Hubbard-type Hamiltonian,
where explicit interaction terms of Hubbard and Hund form are
added. The resulting problem is then e.g. solved within a static
mean field theory within the ``LDA+U'' scheme~\cite{LDA+U-anisimov-1997} or 
within dynamical mean field theory (DMFT) within the combined ``LDA+DMFT'' 
method~\cite{LDA+DMFT-anisimov-1997,LDA+DMFT-licht,kotliar-review-DMFT}.
However, the predictive power of such methods crucially relies
on a reliable assessment of the interactions.
The \emph{ab initio} calculation of these parameters is hence an 
important issue.

Within the last years, tremendous progress has been made 
concerning this question, in particular with the advent of
the so-called ``constrained Random Phase Approximation'' (cRPA)
as introduced in Ref.~[\onlinecite{cRPA-ferdi-2004}]. The cRPA provides a
systematic first principles technique for the construction 
of low-energy Hamiltonians where not only the one-particle 
part of the Hamiltonian but also the interaction part is 
calculated from first principles, that is without adjustable 
parameters. The starting point is the choice of an effective 
low-energy Hilbert space, and the cRPA aims at constructing
the {\it partially screened} interaction that should be
used as the {\it bare} interaction within this space.
The procedure can be viewed as the analogue for the
interaction term of the familiar ``downfolding'' techniques
used for the one-body part of the Hamiltonian
\cite{downfolding-lowdin,NMTO_andersen}. 
Matrix elements of the effective partially screened 
interaction 
are identified with the Hubbard and Hund's interaction
matrices for the effective low-energy Hamiltonian, and
the corresponding parametrizations in terms of 
Hubbard $U$ and Hund's rule $J$ can be explicitly
constructed.

In this work, we have implemented the constrained random phase 
approximation (\crpa)~\cite{cRPA-ferdi-2004} 
within the full-potential augmented plane wave (\LAPWlo) 
framework into the popular density functional 
code \textsc{Wien2k}~\cite{blaha_wien2k}. 
Our approach allows for the calculation of the Coulomb interaction 
matrix elements in a localized basis set, and can be combined 
with many-body techniques working in this framework, such as the
recent implementation of \LDADMFT\ within the same basis 
set~\cite{cRPA-DMFT-LaOFeAs-markus}.
We illustrate the power of our approach on the 
3\dd\ and 4\dd\ transition metal oxides  
\srmo\ (M = V, Cr, Mn, Nb, Mo, Tc) and 
\srmol\ (M = Mo, Tc, Ru, Rh).

It has been argued early on~\cite{U-xray-sawatzky-1977,NiO-sawatzky} that
in systems where the behavior of the correlated orbitals 
is sufficiently close to a purely atomic description, 
the Hubbard and Hund interactions acquire the status
of physical observables, directly measurable e.g. in spectroscopy
experiments.
As a consequence of negligeable interorbital hybridizations,
a natural basis set for the definition of $U$ and $J$
(the atomic one) emerges from the physical situation in these
cases.
Here, we adopt a more general point of view, defining 
Hubbard and Hund interactions even for materials where
hybridization between states of different orbital character
is too strong to allow for uniquely defined atomic orbitals
to emerge.
Even in this general case, Hubbard and Hund interactions can 
still be defined as useful auxiliary quantities for many-body 
calculations, but are no longer direct physical observables.
We illustrate this issue on the examples of 3\dd\ and 4\dd\
transition metal oxides (ranging from correlated metals
to Mott insulators), for which we explicitly construct two
different many-body models, demonstrating the
dependence of $U$ and $J$ on this choice.
We stress that, as a direct consequence of this issue, 
statements on {\it values} of $U$ and $J$ for a given material
are not meaningful pieces of information unless supplemented
by the complete description of the model to which they
apply, both concerning the choice of the low-energy subspace 
and of the correlated orbitals.

The paper is organized as follows: in section II, we discuss
the context of our work with respect to previous methods
for the determination of Hubbard and Hund interaction parameters.
A general description of \crpa\ and of its implementation
within the \LAPWlo\ framework is given 
in section III, supplemented by technical issues in Appendix A and a
discussion of a parametrization of the Hubbard interaction matrix in
terms of Slater parameters in Appendix B.
The results of \crpa\ calculations on the 3\dd\ and 4\dd\ series of 
ternary transition metal oxides \srmo\ with (M=V, Cr, Mn) and 
(M=Nb, Mo, Tc) and on the 4\dd\ layered perovskites \srmol\
with (M=Mo, Tc, Ru, Rh) are described respectively
in Sections IV and V.
In Section VI, we discuss 
perspectives for fully first principles many-body calculations for
correlated materials. We conclude with a brief summary of key results
in Section VII.

\section{Hubbard and Hund interactions: from spectroscopy to first principles calculations}

\subsection{Hubbard $U$: fitting parameter or calculated quantity?}

Historically, the Coulomb energy cost for placing two electrons 
on an atomic site within a solid - the Hubbard interaction $U$ - 
was introduced in the single-orbital Hubbard-Kanamori-Gutzwiller 
model~\cite{Hubbard_model_1963,Kanamori_paper,Gutzwiller_1963}.
$U$ corresponds to the difference between the electron affinity and 
the ionization energy 
when respectively adding and removing an electron on 
the correlated shell (e.g. the $d$-shell) of a given atom:
\begin{eqnarray}
U &=& E(d^{n+1}) + E(d^{n-1}) - 2E(d^{n}),
\end{eqnarray}
where $E(d^{n})$ is the total energy of a system for which $n$ 
electrons fill a given $d$ shell on a given atom. 
When the orbital character of the additional or missing
charge is specified also, an interaction {\it matrix}
$U$ can be defined.
Depending on the degree to which the correlated $d$-shell retains
its atomic character even in the solid, its elements 
can be parametrized by a small number of (Slater) parameters.

An important step in the determination of $U$ and $J$ was the 
pioneering work of Sawatzky and 
co-workers~\cite{U-xray-sawatzky-1977,NiO-sawatzky,U-xray-deboer,CuO-sawatzky,PhD-vandermarel} 
who fitted the multiplet structure measured in x-ray photoemission, 
absorption and Auger spectra with configuration-interaction cluster models 
involving a set of Slater integrals $(\textrm{F}^{k})$~\cite{judd,Sugano_book}. In a very simplified way, $U=\textrm{F}^{0}$ can be obtained from the 
Mott gap whereas the other Slater integrals $\textrm{F}^{k},k>0$ can
be deduced 
by fitting atomic Hartree-Fock-like calculations to experimental
spectra.

Attempts to calculate $U$ from first principles typically rely on 
constrained density functional approaches~\cite{cLDA-Ce-dederichs}. 
Such ``constrained LDA'' (\clda) 
approaches~\cite{cLDA-mcmahan-La2CuO4,cLDA-hybertsen-La2CuO4,
cLDA-andersen,cLDA-gunnarsson,cLDA-anisimov-gunnarsson} 
are based on the observation that the energy of a system with
increased or reduced particle number is in principle accessible
within density functional theory.
$U^{\cldam}$ is then defined as the derivative of the total energy 
while constraining the occupation on a given shell. 
\clda-type schemes have been  
implemented e.g.~within the linear muffin-tin orbital (LMTO) 
framework assuming that the LMTO basis functions 
used in a multi-orbital Hubbard model
give a good representation of the localized degrees of freedom.
Later on, implementations of \clda\ were worked out within various
other electronic structure codes, for example, in the
\LAPWlo\ framework~\cite{cLDA-LAPW-madsen}, or in a basis of maximally 
localized Wannier functions (MLWF)~\cite{cLDA-MLWF-nakamura}. 
We mention moreover the development of alternative linear response 
formalisms assessing the screened interaction by explicitly perturbing
the system~\cite{linearresponse-pickett,linearresponse-cococcioni}. 

In 2004, Aryasetiawan and co-workers~\cite{cRPA-ferdi-2004} proposed
a scheme nowadays known under the name of ``constrained Random Phase 
Approximation (\crpa)''. The \crpa\ is an 
approximation to a
systematic Wilson-like procedure 
to downfold a system to a low-energy Hamiltonian. 

The \crpa\ method has been implemented within the LMTO in the 
atomic sphere approximation (ASA) 
framework, employing the heads of the LMTOs as local 
orbitals~\cite{cRPA-ferdi-2004,cRPA-solovyev-2005} and within the MLWF 
framework using the full-potential~(FP) LMTO~\cite{cRPA-takashi-2008} 
and the \LAPWlo\ basis~\cite{cRPA-friedrich}. 
Materials-wise, the Coulomb interactions in several families of oxypnictides,
parent compounds of the new Fe-based superconductors have been investigated 
within \crpa\ in the MLWF framework~\cite{cRPA-pnictides-takashi}. 
3\dd, 4\dd\ and 5\dd\ transition metals have been studied 
in Refs.~[\onlinecite{Ufirstprinciples-ferdi-2006,
cRPA-takashi-2009,cRPA-friedrich}]. 
Interestingly, even the pressure dependence of the Coulomb 
interactions becomes accessible within \crpa\ 
\cite{cRPA-pressure-jan}. 
A recent attempt to go beyond the \crpa\ method uses an \LDAU\
electronic structure instead of the LDA one as input to the \crpa\
scheme~\cite{cRPA-LDA+U-ferdi}. This
technique has been applied to NiO and elemental Gd.
 
In recent years, several combined \LDADMFT\ studies have 
appeared, which use the {\it ab initio} Hubbard and Hund
interactions calculated from \crpa. More specifically,
the spectral properties of the parent compounds of iron-based 
superconductors 
LaOFeAs~\cite{cRPA-DMFT-LaOFeAs-markus} and FeSe~\cite{cRPA-DMFT-FeSe-markus},
and of the layered perovskite Sr$_{2}$RuO$_{4}$~\cite{Sr2RuO4-jernej} 
and the spin-orbit Mott insulator Sr$_2$IrO$_4$ 
as well as the related compound Sr$_2$RhO$_4$~\cite{Sr2IrO4-cyril}
have been investigated.
Except for the cases of \srirol and \srrhol\, these works however evaluated the 
Coulomb interactions in the MLWF basis set
whereas the \LDADMFT\ calculations were performed in
a different basis set.

\subsection{Hubbard $U$: physical or auxiliary?}

As alluded to above, the definition of $U$ as the difference
between electron addition and removal energies, directly
accessible in spectroscopy measurements, suggested to consider
the local Hubbard interaction as the central physical quantity
characterising Coulomb interactions in the solid.
A change of paradigm was triggered in 2003, when the proposal
of a combined GW and dynamical mean field scheme (``GW+DMFT'')
\cite{GW+DMFT-biermann,GW+DMFT-kotliar} led to a generalisation of the notion of $U$,
degradating it at the same time -- at least from a conceptual
point of view -- to a purely auxiliary quantity.
Indeed, in the GW+DMFT scheme, a dynamical interaction function
$U_{local}(\omega)$ 
is introduced and (in principle
self-consistently) determined such that the resulting
fully screened Coulomb interaction $W$ takes it physical
value. This is akin to representing the density -- within
density functional theory -- by an auxiliary (Kohn-Sham-)
potential, or the local Green's function in dynamical
mean field schemes by a local impurity bath Weiss function.
Formally, this leads to the interpretation of screening
as a two-step process:
$U$ is interpreted as
the {\it bare} interaction within an effective subsystem
with reduced number of degrees of freedom (in the case
of GW+DMFT, a dynamical local impurity problem).
The latter is
such as to generate the physical screened Coulomb
interaction $W$. 
In the same sense, the cRPA as proposed in 
Ref.~[\onlinecite{cRPA-ferdi-2004}], defines $U$ by the
requirement that the corresponding fully screened
interaction 
\begin{eqnarray}
W= \frac{U}{1-P^{sub}U}
\end{eqnarray} 
takes its physical value, and $P^{sub}$ is the
polarization of the chosen subsystem.
The original cRPA calculates both $W$ and $P^{sub}$
from the random phase approximation, but the
general treatment in Ref.~[\onlinecite{DownfoldedSelfEnergy_Ferdi}]
for example, makes it clear that this is not
an essential ingredient. 
Other choices for $P^{sub}$ have been explored
e.g. in Refs.~[\onlinecite{scGW-kutepov}, \onlinecite{review-imada-takashi},
\onlinecite{PhD-Vaugier},
\onlinecite{nomura-imada-12}].

\subsection{Constrained Screening Approaches (CSA)}

In this paragraph, we give a unified description of
the above type of approaches, which we will
refer to as ``constrained screening approaches'' (CSA).
They have in common to consider Hubbard
interactions $U$ as auxiliary quantities, constructed
in such a way that the physical fully screened Coulomb
interaction $W$ of the original Coulomb problem in the
continuum, possibly within some approximation,
coincides with the fully screened
Coulomb interaction of a chosen subspace when $U$ is
applied as a bare interaction to this subspace.
The first step is thus the choice of a 
subsystem with reduced number of degrees of freedom. 
In the ``GW+DMFT'' approach or in the approach
of Ref.~[\onlinecite{nomura-imada-12}] which directly constructs
an impurity problem, this is a 
local problem. In the 
constrained random phase approximation (\crpa), as 
proposed in Ref.~[\onlinecite{cRPA-ferdi-2004}], this
is a low-energy subspace for which a Hubbard model
is to be defined. Other choices, as for example
subspaces of reduced dimension are also possible;
see e.g. Ref.~[\onlinecite{review-imada-takashi}]
for dimensional downfolding to a 2d-lattice model
or Refs.~[\onlinecite{PhD-Vaugier,nomura-imada-12}] for
the direct construction of a local model.

The requirement of the physical $W$ be represented
by this effective problem (either its local part
only, as in GW+DMFT, or its full momentum-dependence
as in the case of the lattice cRPA) leads to 
a formal relation of $U$ and $W$
involving the polarization $P^{sub}$ of the subsystem:
$U^{-1}- W^{-1}=P^{sub}$, that is, $U$ is the partially screened Coulomb
interaction, screened by all processes present in the
original system except those contained in $P^{sub}$.
For the GW+DMFT scheme, one simply has $P^{sub}= P^{imp}$,
the polarization of the dynamical impurity model. In 
constrained screening approaches geared at the construction
of a lattice model (rather than a local one) 
one sets $P^{sub}$ equal to the polarization $P^{d}$ of 
a low-energy correlated subspace.
For obvious reasons, in practical calculations equivalent
approximations should be made on $W$ and $P$.
The constrained random phase approximation \cite{cRPA-ferdi-2004}, can thus be
viewed as a special case, in which both, $W$ and $P^{sub}$
are calculated within the random phase approximation.

In practice, we start by choosing a set of degrees 
of freedom around the Fermi level in order to generate a correlated 
subspace $\CC$ of the full Hilbert space. 
The conceptual division of the Hilbert space into 
a reference space $\CC$ and the remaining degrees of freedom leads 
to the following decomposition of the response of the system to
an external perturbation, expressed by the polarization $P$:
\begin{eqnarray} 
  P&=& 
P^{sub}+P^{r},
\label{eq:Pcrpa1}
\end{eqnarray}
Here, $P^{sub}$ denotes the polarization within the correlated subspace 
$\CC$, and, by definition, $P^{r}=P-P^{sub}$ is a constrained polarization, 
in which the contributions of the target degrees of freedom have been 
projected out. The constrained polarization leads to the partial 
dielectric function $\epsilon^{r}$ 
\begin{eqnarray} \label{eq:epsr}
  \epsilon^{r}(1,2) &=& \delta(1-2) - \int d3 \, P^{r}(1,3)v(3,2).
\end{eqnarray}
We use here a shorthand notation that consists in representing time 
and
possible spatial degrees of freedom by a number. In the case of
the usual lattice cRPA, for example, this number would thus
represent ($\rrr \, \tau$), in the case of local subsystems
it would be the time variable $\tau$ only.
The \emph{partially} screened interaction $W^{r}$ can then be 
defined as follows: 
\begin{eqnarray} \label{eq:Wr}
  W^{r}(1,2) &\equiv& \int d3 \, \epsilon_{r}^{-1}(1,3) v(3,2).
\end{eqnarray}
A simple algebraic manipulation 
shows that further screening of $W^{r}$ by the polarization $P^{sub}$,
or, equivalently, by the low-energy effective dielectric
function
\begin{eqnarray} \label{eq:epsd}
  \epsilon_{sub}(1,2) &=& \delta(1-2) - \int d3 \, P^{sub}(1,3)W^{r}(3,2).
\end{eqnarray}
allows to retrieve the fully screened interaction $W$:
\begin{eqnarray} \label{eq:Wcrpa2}
  W(1,2) &\equiv& \int d3 \, \epsilon_{sub}^{-1}(1,3) W^{r}(3,2).
\end{eqnarray}
This property of $W^{r}$ suggests that it can be identified as 
the effective bare interaction
within the 
low-energy subspace.
More precisely, matrix elements of the static value of
$W^{r}$ in the localized basis 
set can be interpreted as forming the interaction 
matrices~:
\begin{eqnarray}
U_{....} & \equiv & \langle .. | W^{r}|.. \rangle.
\end{eqnarray}

The following remark is in order here: in fact, 
the obtained interactions are \emph{frequency-dependent} because of 
the frequency dependence of the polarization $P^{r}$ (Eq.~\ref{eq:Wr}). 
This is a physical effect, since the response of the electrons in the 
solid to an external perturbation depends on the frequency of the latter. 
In particular, the electrons do not respond to a high-frequency 
oscillating electric field whose frequency exceeds any electronic 
energy scale, that is, the electronic polarizability vanishes at high 
frequency and the partially screened Coulomb interaction $W^r$ coincides, 
in this limit, with the bare Coulomb interaction. 
\crpa\ in principle provides us with the spectrum of the dynamical 
interaction whose combination with many-body solvers 
such as extended dynamical mean field theory
is currently receiving much attention~\cite{udyn-michele,udyn-werner, udyneff-michele}. 
In particular, it was shown how inclusion of the high-energy tail
of the energy-dependent Coulomb interaction leads, in the spectral
functions, to pronounced shifts of weight to higher energies and
a concomitant additional renormalization at low energies
\cite{udyn-michele,udyn-werner}. 
Recently, an explicit expression for these renormalisations was
derived \cite{udyneff-michele}, yielding a prescription for the
construction of a low-energy Hubbard-like model with static Hubbard
and Hund interactions where the high-energy screening effects are
already incorporated. Specifically, it was argued that the static
interaction matrices to be used in such a construction are
precisely the ones obtained from constrained screening
approaches in the static limit ($\omega \rightarrow 0$).
These are the subject of the present work.

\section{The constrained Random Phase Approximation}

\subsection{Constrained Polarization}
While the conceptual construction of the Hubbard interactions is
general and in principle
independent of the specific choice of the subsystem,
the construction acquires its sense from a Wilson-like
argument, when
the reference system is a low-energy subspace:
$P^{sub}=P^{d}$, where the subscript $d$ refers to a low-energy
correlated target space.
The constrained random phase approximation then corresponds
to the choice of calculating both, the polarization of this subspace
and the total polarization within the RPA approximation.
If we 
assume that the low-energy bands that span the correlated subspace 
do not energetically overlap with the remaining bands, the $\CC$-restricted 
polarization reads 
\begin{eqnarray}
&&P^{d}(\rrr,\rrr';\omega)\label{Pcrpa2}
   =\sum_{\kkk, d}^{\occ} \sum_{\kkk', d'}^{\unocc} \psi_{d \kkk}^{*}(\rrr) 
\psi_{d' \kkk'}(\rrr) \psi_{d' \kkk'}^{*}(\rrr') \psi_{d \kkk}(\rrr') \nonumber \\
&& ~~~\times \left\lbrace 
          \frac{1}{\omega - \epsilon_{d' \kkk'}+\epsilon_{d \kkk} +i\eta} 
        - \frac{1}{\omega+\epsilon_{d' \kkk'}-\epsilon_{d \kkk}-i\eta} 
     \right\rbrace,
\end{eqnarray}
where $\epsilon_{d \kkk}$ are the energies of the Bloch states, 
$\psi_{d \kkk}(\rrr)$, that span $\CC$. 
The polarization $P^{r}$ is then deduced from Eq.~(\ref{eq:Pcrpa1})
with $P^{sub}= P^{d}$.

\subsection{Definition of the correlated subspace}

The choice of the ``target'' degrees of freedom -- or low-energy 
subspace -- is obviously not unique. Two different ``downfolded'' 
Hamiltonians will not lead to the same effective interactions. 
However, both choices should yield the same results for physical 
observables at the end, under the condition that both models 
are appropriate for catching the relevant physics of the system. 

In the following, we assume that the band structure -- as in the 
perovskite
materials considered in this paper -- is such that the orbital
character of the ``correlated species'' (here the 3$d$ or 4$d$ orbitals)
is dominantly spread over a subset of bands that are not entangled 
with bands of other orbital characters.
In other words, the Hilbert space 
can be split into a correlated and an itinerant subspace that 
-- for every k-point in the first Brillouin zone -- are separated
in energy.
The situation of entangled subspaces requires 
specific treatments that will be reported elsewhere 
(see also Refs.~[\onlinecite{cRPA-takashi-2009,cRPA-friedrich,
PhD-Vaugier}]).

Our implementation is based on the full-potential \LAPWlo\ code
\textsc{Wien2k} \cite{blaha_wien2k}, and 
the construction of localized orbitals according to the procedure
implemented 
in Ref.~[\onlinecite{cRPA-DMFT-LaOFeAs-markus}]. 
In the case of separated bands in the sense described above, 
this construction results in a set
of Wannier functions $\{ |\phi_{\kkk m}^{\alpha \sigma} \rangle\}$ 
that span the ``target'' 
space $\CC$.
The index $m$ denotes an orbital quantum number, $\alpha$ is an atomic 
index inside a given unit cell centered at
$\RRR$ and $\sigma$ is the spin degree 
of freedom. 

The practical construction of the orbitals
$\{ |\phi_{\kkk m}^{\alpha \sigma} \rangle\}$ 
proceeds as follows: 
one starts by choosing the dominant orbital character
of the states that are to be considered as correlated,
as well as an energy window $\WW$ that contains at least all
bands whose majority character is the correlated one.
One then uses a set of localized atomic orbitals of the
chosen character 
$\{|\chi_{\kkk m}^{\alpha \sigma}\rangle \}$ 
and calculates their projection onto the
Bloch states 
that lie in the energy window $\WW$:
\begin{eqnarray}
|\tilde{\chi}_{\kkk m}^{\alpha \sigma} \rangle 
&=& \sum_{\nu \in \WW} \langle \psi_{\kkk\nu}^{\sigma}|
\chi_{m}^{\alpha \sigma} \rangle |\psi_{\kkk \nu}^{\sigma} \rangle.
\label{Wproj1}
\end{eqnarray}
The set of wave functions 
$\{|\tilde{\chi}_{\kkk m}^{\alpha \sigma}\rangle \}$ 
is then subject to an orthonormalization procedure,
yielding the 
orthonormal Wannier basis
$\{ |\phi_{\kkk m}^{\alpha \sigma} \rangle\}$. 
It is an easy task to verify that this set is indeed
complete within the selected subspace.

In order to be able to transform back and forth between
the Kohn-Sham and the Wannier basis, a set of 
\emph{``Wannier'' projectors} $P_{m \nu}^{\alpha \sigma}(\kkk)$ 
are defined as follows~:
\begin{eqnarray}
  P_{m, \nu}^{\alpha \sigma}(\kkk) \equiv \langle \phi_{\kkk m}^{\alpha \sigma} | \psi_{\kkk \nu}^{\sigma} \rangle,
\label{Cprojector1}
\end{eqnarray} 
and with simplified notations, $L=(m,\alpha,\sigma)$, 
\begin{eqnarray}
|\phi_{\kkk L} \rangle &=& \sum_{\nu \in \WW} P_{L \nu}^{*}(\kkk) | \psi_{\kkk \nu}\rangle.
\label{Cprojector2}
\end{eqnarray}
The atom-centered Wannier functions in which the Hubbard model
is formulated are finally obtained from a simple
Fourier transformation from the reciprocal to the direct space:
\begin{eqnarray}
|\phi_{\RRR L} \rangle &=& \frac{1}{\sqrt{\NN}} \sum_{\kkk} \expo^{-i\kkk \cdot \RRR} |\phi_{\kkk L} \rangle,
\label{Wfourier}
\end{eqnarray}
where $\NN$ corresponds to the number of $\kkk$ vectors inside the first Brillouin zone. 

As mentioned before, in general the choice of the different subspaces
is not unique. Depending on the material and the physical quantities
considered, one has to decide for example whether to include ligand
states in the model or not.
For the oxide compounds considered in this work, two natural choices
appear.
We use the notations previously introduced in Ref.~[\onlinecite{cRPA-LaOFeAs-miyake}]. 

\textbf{``\t2g-\t2g\ Hamiltonian''}: 
Due to the 
octahedral crystal field of the oxygen ligands, the \dd\ states of 
the metal ion M$^{4+}$ in \srmo\ are split into \t2g\ and \eg\ bands. 
The \t2g\ bands are partially filled (with one electron in \srvo\,
three electrons in \srmno) whereas the \eg\ states are empty. 
In a quite intuitive way, one can choose the \t2g\ bands as the 
low-energy degrees of freedom in the \crpa\ downfolding procedure. 
The correlated subspace $\mathcal{C}_{\t2g}$ thus includes only the 
\t2g\ degrees of freedom.
To construct this subspace explicitly, a set of Wannier functions with the \t2g\ character is constructed out of Kohn-Sham bands within the energy window $\WW_{\t2g}$. The energy window $\WW_{\t2g}$ is chosen such that only \t2g\ bands are included. To calculate the constrained polarization $P^{r}$,
transitions from and to the target \t2g\ bands included in $\WW_{\t2g}$ are removed from the total polarization
(Eq.~\ref{eq:Pcrpa1}).
A method based on the Kohn-Sham indices or on an energy window can be employed to label the transitions that have to be eliminated, since the \t2g\ bands do not energetically overlap with \pp\ and \eg\ ones.

We finally obtain the interaction parameters that correspond to a $\CC_{\t2g}$-restricted lattice Hamiltonian.
We call this model the ``\t2g-\t2g\ Hamiltonian'', where the first index
refers to the majority character of the bands that are cut out
from the screening process when going from $P$ to $P^{r}$, and
the second refers to the majority character of the bands that
are used to construct the Wannier functions in which matrix
elements are taken.

\textbf{``\dd-\dd\pp\ Hamiltonian''}: Alternatively, one can choose as
correlated subspace $\mathcal{C}_{dp}$ the space that contains also
the \eg\ degrees of freedom in addition to the \t2g\ ones. In an
octahedral crystal field, however, the (\dz2,\dx2y2) orbitals strongly
hybridize with the ligands, forming bonding and anti-bonding bands.
The former are dominantly of ligand-\pp\ character, with substantial
weight of the transition metal $d$ contribution, however, while the
latter are dominantly formed by the \eg\ states, with admixture
from ligand $p$ states. This situation suggests to use for the
construction of the Wannier functions an extended \dd\pp\ energy
window $\WW_{dp}$ that includes the whole \dd\ manifold as well as
the \pp\ one.
It could thus appear natural in this context to construct a low-energy
effective Hamiltonian, where both, $d$- and $p$-states, are treated as
correlated states. Treating Coulomb interactions between the in
general rather extended $p$ electrons as local, however, seems to
be a more drastic approximation than treating these interactions
in a static mean field fashion.
In practice, it is thus more common to describe oxides within a
hybrid model, where the Wannier functions are constructed within
an extended \dd\pp\ window, while applying Hubbard interaction
terms only to the $d$-manifold.
Analogous constructions have for example been performed for
iron pnictide compounds
\cite{cRPA-LaOFeAs-miyake}
where the Fe-$d$ states were considered as correlated while ``uncorrelated''
pnictogen or oxygen states had been included for the sake of the
construction of sufficiently localized Wannier functions.
The philosophy that is currently taken in this context is
then to construct the constrained polarization $P^{r}$
from the total polarization by cutting off the transitions
from and to the target \dd\ bands.
The low-energy subspace described by the Hamiltonian is
however composed of the full $d$- and $p$-manifolds.
Following the notation of reference
\cite{cRPA-LaOFeAs-miyake}
we call this hybrid model the ``\dd-\dd\pp\ Hamiltonian''.

\subsection{Calculated quantities}

\subsubsection{Definition of matrix elements}

The \crpa\ method allows to calculate the effective interaction matrix elements of a 
low-energy lattice Hamiltonian expanded in the localized basis set, 
$\{ |\phi_{\RRR L} \rangle \}$~:
\begin{eqnarray}
&& U_{L_{1}L_{2}L_{3}L_{4}}^{\RRR_{1}\RRR_{2}\RRR_{3}\RRR_{4}}(\omega) \equiv \langle \phi_{\RRR_{1}L_{1}} \phi_{\RRR_{2}L_{2}}|W^{r}(\omega) | \phi_{\RRR_{3}L_{3}} \phi_{\RRR_{4}L_{4}}\rangle \nonumber \\
&& \\
&& = \iint d\rrr d\rrr' \phi_{\RRR_{1}L_{1}}^{*}(\rrr) \phi_{\RRR_{3}L_{3}}(\rrr) W^{r}(\rrr,\rrr';\omega) \nonumber \\
&& ~~~~~~~~~~~~~~~~~~~~~~~~~ \times \phi_{\RRR_{2}L_{2}}^{*}(\rrr') \phi_{\RRR_{4}L_{4}}(\rrr').
\label{Wcrpa3}
\end{eqnarray} 
By expanding the localized orbitals into the Kohn-Sham states within the energy window $\WW$, the interaction matrix elements read (Appendix A)~:
\begin{eqnarray}
&& U_{L_{1}L_{2}L_{3}L_{4}}^{\RRR_{1}\RRR_{2}\RRR_{3}\RRR_{4}}(\omega) = \NN^{-2}\sum_{\kkk_{1}\kkk_{2}\kkk_{3}\kkk_{4}}\expo^{i(\kkk_{1}\RRR_{1}-\kkk_{3}R_{3}+\kkk_{2}\RRR_{2}-\kkk_{4}\RRR_{4})} \nonumber \\
&&  \times ~~~~~~~~~~~~~~~~~~~~~~~~~~ \sum_{\nu_{1}\nu_{2}\nu_{3}\nu_{4} \in \WW} P_{L_{1}\nu_{1}}(\kkk_{1})[P_{L_{3}\nu_{3}}(\kkk_{3})]^{*} \nonumber \\
&&  \times \langle \psi_{\kkk_{1}\nu_{1}}\psi_{\kkk_{2}\nu_{2}}|W^{r}(\omega)| \psi_{\kkk_{3}\nu_{3}}\psi_{\kkk_{4}\nu_{4}}\rangle [P_{L_{4}\nu_{4}}(\kkk_{4})]^{*}P_{L_{2}\nu_{2}}(\kkk_{2}). \nonumber \\
&& 
\label{Wcrpa4}
\end{eqnarray}
Analogously, one can compute the fully screened interaction, 
\begin{eqnarray}
 W_{L_{1}L_{2}L_{3}L_{4}}^{\RRR_{1}\RRR_{2}\RRR_{3}\RRR_{4}}(\omega) &\equiv& \langle \phi_{\RRR_{1}L_{1}} \phi_{\RRR_{2}L_{2}}|W(\omega) | \phi_{\RRR_{3}L_{3}} \phi_{\RRR_{4}L_{4}}\rangle, \nonumber \\
&&
\end{eqnarray}
and the bare interaction, 
\begin{eqnarray}
v_{L_{1}L_{2}L_{3}L_{4}}^{\RRR_{1}\RRR_{2}\RRR_{3}\RRR_{4}} &\equiv &\langle \phi_{\RRR_{1}L_{1}} \phi_{\RRR_{2}L_{2}}|v | \phi_{\RRR_{3}L_{3}} \phi_{\RRR_{4}L_{4}}\rangle.
\end{eqnarray}

While the procedure is in principle directly applicable also to 
non-local interactions, we will restrict the discussion
in the following to local (that is intra-atomic) interactions 
($\RRR_{1}=\RRR_{2}=\RRR_{3}=\RRR_{4}$). 
The interaction matrix elements are denoted 
$U_{m_{1}m_{2}m_{3}m_{4}}^{\mathcal{S}}$ where $m$ refers to angular 
quantum numbers indicating the dominant orbital character
of the local basis set
and $\mathcal{S}$ to the symmetry of the crystal field used 
for constructing the localized basis $\{ |\phi_{\RRR L}\rangle\}$
(in our case, cubic harmonics or spherical harmonics).
The four-index interaction matrix within the cubic harmonics,
$U_{m_{1}m_{2}m_{3}m_{4}}^{\textrm{cubic}}$, includes in principle
not only density-density interaction terms.
Reduced interaction matrices, $U^{\sigma \bar{\sigma}},U^{\sigma \sigma}$, for the density-density interaction with cubic symmetry between opposite spin and parallel spin electrons respectively, can be defined as follows~:
\begin{eqnarray}
&& U_{mm'}^{\sigma \bar{\sigma}} \equiv U^{\textrm{cubic}}_{mm'mm'} = \langle \phi_{m} \phi_{m'} | W^{r}(0) | \phi_{m} \phi_{m'}\rangle \label{reducedU-1}\\
&& J_{mm'}  \equiv U^{\textrm{cubic}}_{mm'm'm} = \langle  \phi_{m} \phi_{m'} | W^{r}(0) | \phi_{m'} \phi_{m}\rangle  \label{reducedU-2}  \\
&& U_{mm'}^{\sigma \sigma} \equiv U^{\textrm{cubic}}_{mm'mm'} - J_{mm'}.  \label{reducedU-3}
\end{eqnarray}

For practical reasons, and in order to simplify its further
use within many-body techniques, it is convenient to 
parametrize these matrices by a reasonable number of parameters. 
Common practice and physical insight suggests the popular
choice of the Slater parametrization. 
Based on the assumption of spherical symmetry, this form is strictly 
valid only for atoms, but yields -- for sufficiently localized 
atomic-like states -- a good approximation for the correlated
shells of a solid.

\subsubsection{Slater integrals}
The parametrization by Slater integrals~\cite{judd,Slater_book,Sugano_book} 
makes use of the specific form of the angular momentum integrals in the
atomic case, reducing the problem to the calculation of a restricted
number of radial integrals (3 in the case of a $d$ shell, 4 in the case of a $f$ shell).
Useful relations are summarized in Appendix B. 
The Slater integrals $\textrm{F}^{k}(\omega)$ can be employed 
for building the symmetrized interaction matrix 
$\bar{U}_{m_{1}m_{2}m_{3}m_{4}}^{\textrm{cubic}}(\omega)$ with cubic symmetry 
as follows:
\begin{eqnarray}
&& \bar{U}_{m_{1}m_{2}m_{3}m_{4}}^{\textrm{cubic}}(\omega) = \sum_{m_{1}'m_{2}'m_{3}'m_{4}'}\mathcal{S}_{m_{1}m_{1}'} \mathcal{S}_{m_{2}m_{2}'} \nonumber \\
&& \times \bigg\{ \sum_{k=0}^{2l} \alpha_{k}(m_{1}',m_{2}',m_{3}',m_{4}') \textrm{F}^{k}(\omega) \bigg\} \mathcal{S}_{m_{3}'m_{3}}^{-1} \mathcal{S}_{m_{4}'m_{4}}^{-1}, \nonumber \\
&&
\label{slaterFU2}
\end{eqnarray}
where $\alpha_{k}$ are the Racah-Wigner numbers (Appendix B) 
and $\mathcal{S}$ is the transformation matrix from spherical 
harmonics to cubic ones. 

Different conventions exist for the Hubbard $U$ parameter.
We follow
the convention of identifying $U$ with the
Slater integral $\textrm{F}^{0}$ (Refs.~[\onlinecite{LDA+U-anisimov-NiO,LDA+U-anisimov-1997}]) 
related to 
matrix elements of the interaction as a simple average over
all possible pairs of orbitals:
\begin{eqnarray}
U &\equiv& \frac{1}{(2l+1)^{2}} \sum_{mm'} \sum_{k=0}^{2l} \alpha_{k}(m,m',m,m') \textrm{F}^{k} \\
 &=& \textrm{F}^{0}. \label{UF0}
\end{eqnarray}
Hund's exchange $J$ is given by \cite{LDA+U-anisimov-NiO}~:
\begin{eqnarray}
J &\equiv& \frac{1}{2l(2l+1)} \sum_{m \neq m'} \sum_{k} \alpha_{k}(m,m',m',m) \textrm{F}^{k}. 
\end{eqnarray} 
Using Slater integrals manipulations, it can be shown that~:
\begin{equation}
J = \left\{ \begin{array}{ll}
                     (\textrm{F}^{2}+\textrm{F}^{4})/14 & \quad \textrm{(for \dd\ shells)} \label{JF2F4} \\
                     (286\textrm{F}^{2}+195\textrm{F}^{4} + 250\textrm{F}^{6})/6435 & \quad \textrm{(for \ff\ shells),}
                 \end{array} \right.
\end{equation}

Analogously, one can define bare parameters, $v$ and 
$J_{\textrm{bare}}$, and fully screened ones, $W$ and 
$J_{\textrm{screened}}$, using the appropriate Slater integrals. 
The underlying assumption of this construction is that
all three, bare, partially screened and fully screened
interactions are spherically symmetric, that is 
keep the form imposed by the purely
atomic angular momentum integrals. In general, this
can be expected to be a good approximation for the
bare interaction if the orbitals are sufficiently
localized and thus atomic-like. In how far it is a
good description for the partially and fully screened
interaction depends on in how far screening acts as
to break the spherical symmetry,
inducing additional anisotropies through hybridizations
in the local environment in the solid.

Based on Eq.~\ref{slaterFU2}, one can extract Slater-symmetrized effective interactions for the whole \dd\ shell, corresponding to density-density interactions within the \t2g\ subspace, the \eg\ subspace and also between the \t2g\ and \eg\ \cite{Sugano_book}. For further comparison with the interactions between \t2g\ orbitals but within the \t2g-\t2g\ model, we will restrict ourselves to the Slater-symmetrized interactions for the \t2g\ subspace only~:  
\begin{eqnarray}
\bar{U}_{mm} &=& \textrm{F}^{0} + \frac{4}{49}\textrm{F}^{2} + \frac{4}{49}\textrm{F}^{4} \label{slater4} \\
\bar{U}_{mm'} &=& \textrm{F}^{0} - \frac{2}{49}\textrm{F}^{2} -\frac{4}{441}\textrm{F}^{4} \label{slater5} \\
\bar{J}_{m} &\equiv& \frac{1}{2}(\bar{U}_{mm} - \bar{U}_{mm'}) = \frac{3}{49}\textrm{F}^{2} + \frac{20}{441}\textrm{F}^{4}. \label{slater6}
\end{eqnarray}  

\subsubsection{Hubbard-Kanamori parameters}

In the case of large crystal field splittings, the assumption of 
spherical symmetry of the correlated states is no longer valid,
and
the appropriate parameters for constructing the \t2g-restricted lattice Hamiltonian (\t2g-\t2g) are the Hubbard-Kanamori parameters \cite{Sugano_book}. They are directly calculated from the interaction matrix elements in the $\CC_{\t2g}$ subspace as follows~:
\begin{eqnarray}
\mathcal{U} &=& \frac{1}{N} \sum_{m=1}^{N=3} U^{\textrm{cubic}}_{mmmm} \label{slater7} \\
\mathcal{J} &=& \frac{1}{N(N-1)} \sum_{m \neq m'}^{N=3} U^{\textrm{cubic}}_{mm'm'm} \label{slater8} \\
\mathcal{U}' &=& \frac{1}{N(N-1)} \sum_{m \neq m'}^{N=3} U^{\textrm{cubic}}_{mm'mm'} \label{slater9},
\end{eqnarray}
Analogously, one can define bare parameters, $(\mathcal{V},\mathcal{J}_{\textrm{bare}})$, and fully screened ones, $(\mathcal{W},\mathcal{J}_{\textrm{screened}})$, by considering the bare and fully screened interaction matrices, $v^{\textrm{cubic}}$ and $W^{\textrm{cubic}}$, respectively.

\subsubsection{Calculated quantities: summary of notations}

The Hubbard interaction matrix elements with cubic symmetry are denoted $U_{m_{1}m_{2}m_{3}m_{4}}^{\textrm{cubic}}$ (Eq.~\ref{Wcrpa4}), and the corresponding reduced interaction matrices, $U_{mm'}^{\sigma \bar{\sigma}},U_{mm'}^{\sigma \sigma}$ (Eqs. \ref{reducedU-1} and \ref{reducedU-3}). 

Within the \dd-\dd\pp\ Hamiltonian, the average interaction matrix elements between \t2g\ orbitals that are directly calculated with cubic symmetry are denoted $U_{mm}, U_{mm'}$ and $J_{m}$ for the on-site intra-\t2g, inter-\t2g\ and exchange interaction, respectively. Analogous quantities, but for the \emph{bare} situation, $v_{mm}, J^{\textrm{bare}}_{m}$, and for the \emph{fully screened} situation, $W_{mm}, J^{\textrm{screened}}_{m}$, can be introduced. 

We deduce from the Slater integrals parametrization, the Hubbard parameter $U\equiv \textrm{F}^{0}$ (Eq. \ref{UF0}) and Hund's exchange $J \equiv (\textrm{F}^{2}+\textrm{F}^{4})/14$ for the $d$ shell (Eq. \ref{JF2F4}). They are the parameters that should be considered for constructing the interaction Hamiltonian downfolded into the \dd-\dd\pp\ low-energy subspace chosen (Eq. \ref{slaterFU2}). Analogous parameters but within \emph{bare} (unscreened) repulsions, 
$v$ and $J_{\textrm{bare}}$, or within \emph{fully screened} repulsions, $W$ and 
$J_{\textrm{screened}}$, are introduced. 

Employing Eqs. \ref{slater4}, \ref{slater5} and \ref{slater6}, Slater-average on-site, $\bar{U}_{mm}$, and exchange, $\bar{J}_{m}$, interactions between \t2g\ orbitals within the \dd-\dd\pp\ Hamiltonian, can be extracted -- and analogously for the \emph{bare}, $\bar{v}_{mm},\bar{J}_{m}^{\textrm{bare}}$ and the \emph{fully screened} $\bar{W}_{mm},\bar{J}_{m}^{\textrm{screened}}$ cases. They correspond to a part only -- the \t2g\ one -- of the \dd-\dd\pp\ low-energy Hamiltonian and therefore, should not be taken for parametrizing this Hamiltonian.

Within the \t2g-\t2g\ Hamiltonian, the Hubbard-Kanamori terms, $\mathcal{U},\mathcal{U}',\mathcal{J}$ (Eqs. \ref{slater7}, \ref{slater8} and \ref{slater9}), refer to the interactions between \t2g\ orbitals within the \t2g-restricted Hamiltonian. \uk\ corresponds to the intra-orbital Coulomb repulsion, whereas $\mathcal{U}'(=\mathcal{U}-2\mathcal{J}$ with cubic symmetry$)$ is the inter-orbital interaction which is reduced by Hund's exchange, $\mathcal{J}$. They are the appropriate parameters for mapping the low-energy Hamiltonian downfolded into the \t2g\ subspace. Analogous parameters but within \emph{bare} repulsions, $\mathcal{V},\mathcal{J}_{\textrm{bare}}$, or within \emph{fully screened} 
repulsions, $\mathcal{W},\mathcal{J}_{\textrm{screened}}$, are also introduced. 

\begin{table}
\caption{Lattice parameters used for cubic perovskites \srmo\ and energy windows $\WW$ (in eV) for the \dd-\dd\pp\ and the \t2g-\t2g\ models. \dd\ and \t2g\ Wannier-like functions are constructed out of the Kohn-Sham states included in $\WW$. Because of the entanglement of the \eg\ states with Sr-like \dd\ states, the \dd-\dd\pp\ model is not considered for 4\dd\ \srmo.}
\label{tab:windows113}
\begin{center}
\begin{tabular*}{0.48\textwidth}{@{\extracolsep{\fill}}lccc}
\hline \hline
              & a (\AA) &  $\WW_{dp}$ & $\WW_{\t2g}$ \\ \hline 
SrVO$_{3}$    &  3.842  &  $[-7.5,5.5]$  & $[-1.8,1.8]$   \\ 
SrCrO$_{3}$   &  3.820  &  $[-7.5,4.7]$  & $[-1.7,1.0]$      \\
SrMnO$_{3}$   &  3.805  &  $[-7.5,4.2]$  & $[-1.7,1.0]$  \\  \hline
SrNbO$_{3}$   &  3.997  &                       & $[-3.0,2.8]$ \\
SrMoO$_{3}$   &  3.976  &                       & $[-3.0,2.0]$ \\ 
SrTcO$_{3}$   &  3.950  &                       & $[-2.6,1.3]$ \\ \hline \hline
\end{tabular*}
\end{center}
\end{table}

\begin{figure*}
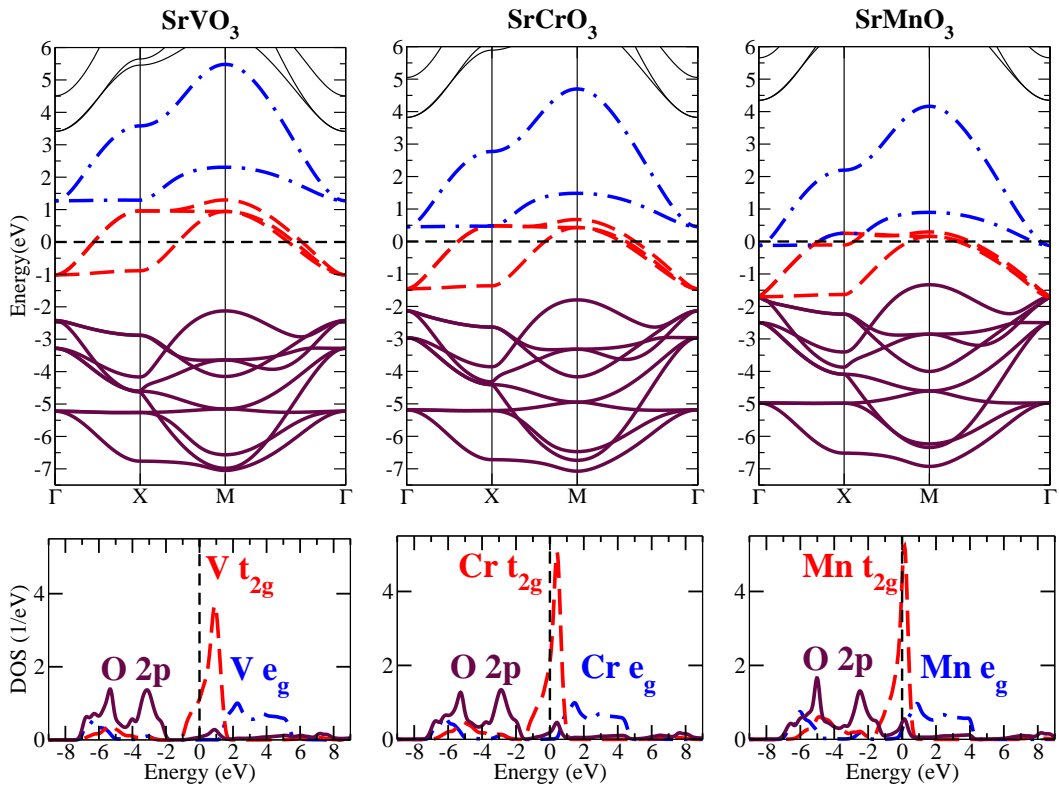

\begin{center}
\includegraphics[width=1.8in,clip]{Fig1a.eps}
\includegraphics[width=1.8in,clip]{Fig1b.eps}
\includegraphics[width=1.8in,clip]{Fig1c.eps}
\end{center}
\begin{center}
\includegraphics[width=1.8in,clip]{Fig1d.eps}
\includegraphics[width=1.8in,clip]{Fig1e.eps}
\includegraphics[width=1.8in,clip]{Fig1f.eps}
\end{center}
\caption{(color online) Electronic band structures (top) and projected
  density of states (bottom) of \srmo\ (M = V, Cr, Mn from left to
  right) obtained from LDA in the paramagnetic phase. The 3\dd\ \t2g\
  states are highlighted in red (dash), the 3\dd\ \eg\ in blue
  (dash-dot) and the oxygen \pp\ states in maroon (solid).}
\label{fig:3dKS}
\end{figure*}

\section{Hubbard interactions in perovskite transition-metal oxides  \srmo\ (M = V, Cr, Mn, Nb, Mo, Tc)}

\subsection{General trends}

\begin{figure*}
\begin{center}
\includegraphics[width=1.8in,clip]{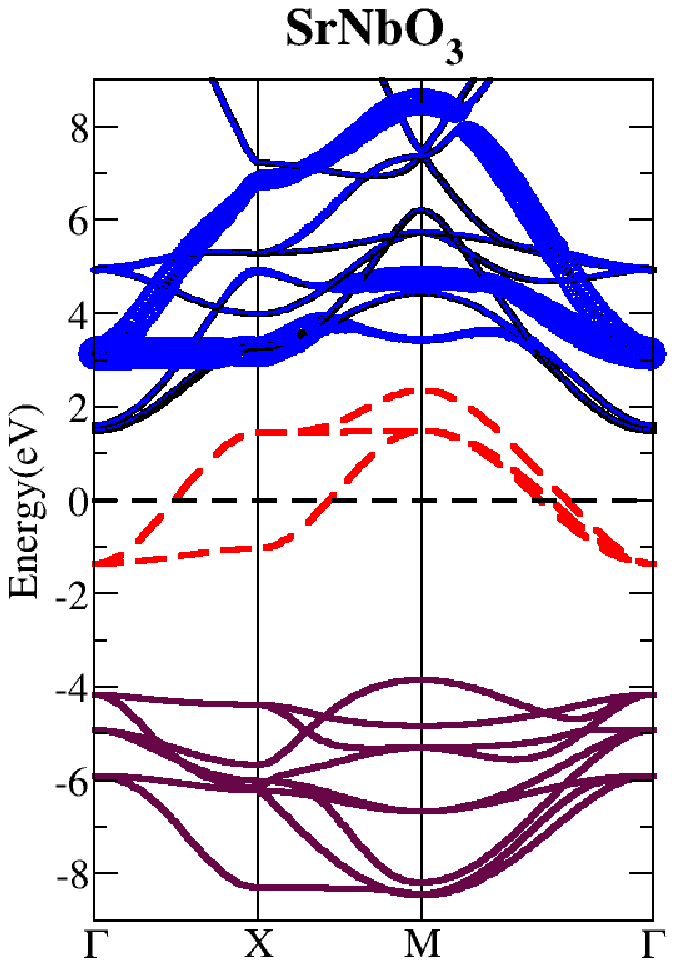}
\includegraphics[width=1.8in,clip]{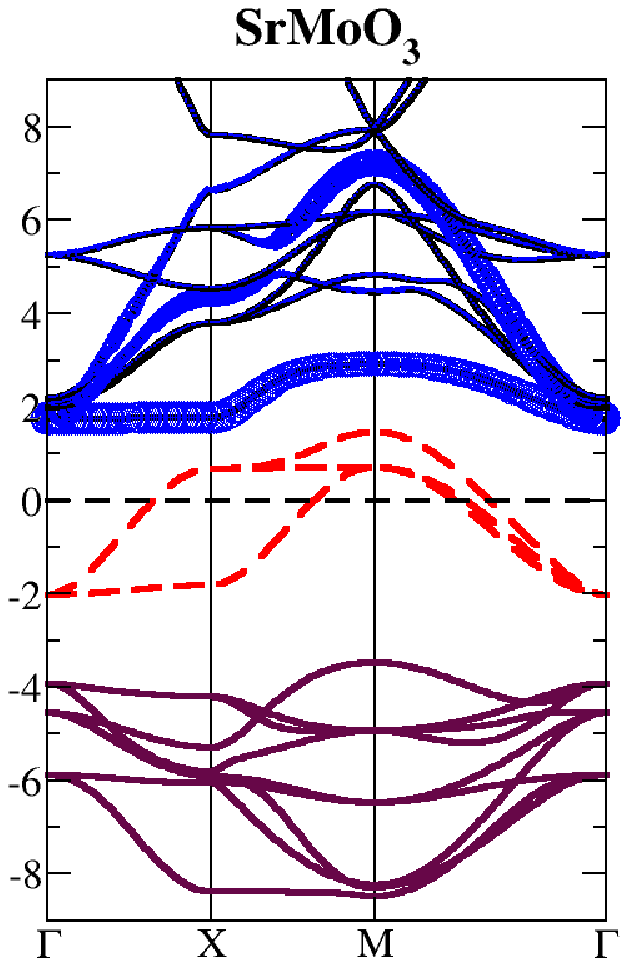}
\includegraphics[width=1.8in,clip]{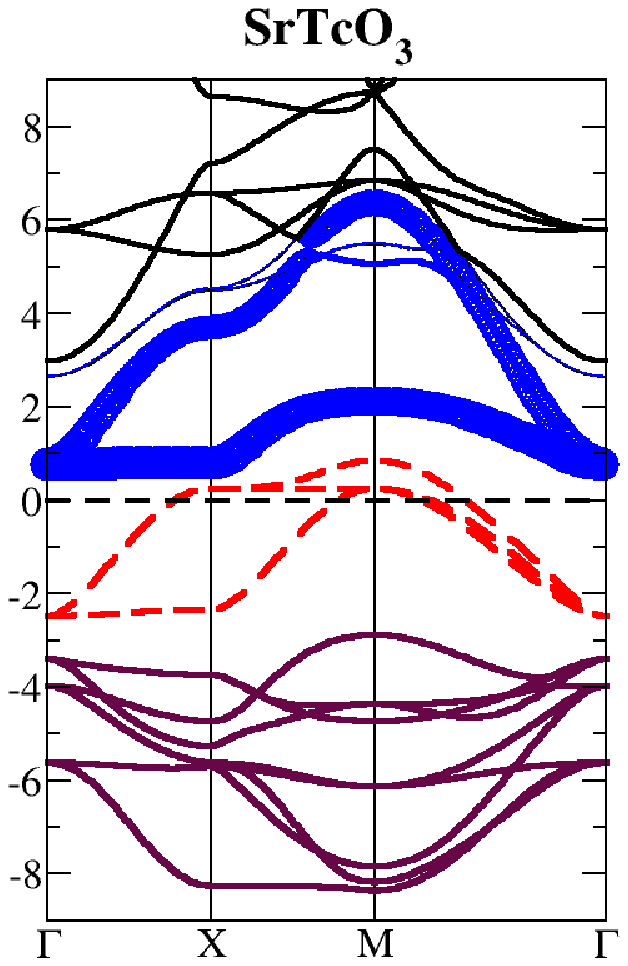}
\includegraphics[width=1.8in,clip]{Fig2d.eps}
\includegraphics[width=1.8in,clip]{Fig2e.eps}
\includegraphics[width=1.8in,clip]{Fig2f.eps}
\caption{\footnotesize{(color online) 
Electronic band structures (top) and projected density of states
(bottom) of \srmo\ (M = Nb, Mo, Tc from left to right) obtained from
LDA in the paramagnetic phase. The 4\dd\ \t2g\ states are highlighted
in red (dash), the 4\dd\ \eg\ states in blue (dash-dot) and the oxygen
\pp\ states in maroon (solid).}}
\label{fig:4dKS}
\end{center}
\end{figure*}

We first consider the 
ternary transition metal oxides SrMO$_{3}$ 
(M=V, Cr, Mn, Nb, Mo, and Tc) in an undistorted 
cubic perovskite structure.
This is the stable structure for \srvo,
while for the other compounds this is an
idealisation.
In this structure, 
the transition metal ion is octahedrally coordinated
with oxygen ligands,
inducing a splitting of the \dd\ orbitals of the 
transition metal into \t2g\ and \eg, where the 
\eg\ form bonding and anti-bonding states with the 
oxygen \pp. The \t2g\ states form a relatively narrow 
partially filled band, separated in energy
both from the \eg\ and the \pp\ states.
There are two important factors that 
characterise trends in
the electronic structure of transition 
metal (TM) perovskites: 
1) the localization of TM-$d$ orbitals, which increases 
as the atomic number increases within the same period of 
the periodic table, and decreases as the atomic number 
increases in the same group, and 
2) the hybridization between transition metal $d$ and 
O-$2p$ states, which depends on the localization of the 
$d$ orbitals as well as on the energetic splitting between 
$d$ and O-$2p$, the charge transfer energy and ligand field 
$\Delta_{pd}$. The interplay of these two factors, together 
with the filling of the $d$ manifold, gives rise to extremely 
intriguing physics and chemistry. The 
DFT-LDA band structures of the 3\dd\ and 4\dd\ transition metal 
perovskites 
in their paramagnetic phases are shown in Figs. \ref{fig:3dKS} 
and \ref{fig:4dKS}, respectively. The lattice parameters used 
in the calculations are listed in Table ~\ref{tab:windows113}. 

Due to the contraction of the \dd\ orbitals with increasing atomic 
number, the hybridization between \dd\ and \pp\ orbitals lessens 
from the early to the late oxides. Similarly, the charge transfer 
energy between \dd\ and \pp\ bands, $\Delta_{pd}$, also decreases, 
in qualitative agreement with optics 
experiments~\cite{Lee-SrMO3,torrance-deltapd}. On the other hand, 
the charge transfer energy is larger in 4\dd\ than in 3\dd, due to 
the larger orbital extension of the former (Fig.~\ref{fig:4dKS}). 

The narrowing of the 3\dd\ bands around the Fermi level makes the 
corresponding oxides prototypical for the 
interplay between electronic itineracy and 
localization~\cite{review-imada,georges-DMFT-2004}. The 3\dd-based materials are usually considered as more ``correlated'' than their 4\dd\ analogues due to the more localized character of the 3\dd\ orbitals. Within the Zaanen-Sawatzky-Allen classification~\cite{TMO-zaanen} and the evolution of $\Delta_{pd}$ through the series, the early transition metal oxides are more prone to Mott localization whereas charge transfer physics is dominant in the late ones. As commonly found in the literature
-- although never calculated yet  -- 
the local intra-orbital interactions are believed to increase with the atomic number in a similar way as in atomic systems. 
We show below that the trend for these interactions can differ from the one for atoms and we rationalize this discrepancy within the antagonism between the effects of the orbital localization and the screening.  

\subsection{3\dd\ perovskites: experimental facts}

We now give a brief -- and by no means exhaustive --
summary of experimental results on the oxide perovskite 
families, SrMO$_{3}$ (M= V, Cr, Mn).

\textbf{\srvo:}
Among the chosen oxides, \srvo\ has been intensively investigated both 
experimentally and theoretically as a prototypical TM oxide that falls 
in the intermediately correlated regime. 
Experimentally it is found to be an undistorted paramagnetic metal
which has been well characterized by experiments like optics,
thermodynamical measurements, transport or angle-integrated and
angle-resolved photoemission spectroscopy \cite{SrVO3-sekiyama,
  SrVO3-eguchi, Mott-Hubbard-fujimori, SrVO3-yoshida, SrVO3-morikawa,
  review-imada}. Photoemission spectra
(Refs[~\onlinecite{Mott-Hubbard-fujimori, SrVO3-yoshida}]) show a
lower Hubbard band at about $1.7$ eV binding energy and a
quasiparticle peak characterized by the renormalization factor $Z
\approx 0.6$. An electron addition peak $d^{1} \rightarrow d^{2}$ has
been identified by inverse photoemission at about $2.5-3$ eV \cite{SrVO3-morikawa}. DFT-LDA can give a qualitatively correct description of the band structure of \srvo, as shown in Fig.~\ref{fig:3dKS}, but a quantitative description including the
band renormalisation requires a more accurate treatment of the correlation
effects in the $d$-manifold~\cite{review-imada-takashi, SrVO3-sekiyama,Wannier-lechermann-2006,cRPA-DMFT-LaOFeAs-markus}.

\textbf{\srcro:} Because of difficulties in the synthesis of \srcro, only few studies have been conducted. Early work on single crystals with cubic structure \cite{SrCrO3-chamberland} shows a metallic behavior with a Pauli paramagnetic susceptibility. This is in disagreement with a more recent study on polycrystals reporting transport, thermal conductivity and magnetic susceptibility in favor of a non-magnetic insulating state \cite{SrCrO3-zhou}. 
\srcro\ crystals have been also recently reinvestigated within x-ray diffraction studies \cite{ortegaPRL07}. The common belief found in the literature -- supported by density functional calculations -- is that a structural transition from a non-magnetic orbitally-degenerate cubic to a distorted tetragonal- maybe antiferromagnetic- structure with orbital ordering could appear below 70 K \cite{ortegaPRL07,SrCrO3-pickett,SrCrO3-qian}.

\textbf{\srmno:} At room temperature, \srmno\ is found in an hexagonal phase \cite{SrMnO3-daoud,SrMnO3-takeda,SrMnO3-sondena,SrMnO3-Neel} but a cubic phase can be quenched and stabilized in a metastable state down to low temperatures. 
Both phases are deeply insulating. A G-type antiferromagnetic ordering with a magnetic moment around 2.6 $\mu_{B}$ emerges below the N\'eel temperature $T_{N} \sim 260 \textrm{K}$~\cite{SrMnO3-takeda}. 
According to x-ray photoemission and absorption \cite{saitoh-earlyTM,bocquet-earlyTMO,SrMnO3-kang},
the spectroscopic properties of \srmno\ are rather involved.
Indeed, the O-$p$ states are strongly entangled with the
lower Hubbard band of \t2g\ character, and the conduction
band has been proposed to be of \eg\ character \cite{saitoh-earlyTM}.
These facts strongly question the validity of a pure
\t2g\ model for the description of the low-energy spectra.
We nevertheless present both, a \t2g\ and \dd-\dd\pp\ model, for
the sake of comparison with work done in the literature
and assessing trends along the perovskite series.

\subsection{Hubbard parameters within the \dd-\dd\pp\ Hamiltonian}
We first consider the \dd-\dd\pp\ low-energy Hamiltonian for the 3\dd\ series. The Slater integrals (Appendix B, Eq.~\ref{slaterFU1}) for the screened interactions, $\textrm{F}^{k}$, and for the bare interactions, $\textrm{F}^{k}_{\textrm{bare}}$ are collected in Table~\ref{tab:slaterTMddp}. The monopole part, $\textrm{F}^{0}$, is more efficiently screened than the multipole parts, ($\textrm{F}^{2},\textrm{F}^{4}$) \cite{U-xray-sawatzky-1977,U-xray-sawatzky-1984,PhD-vandermarel}.
The bare ratio $\textrm{F}^{4}/\textrm{F}^{2}|_{\textrm{bare}}$ is close to the atomic value around $0.63$. In contrast, the partially screened ratio $\textrm{F}^{4}/\textrm{F}^{2}$ deviates from this limit. This illustrates the importance of calculating the \emph{whole set of three} Slater integrals for accurately parametrizing the four-index Hubbard interaction matrix, rather than deducing $\textrm{F}^{0}$, $(\textrm{F}^{2}+\textrm{F}^{4})/14$ from two independent relations and then assuming $\textrm{F}^{4}/\textrm{F}^{2}=0.63$ as a third one.     

\begin{table}
\caption{ (top) Slater integrals (in eV) for the \dd-\dd\pp\ model of
  \srmo\ (M=V, Cr, Mn) corresponding to screened ($W^{r}$) and bare
  ($v$) Coulomb interaction. (bottom) Slater-symmetrized effective
  interactions (in eV) between \t2g\ orbitals.}
\begin{tabular*}{0.48\textwidth}{@{\extracolsep{\fill}}lcccc|cccc}
\hline \hline
(eV)  &    F$^{0}$  & F$^{4}$/F$^{2}$ & $J$ &&&   F$^{0}_{\textrm{bare}}$  & F$^{4}$/F$^{2}|_{\textrm{bare}}$ & $J_{\textrm{bare}}$      \\ 
\hline
\srvo      &    3.2       &  0.795      &  0.85  &&& 19.5       &  0.652      &  1.06\\
\srcro     &    2.9       &  0.781      &  0.85  &&& 20.1       &  0.628      &  1.06 \\
\srmno     &    2.8       &  0.774      &  0.89  &&& 21.2       &  0.625      &  1.11  \\  
\end{tabular*} 
\begin{tabular*}{0.48\textwidth}{@{\extracolsep{\fill}} lccc|ccc}
\hline \hline 
      &    $\bar{U}_{mm}$    & $\bar{U}_{mm'}$    &  $\bar{J}_{m}$   & $\bar{v}_{mm}$  & $\bar{v}_{mm'}$  &  $\bar{J}_{m}^{\textrm{bare}}$      \\ \hline 
\srvo      &   4.1      &  2.8     & 0.65    &   20.7  & 19.1 & 0.81 \\ 
\srcro     &   3.9      &  2.6     & 0.65    &   21.4  & 19.7 & 0.82  \\
\srmno     &   3.8      &  2.4     & 0.68    &   22.5  & 20.8 & 0.86 \\ \hline \hline
\end{tabular*}
\label{tab:slaterTMddp}
\end{table}

The Slater integrals can still be used in order to deduce effective interactions between \t2g\ orbitals (from Eq.~\ref{slater4} to Eq.~\ref{slater6}). The values of these interactions are shown in Table~\ref{tab:slaterTMddp}. The validity of the Slater parametrization can be assessed by comparing these values to the ones obtained from a direct calculation with cubic symmetry (Appendix C). For \srvo, the direct calculation gives for the intra-\t2g\ interaction $U_{mm}=4.0$ eV and $J_{m}=0.60$ eV, hence in reasonable agreement with the values calculated from the Slater integrals in Tab.~\ref{tab:slaterTMddp}. For the \srmo\ series below, we will hence refer either to $U_{mm},J_{m}$ or $\bar{U}_{mm},\bar{J}_{m}$. 

\begin{figure}
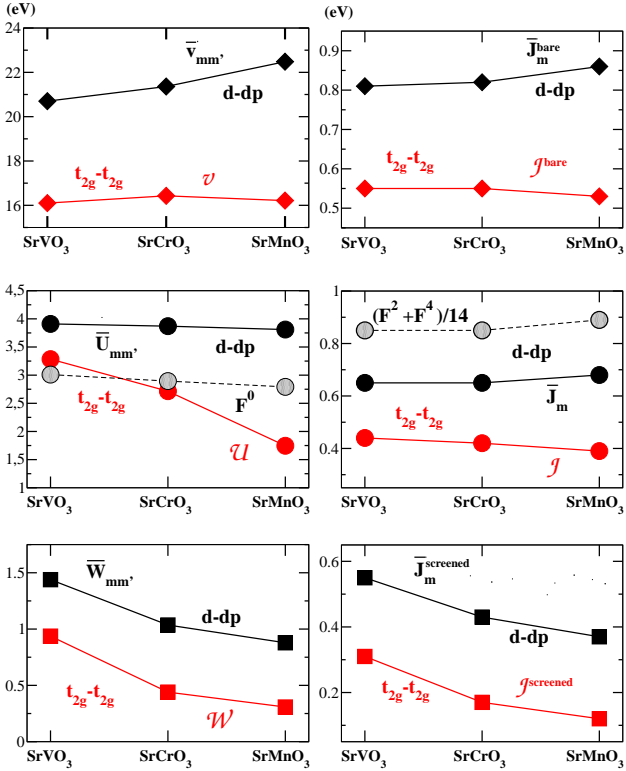

\begin{center}
\includegraphics[width=1.6in,clip]{Fig3a.eps}
\includegraphics[width=1.6in,clip]{Fig3b.eps}
\end{center}
\begin{center}
\includegraphics[width=1.6in,clip]{Fig3c.eps}
\includegraphics[width=1.6in,clip]{Fig3d.eps}
\end{center}
\begin{center}
\includegraphics[width=1.6in,clip]{Fig3e.eps}
\includegraphics[width=1.6in,clip]{Fig3f.eps}
\end{center}
\begin{center}
\caption{\footnotesize{(color online)  Middle panel: On-site Hubbard interaction $\bar{U}_{mm}$ (left) and exchange interaction $\bar{J}_{mm}$ (right) between \t2g\ orbitals within the \dd-\dd\pp\ (black curve) low-energy Hamiltonians for 3\dd\ \srmo , compared to the on-site $\mathcal{U}$ and $\mathcal{J}$ within \t2g-\t2g\ (red curve).
In dashed line with open circles, the Hubbard parameter $U=\textrm{F}^{0}$ and Hund's exchange $J=(\textrm{F}^{2} + \textrm{F}^{4})/14$ are shown. 
Top panel: Same but for the bare interactions between \t2g\ orbitals. Bottom panel: Same but for the fully screened interactions between \t2g\ orbitals.}}
\label{fig:u3dt2g}
\end{center}
\end{figure}

In Fig.~\ref{fig:u3dt2g}, the \t2g\ interactions for the bare, partially and fully screened cases, are shown. Along the 3\dd\ series, the bare interactions $v_{mm}$ and $J_{m}^{\textrm{bare}}$ increase with the number of electrons in the \dd\ manifold. This is due to the increasing localization of the Wannier \dd\ orbitals from \srvo\ to \srmno\ within the downfolded \dd-\dd\pp\ Hamiltonian. A similar behavior is expected for hydrogenoid systems within the Slater rules, for which the atomic \dd\ radial extension decreases from V$^{4+}$ to Mn$^{4+}$. The less extended the orbitals, the higher the Coulomb repulsion.

The effective interactions, $U_{mm}$ and $W_{mm}$, differs from the bare interaction $v_{mm}$ because of the screening that arises from the electronic polarizability and lowers the Coulomb repulsion. Interestingly, $W_{mm}$ {\it decreases} with the atomic number as a consequence of increasing screening. 
Screening therefore counteracts the trend of increasing orbital localization
that tends to increase the interaction matrix elements towards the end
of the 3d series.
The ratio $W_{mm}/v_{mm}$ quantifies the increase of the screening from 
\srvo\ to \srmno. The smaller this ratio the stronger is the screening. 
We obtain $6.7/100$ in \srvo, $4.6/100$ in \srcro\ and $4/100$ in \srmno.  

Electronic screening is mediated by the creation of particle-hole and 
plasma excitations. At the RPA level, its strength varies with the inverse 
of the energy difference between occupied and empty states. 
The knowledge of the DFT-LDA band structure (Fig.~\ref{fig:3dKS})
thus allows us to analyse the efficiency of the different screening channels.
The charge transfer energy between the oxygen \pp\ and the transition metal \dd\ states decreases from SrVO$_{3}$ to SrMnO$_{3}$. The Kohn-Sham states with ligand \pp\ character 
go up toward the \t2g\ ones. 
Furthermore, the crystal field splitting to the \eg\ states 
decreases.
While the absolute magnitude of the energy difference between
$p$- and \eg- states may depend on the approximation used in the
band calculation (and in particular, is slightly underestimated
in the LDA), the overall trend is 
not an artifact of DFT-LDA
\cite{torrance-deltapd,review-imada}. Experimentally, it has been 
evidenced e.g. in optical measurements~\cite{Lee-SrMO3}. 
As a consequence, \dd\dd\ and \pp \dd\ screening channels give
stronger contributions in SrMnO$_{3}$ than in SrVO$_{3}$. 
However, 
these screening channels are only part of the multiple screening 
channels that contribute to the reduction of the Coulomb repulsion from $v_{mm}$ to $W_{mm}$. For example in \srvo, $v_{mm}=20.7$ eV whereas $W_{mm}=1.4$ eV. As already pointed out in Ref.~[\onlinecite{Ufirstprinciples-ferdi-2006}], the effects of the screening channels are not additive.
Nevertheless, their contribution to the total polarizability can be estimated by evaluating partially screened interactions. 

The evaluation of the quantity $U_{mm}/v_{mm}$ and the comparison with $W_{mm}/v_{mm}$, quantify the weight of the \dd\dd\ transitions in the total polarization. 
For example in \srvo, $U_{mm} /v_{mm}$ equals $19.8/100$ whereas $W_{mm} /v_{mm}$ equals $6.7/100$. In \srcro, $U_{mm} /v_{mm} = 18.2/100$ and $W_{mm}/v_{mm} = 4.6/100$, while in \srmno, $U_{mm}/v_{mm} = 16.9/100$ and $W_{mm}/v_{mm} = 4.0/100$.
The partial screening \emph{without the \dd\dd\ channels}, increases more slowly through the early 3\dd\ TM oxides than the total screening. The interactions $U_{mm}$ thus correspond to an intermediate situation between $v_{mm}$ and $W_{mm}$, where the increase in partial screening counterbalances the increasing localization of the orbitals. 
As a result, even though $W_{mm}$ decreases and $v_{mm}$ 
increases with the atomic number,
$U_{mm}$ depends barely on it.

The values of $U_{mm}$
indicate how strong the contribution of the \dd\dd\ screening channels is to the total polarizability. These channels are only partly responsible for the screening that reduces $v_{mm}$. The role of the \pp\dd\ screening channels can also be estimated. By computing the partially screened interaction after the elimination of all \pp\dd\ transitions (as well as \dd\dd), one gets $11$ eV in \srvo, $10.8$ eV for \srcro\ and $10.7$ eV for \srmno. Screening without these channels has thus reduced the interaction by roughly a factor two, compared to the bare interaction $v_{mm}$. Low-energy screening channels, such as \dd\dd\ or \pp\dd, are thus responsible for about half of the total screening. On the other side, the transitions at higher energies also participate to the total polarizability leading to the strong reduction of the bare Coulomb repulsion. 

Hund's exchange $J$ is deduced from the Slater integrals F$^{2}$ and F$^{4}$ (Tab.~\ref{tab:slaterTMddp}). In the 3\dd\ \srmo\ series, $J$ goes from about $0.8$ eV to $J_{\textrm{bare}}\approx 1.0$ eV. The corresponding exchange interaction between \t2g, $J_{m} \approx 0.6-0.7$ eV, is smaller than $J$ and the values agree with the ones usually employed in models for such oxides.

The fully screened exchange interaction $J_{m}^{\textrm{screened}}$, 
calculated from the total polarization, can be evaluated for the series. 
$J_{m}^{\textrm{screened}}$ slightly decreases with the \dd\ electron number 
in contrast to the unscreened Hund's exchange $J_{m}^{\textrm{bare}}$. 
Cutting off transitions from and to the \dd\ Kohn-Sham bands 
within $\WW_{\t2g}$, $J_{m}$ still increases from \srvo\ to \srmno. 
In contrast to $U_{mm}$, $J_{m}$ hence reproduces the atomic-like trend. 
The weaker dependence of the exchange interactions on the screening 
results in $J_{m}$ varying in a similar way as $J_{m}^{\textrm{bare}}$.

\subsection{Hubbard parameters within the \t2g-\t2g\ Hamiltonian}

Within the \t2g-\t2g\ Hamiltonian, the \t2g-projected local orbitals within the energy window $\WW_{\t2g}$ (Tab.~\ref{tab:windows113}) lead to ``extended'' \t2g\ Wannier orbitals. 
The charge transfer energy and the hybridization between the \t2g\ and the oxygen ligand \pp\ bands, are responsible for the finite weight of the \t2g\ Wannier functions on the oxygen atomic sites. The tail and hence the extension of the so constructed local orbitals, increases when the \pp\dd\ charge transfer energy becomes smaller, as happens for the 3\dd\ \srmo\ series. 

The \t2g\ local orbitals are hence more extended in \srmno\ than in \srvo, in contrast to the atomic orbitals. It implies that i) the unscreened interaction $\mathcal{V}$ within the \t2g-\t2g\ model does not increase with the atomic number as $v_{mm}$ in the previous \dd-\dd\pp\ model (Fig.~\ref{fig:u3dt2g} and Tab.~\ref{tab:slaterTMddp}) and ii) the values of \vk\ are smaller than the values of $v_{mm}$ within \dd-\dd\pp. 

\begin{table} 
\caption{\footnotesize{Hubbard-Kanamori \uk, bare \vk\ and fully
    screened \wk\ interactions (in eV) and corresponding exchange interactions, \jk, $\mathcal{J}_{\textrm{bare}}$ and $\mathcal{J}_{\textrm{screened}}$ between \t2g\ orbitals within the \t2g-\t2g\ downfolded Hamiltonian for the early 3\dd\ series \srmo\ (M= V, Cr, Mn) and 4\dd\ (M= Nb, Mo, Tc). The inter-orbital interactions $\mathcal{U}'$ coincide with $\mathcal{U}-2\mathcal{J}$.}}
\begin{center}
\begin{tabular*}{0.48\textwidth}{@{\extracolsep{\fill}}lccc|ccc} \hline \hline
(eV)            & V      &  Cr    & Mn    & Nb      & Mo     & Tc             \\ \hline 
\uk     & 3.2   &  2.7   & 1.8   &  3.0   &  3.0   & 2.9              \\
\jk      & 0.44  & 0.42 & 0.39  & 0.29  &  0.31 & 0.31           \\ \hline
\vk      & 16.1  & 16.4  & 16.2 &  10.7  & 11.6  & 11.8           \\
$\mathcal{J}_{\textrm{bare}}$       & 0.55  & 0.55 & 0.53  & 0.38  &  0.40 & 0.39          \\ \hline
\wk     & 0.9   &  0.4   & 0.3   &  0.9   &  0.5   & 0.4              \\  
$\mathcal{J}_{\textrm{screened}}$  & 0.30  & 0.17 & 0.12  & 0.24  &  0.19 & 0.16   \\ \hline \hline
$\mathcal{U}/\mathcal{V} \times 100$ & $19.8$  & $16.4$  & $11.1$  &  $28$ &  $25.8$  & $24.6$     \\
$\mathcal{W}/\mathcal{V} \times 100$     & $5.6$ & $2.4$  & $1.8$ & $8.4$  &  $4.3$ & $3.4$ \\ \hline \hline
\end{tabular*}
\end{center}
\label{tab:u3dTM}
\end{table}

The interaction values that would be appropriate for the \t2g-restricted
Hubbard Hamiltonian are given in Tab.~\ref{tab:u3dTM}. Their evolution
along the series is illustrated in Fig.~\ref{fig:u3dt2g}.  
A comparison with the \crpa\ values from the literature but with different frameworks for constructing \t2g\ local orbitals, is in order here: In \srvo, Aryasetiawan and co-workers \cite{Ufirstprinciples-ferdi-2006} report $\mathcal{U}=3.5$ eV within the head of LMTO's, while Miyake and Aryasetiawan \cite{cRPA-takashi-2008} give $\mathcal{U}=3.0$ eV,$ \mathcal{J}=0.43$ eV within MLWF. We obtain $\mathcal{U}=3.2$ eV and $\mathcal{J}=0.46$ eV. 

We now turn to the values through the early series of the 3\dd\ TM oxides. \uk\ significantly lessens from \srvo\ to \srmno\ and the decrease of \uk\ is more pronounced than $U_{mm}$ within the \dd-\dd\pp\ model. The decrease of \uk\ is still due to the screening which gets larger with the atomic number. More precisely, the \pp $\rightarrow$ \t2g\ and \t2g $\rightarrow$ \eg\ channels contribute more and more to the screening from \srvo\ to \srmno, because of the \pp\ and \eg\ Kohn-Sham bands that go closer and closer to the Fermi level. Quantitatively, the ratio $\mathcal{U}/\mathcal{V}$ is about twice larger in \srvo\ than in \srmno\ (Tab.~\ref{tab:u3dTM}). The fully screened interactions, $\mathcal{W}$, on the other hand, decrease through the series.

In contrast to $J_{m}$ and $J_{m}^{\textrm{bare}}$ within the \dd-\dd\pp\ model, the exchange interactions, $\mathcal{J}$ and $\mathcal{J}_{\textrm{bare}}$ within the \t2g-\t2g\ model slightly decrease with the number of \dd\ electrons. Considering the total polarization leads to a significant decrease of $\mathcal{J}_{\textrm{screened}}$ through the series. Since the reduction is weaker for \jk\ than for $\mathcal{J}_{\textrm{screened}}$, the screening induced by the \t2g\t2g\ transitions must be responsible for the behavior of $\mathcal{J}_{\textrm{screened}}$. 
As \uk, the exchange interactions therefore depend on the extension of the local orbitals: \jk\ is smaller within the 
\t2g-\t2g\ Hamiltonian than $J_{m}$ within \dd-\dd\pp.

\subsection{4\dd\ perovskites: experimental facts}
We now turn to the 4d series
SrMO$_{3}$ (M= Nb, Mo, Tc). The lattice parameters used in the calculations are summarized in Table \ref{tab:windows113}.

\textbf{\srnbo:} This compound is usually obtained with a non-stoechiometric perovskite structure \cite{Hannerz-SrNbO3}. When doping into Sr$_{x}$NbO$_{3}$ with $x>0.80$, a cubic perovskite phase is observed exhibiting a poor paramagnetic metallic behavior at temperatures below 300 K \cite{SrNbO3-isawa}. 

\textbf{\srmoo:} It is, in contrast, an excellent paramagnetic metal- even the 4\dd\ transition metal oxide that displays the highest electrical conductivity \cite{SrMoO3-nagai}. With two electrons on the \t2g\ shell, it is an electronic analogue of SrRuO$_{3}$ that has two holes on the \t2g's but larger correlation effects due to the Van Hove singularity found in its density of states \cite{Sr2RuO4-jernej}. 

\textbf{\srtco:} Because of the radioactivity of technetium, this compound has been less studied, and only structural and magnetic properties are known. A huge N\'eel temperature of $1023$ K -- accompanied by a G-type antiferromagnetic ordering with the magnetic moment $2.1 \mu_{B}$ below $T_{N}$ -- has been recently discovered \cite{SrTcO3-rodriguez}. In particular, this high $T_{N}$ is larger than in the 3\dd\ analogue \srmno. The large $T_{N}$ in \srtco\ was first interpreted within density functional calculations \cite{SrTcO3-rodriguez,FranchiniPRB11,Middey11}, e.g. in terms of the larger covalency of the Tc-O bonding compared to Mn-O \cite{SrTcO3-rodriguez}. Recently, another scenario has been put forward for interpreting the difference of magnitude in $T_{N}$ between \srtco\ and \srmno, based on the proximity of \srtco\ to the Mott transition in the presence of large Hund's exchange at half-filling \cite{SrTcO3-jernej}.

\subsection{Hubbard interactions in 4\dd\ perovskites}

In the 4\dd\ \srmo\ series, the evolution of the Kohn-Sham bands 
from Nb to Tc oxides 
is analogous to the one in the isoelectronic and isostructural 3\dd\ 
compounds, but the trends are less pronounced (Fig.~\ref{fig:4dKS}). 
The \pp\dd\ charge transfer energy decreases more slowly in the 4\dd\ series,
and so do the crystal and ligand fields that split the \t2g\ and \eg\ bands. 
As a result, the transition metal \eg\ states overlap with 
strontium 5\sss/4\dd. For this reason, only the \t2g\ Hamiltonian is 
considered here. 

\begin{figure}
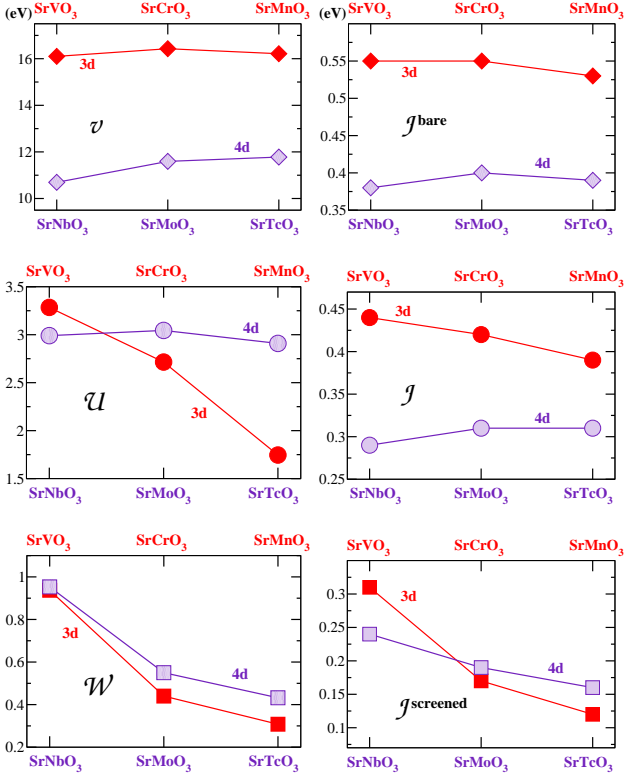

\begin{center}
\includegraphics[width=1.6in,clip]{Fig4a.eps}
\includegraphics[width=1.6in,clip]{Fig4b.eps}
\end{center}
\begin{center}
\includegraphics[width=1.6in,clip]{Fig4c.eps}
\includegraphics[width=1.6in,clip]{Fig4d.eps}
\end{center}
\begin{center}
\includegraphics[width=1.6in,clip]{Fig4e.eps}
\includegraphics[width=1.6in,clip]{Fig4f.eps}
\end{center}
\begin{center}
\caption{\footnotesize{(color online) Middle panel: On-site Hubbard-Kanamori interaction $\mathcal{U}$ (left) and exchange interaction $\mathcal{J}$ (right) between \t2g\ orbitals within the \t2g-\t2g\ low-energy Hamiltonian for 3\dd\ (red curve) and 4\dd\ (indigo curve) \srmo\ (see also Tab.~\ref{tab:u3dTM}). Top panel: Same but for the bare interaction $\mathcal{V}$ and $\mathcal{J}_{\textrm{bare}}$ between \t2g\ orbitals. Bottom panel: Same but for the fully screened interaction $\mathcal{W}$ and $\mathcal{J}_{\textrm{screened}}$ between \t2g\ orbitals.}}
\label{fig:u4dt2g}
\end{center}
\end{figure}

The atomic-like character of the \t2g\ Wannier basis is evidenced by the atomic-like behavior of the bare interaction \vk\ in the 4\dd\ series (Fig.~\ref{fig:u4dt2g}): \vk\ becomes larger with the atomic number, as in 3\dd\ when considering the \dd-\dd\pp\ Hamiltonian. The local basis designed for the \t2g\ degrees of freedom, is atomic-like for 4\dd\ oxides in contrast to the one for \t2g\ in 3\dd. The difference in 4\dd\ comes from the larger charge transfer energy which leads to a smaller tail of the downfolded orbital with \t2g\ character on the oxygen sites.     

However, the orbital extension of the \t2g\ local orbitals has to be larger in 4\dd\ than in 3\dd\ oxides since the bare interaction \vk\ are twice smaller in the former (Tab.~\ref{tab:u3dTM}). This makes sense in an atomic-like basis in which the extension of the 4\dd\ atomic wavefunctions is larger than the extension of 3\dd. 

Since the $pd$ charge transfer energy as well as the \t2g-\eg\ splitting decrease slower in 4\dd\ oxides, the \pp\dd\ and the \t2g\eg\ channels are not as efficient in screening as in the 3\dd\ analogues. This is quantitatively highlighted by the ratios $\mathcal{U}/\mathcal{W}$ (Tab~\ref{tab:u3dTM}), which are larger in 4\dd\ and decrease slower through the 4\dd\ series. 

Consequently, \uk\ in 4\dd\ is almost constant with the \dd\ electron number, like $U_{mm}$ in the \dd-\dd\pp\ model for 3\dd\ TM oxides, whereas \uk\ in 3\dd\ significantly lessens until becoming smaller in \srmno\ than in \srtco\ (Fig.~\ref{fig:u4dt2g}). This is an effect of the screening which has stronger impact in the 3\dd\ extended Wannier basis than in the atomic-like 4\dd\ one.    

The exchange interactions in 4\dd\ oxides behave in a similar way as in 3\dd\ (Tab.~\ref{tab:u3dTM}). The exchange interactions in 4\dd\ are about $0.1$ eV smaller than in 3\dd. This is due to the higher extension of the 4\dd\ Wannier orbitals which is also responsible for the smaller bare interactions \vk. 

\subsection{The influences of structural distortions: the example of \srmno}
\begin{figure}[!h]
\begin{center}
\includegraphics[width=0.25\textwidth]{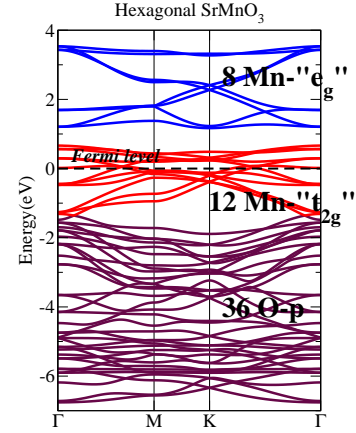}
\caption{\footnotesize{(color online) DFT-LDA band structure of the four-layers hexagonal unit cell of \srmno\ in the paramagnetic phase. Lattice parameters $a=5.461 \textrm{A}^{\circ},c=9.093 \textrm{A}^{\circ}$ are taken from~\cite{SrMnO3-daoud}.}}
\label{fig:SrMnO3-hex}
\end{center}
\end{figure}

At room temperature, \srmno\ crystallises in the four-layers hexagonal structure ($P63/mmc$) with $a=5.461$ \AA, $c=9.093$ \AA~\cite{SrMnO3-daoud}. The DFT-LDA band structure is shown in Fig.~\ref{fig:SrMnO3-hex}. 

The four-times larger unit cell leads to a backfolding of bands in the first Brillouin zone.   
A set of \t2g-like bands (Fig.~\ref{fig:SrMnO3-hex}) can be identified with the twelve bands around the Fermi level and a set of \eg-like bands with the eight ones above. A \dd-\dd\pp\ model is constructed rather than a \t2g\ one. This allows for a direct comparison of the Slater integrals (Eq.~\ref{slaterFU1}) calculated within the \dd-\dd\pp\ model built for the cubic crystal structure (Tab.~\ref{tab:usrmnodp}). 

$U$ as well as $v$ are slightly bigger in the hexagonal phase than in the cubic one. This is an effect of the Wannier orbital localization rather than a screening effect since also the bare interaction $v$ is enhanced. Hund's exchange $J$ does not change much between the two crystal structures.      

\begin{table}
\caption{\footnotesize{Screened and bare Slater integrals (in eV) for the \dd-\dd\pp\ model in cubic and hexagonal \srmno. }}
\label{tab:usrmnodp}
\begin{center}
\begin{tabular*}{0.48\textwidth}{@{\extracolsep{\fill}}lcccc|cccc}
\hline \hline 
(eV)         & F$^{0}$ & $\textrm{F}^{4}/\textrm{F}^{2}$    & \JJ &&& F$^{0}_{\textrm{bare}}$ & $\textrm{F}^{4}/\textrm{F}^{2}|_{\textrm{bare}}$    & $J_{\textrm{bare}}$    \\ \hline
cubic      & 2.8        & 0.774            & 0.9    &&&  21.2  & 0.625            & 1.1      \\ 
hexagonal & 3.1     & 0.777            & 0.9    &&&  21.9  & 0.621            & 1.1      \\
\hline \hline
\end{tabular*}
\end{center}
\end{table}

\section{Hubbard interactions in the layered perovskite oxides \srmol\ (M = Mo, Tc, Ru, Rh)}

\subsection{General features}
The second class of oxides that we have investigated are the layered perovskites Sr$_{2}$MO$_{4}$ (M= Mo, Tc, Ru, Rh) where M$^{4+}$ is a 
4\dd\ transition metal. The lattice parameters used in the electronic structure calculations for the paramagnetic phase are given in Table \ref{tab:windows214}. \newline

\textbf{Sr$_{2}$MoO$_{4}$:} This compound exhibits a metallic behavior over a wide range of temperature between 80 mK and 300 K, with a resistivity increasing between 2 and 10 m$\Omega$.cm  \cite{Sr2MoO4-ikeda}. It is usually investigated under the possibility that it could exhibit analogous electronic properties than Sr$_{2}$RuO$_{4}$, although a superconducting state has not been reported down to 25 mK. 

\textbf{\srtcol:} Due to the radioactivity of technetium elements, only structural properties are known for \srtcol. An undistorted layered perovskite structure is considered below. 

\textbf{\srruol:} The resistivity of this compound obeys a T$^{2}$ law below 30 K, evidencing a Fermi liquid behavior. A strong anisotropy of the transport properties -- due to its layered structure -- was reported~\cite{husseyPRB98}.  
A large mass enhancement and a low coherence scale have been determined experimentally. \srruol\ also exhibits an unconventional superconductivity below 2 K \cite{Sr2RuO4-review}. 
\newline
The three-sheet Fermi surface determined experimentally is reasonably well described by DFT-LDA calculations, but the enhancement and the anisotropy in the mass are missed. The largest mass enhancement surprisingly appears for the widest \dxy\ band as determined by quantum oscillations experiments~\cite{Sr2RuO4-review}. Several \LDADMFT\ calculations have been carried out to reproduce the low-energy properties of the photoemission spectra \cite{Sr2RuO4-jernej,liebschPRL00,anisimovEPL02,pchelkinaPRB07}. 

\textbf{\srrhol:} The symmetry is lowered from the K$_{2}$NiF$_{4}$ class by $11^{\circ}$ rotation around the c-axis of the RhO$_{6}$ octahedra \cite{HuangJSSC112-1994}. It is a paramagnetic metal down to 36 mK \cite{Sr2RhO4-moon}. Spin-orbit coupling (SOC) was found to be relevant in addition to electronic correlations \cite{Sr2RhO4-tamai,Sr2RhO4-haverkort,Sr2RhO4-Liu}. \LDADMFT\ calculations with SOC have been recently performed for the distorted structure and compared to the isoelectronic but Mott insulating \srirol\ \cite{Sr2IrO4-cyril}. In particular, it was shown how the interplay of SOC, correlations and structural distortions leads to a suppress of the effective orbital degeneracy, leaving only two orbitals at the Fermi level. In the following, for computational reasons, the undistorted crystal structure of \srrhol\ is considered.  

\begin{figure}[!h]
\begin{center}
\includegraphics[width=1.6in,clip]{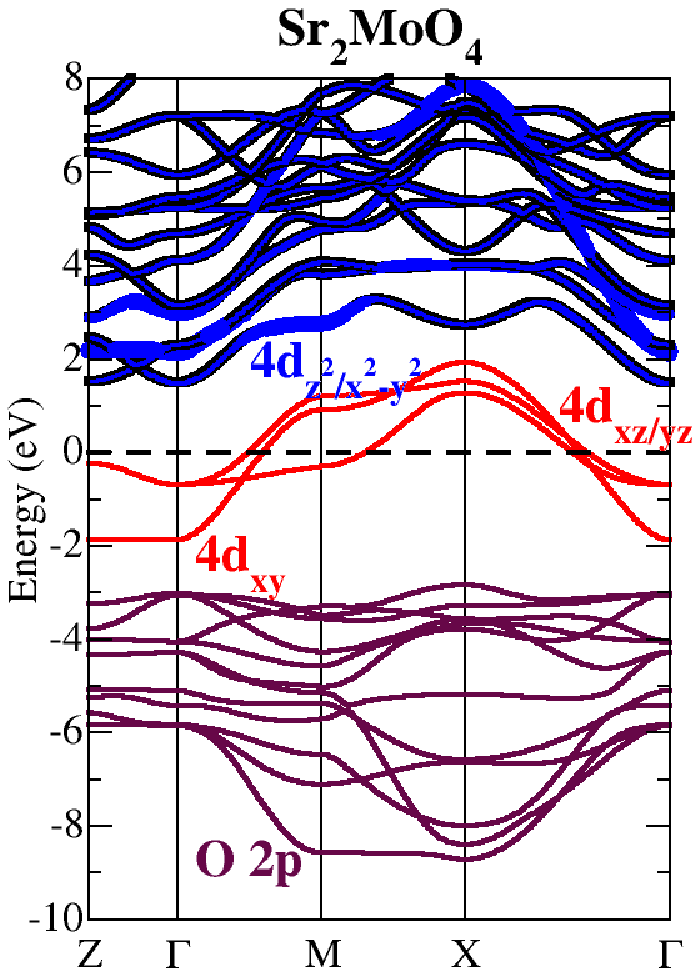}
\includegraphics[width=1.65in,clip]{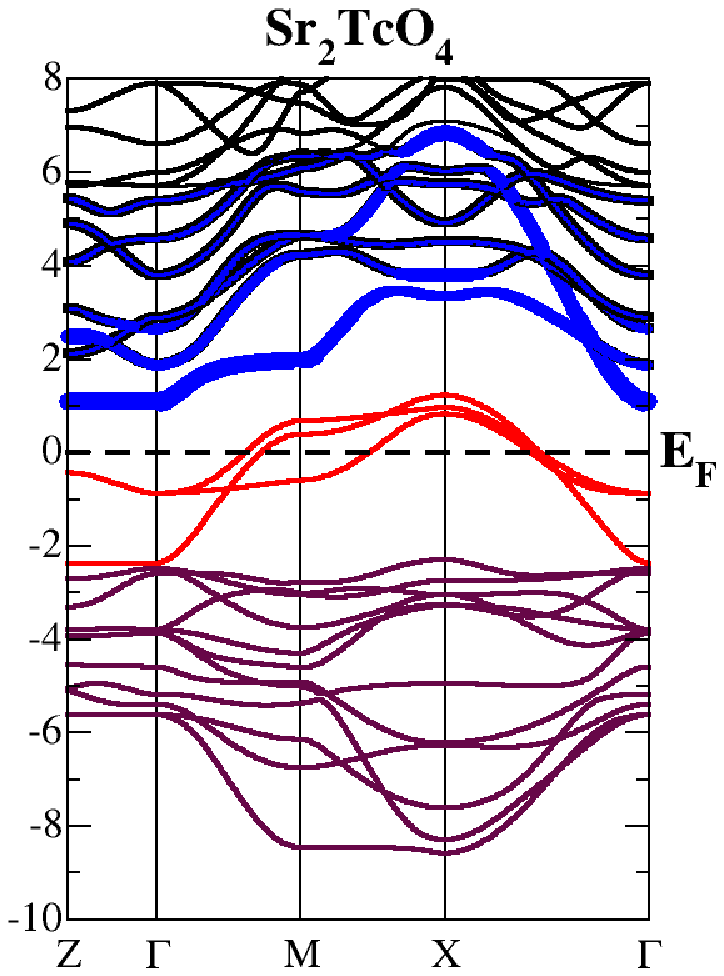}
\includegraphics[width=1.6in,clip]{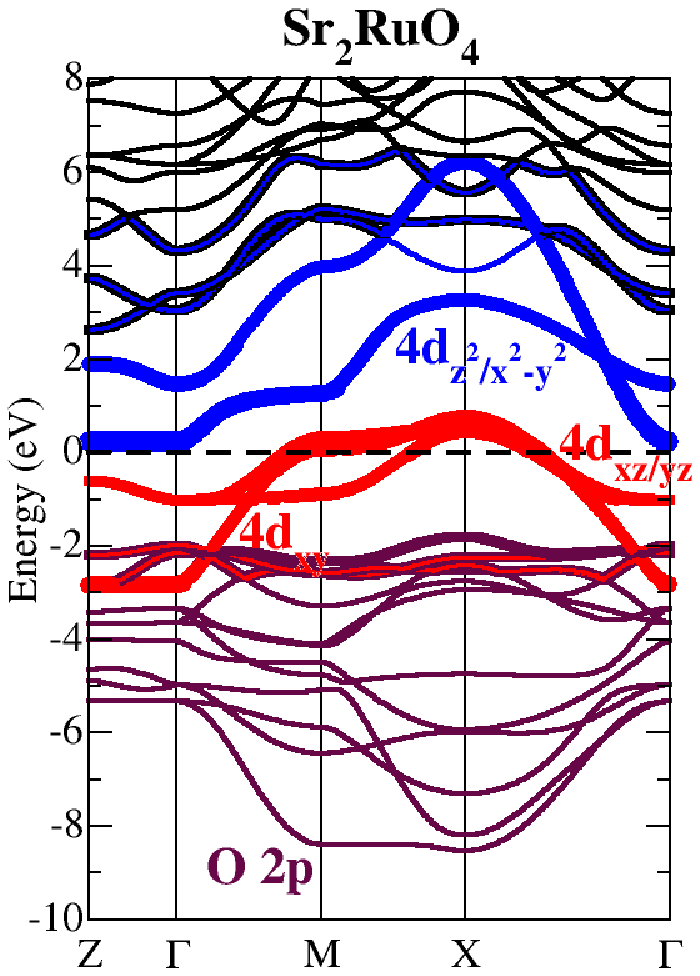}
\includegraphics[width=1.65in,clip]{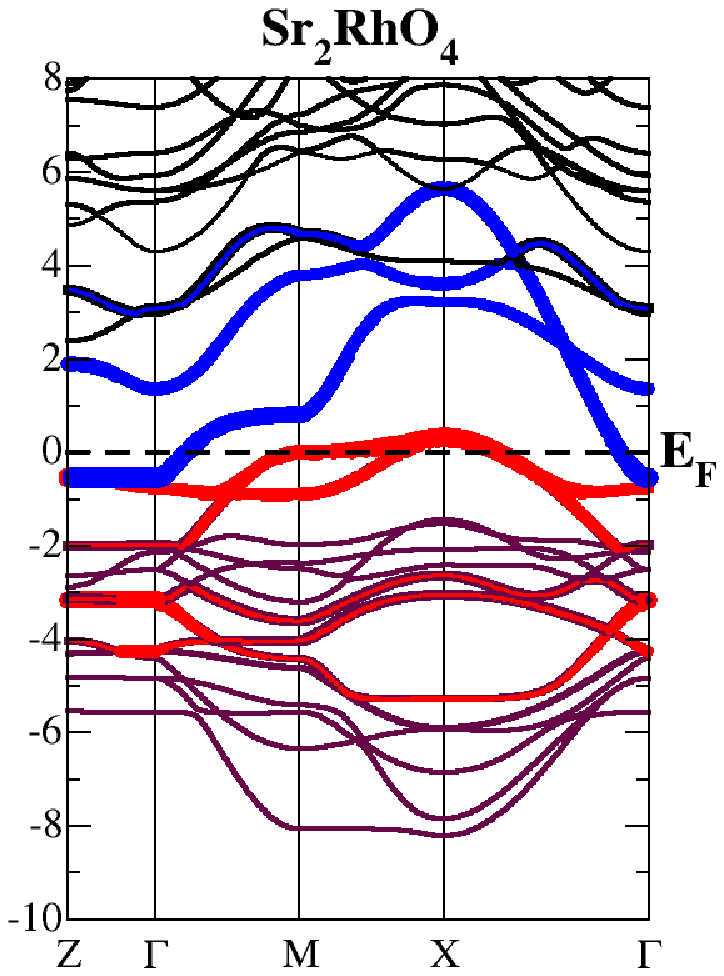}
\caption{\footnotesize{(color online) DFT-LDA paramagnetic band structures for layered perovskites Sr$_{2}$MO$_{4}$. (top): From left to right: M = Mo, Tc. (bottom): From left to right: M = Ru, Rh.}}
\label{fig:214KS}
\end{center}
\end{figure}

The DFT-LDA band structures for the paramagnetic phases are shown in Fig.~\ref{fig:214KS}. Bands with (\dxy,\dxz,\dyz) orbital character emerge around the Fermi level, whereas the (\dz2,\dx2y2) orbital character is found above and the oxygen \pp's lie below. The band with \dxy\ character leads to a quasi-two dimensional Fermi surface whereas the degenerate bands with (\dxz,\dyz) character give rise to a quasi-one dimensional Fermi surface. 
In \srmool\ and \srtcol, the \eg\ states are entangled with the strontium-like \dd's but come closer to the \t2g's when the 4\dd\ electron number increases. On the other hand, the \pp\dd\ charge transfer energy decreases with this number, since the \pp's go up toward the Fermi level. This decrease makes larger the screening from the \pp\ channels, in an analogous way to in the early transition metal oxide series. 

\begin{table}
\caption{\footnotesize{Lattice parameters used for the layered \srmol\ perovskites and energy windows $\WW$ (in eV) for the \dd-\dd\pp\ and \t2g-\t2g\ models.}}
\begin{tabular*}{0.48\textwidth}{@{\extracolsep{\fill}}lcc|cc}
 \hline \hline
                &    a (\AA) & c (\AA)  & $\WW_{dp}$ & $\WW_{\t2g}$  \\ \hline
Sr$_{2}$MoO$_{4}$ & 3.917  & 12.859 & $[-10,8.5]$  &  $[-2.0,2.0]$  \\ 
Sr$_{2}$TcO$_{4}$  & 3.902  & 12.720 & $[-10,7.5]$  &  $[-2.6,1.3]$ \\ 
Sr$_{2}$RuO$_{4}$  & 3.863  & 12.724 & $[-10,6.9]$  &  $[-3.0,1.0]$   \\
Sr$_{2}$RhO$_{4}$  & 3.854  & 12.880 & $[-10,6.0]$  &  $[-3.3,0.5]$ \\ \hline\hline
\end{tabular*}
\label{tab:windows214}
\end{table}

\subsection{Hubbard parameters within the \dd-\dd\pp\ Hamiltonian}
As previously, also for the early TM oxides within the layered perovskite structure \srmol, we consider first the \dd-\dd\pp\ Hamiltonian.  
The 4\dd\ Wannier orbitals are constructed out of the Kohn-Sham bands within the energy window $\WW_{dp}$ (Tab.~\ref{tab:windows214}).   

The Slater integrals are given in Tab.~\ref{tab:slater4d214}. The value of the ratio $\textrm{F}^{4}/\textrm{F}^{2}|_{\textrm{bare}}$ is close to the one calculated for 4\dd\ atoms \cite{haverkort-tesis}. 
A significant deviation is obtained for the screened ratio of the Slater integrals, which is even stronger than the one in the 3\dd\ \srmo\ series. This seems natural, given the anisotropy of the structure and the screening which increases with the orbital extension.

\begin{table}
\caption{\footnotesize{(top) Static and bare Slater integrals (in eV) for the \dd-\dd\pp\ Hamiltonian in 4\dd\ Sr$_{2}$MO$_{4}$. 
(bottom) Slater-symmetrized effective interactions between \t2g\ orbitals.}}
\begin{tabular*}{0.48\textwidth}{@{\extracolsep{\fill}}lcccc|cccc} \hline \hline
(eV) & F$^{0}$ & $\textrm{F}^{4}/\textrm{F}^{2}$    & $J$ &&&   F$^{0}_{\textrm{bare}}$ & $\textrm{F}^{4}/\textrm{F}^{2}|_{\textrm{bare}}$    & $J_{\textrm{bare}}$              \\ \hline 
Mo  & 3.26    & 0.862            & 0.67     &&& 14.50   & 0.684            & 0.86                 \\ 
Tc   & 3.19    & 0.850            & 0.70     &&& 15.25   & 0.673            & 0.90                    \\ 
Ru   & 3.23    & 0.838            & 0.74     &&& 15.97   & 0.669            & 0.94                    \\  
Rh   & 3.44    & 0.820            & 0.78     &&& 16.77   & 0.663            & 0.98                      \\\hline \hline 
\end{tabular*}
\begin{tabular*}{0.48\textwidth}{@{\extracolsep{\fill}}lccc|ccc}
           &   $\bar{U}_{mm}$   &  $\bar{U}_{mm'}$  &  $\bar{J}_{m}$  &  $\bar{v}_{mm}$  &  $\bar{v}_{mm'}$  & $\bar{J}_{m}^{\textrm{bare}}$          \\ 
\hline 
Mo       &  4.0   & 3.0  &  0.50    & 15.5   &  14.1 & 0.66         \\
Tc        &  4.0   & 2.9  &  0.53    & 16.3   &  14.9 & 0.69          \\
Ru        &  4.1   & 2.9  &  0.56    & 17.0   &  15.6 & 0.72       \\ 
Rh        &  4.3   & 3.1  &  0.59    & 17.9   &  16.4 & 0.75        \\ 
\hline \hline
\end{tabular*}
\label{tab:slater4d214}
\end{table}

A cubic approximation can be used for deducing a set of interactions between (\dxy,\dxz,\dyz) local orbitals from the Slater integrals (Eqs.~\ref{slater4}, \ref{slater5} and~\ref{slater6}). The values are shown in Tab.~\ref{tab:slater4d214} and can be compared to the matrix elements calculated directly (Tab.~\ref{tab:Ut2g214} and see also Appendix D). The latter are anisotropic within the plane of the TM and oxygen octahedra. 
The Slater parametrization is more accurate for the late materials, \srtcol\ and \srrhol, for which the spherical approximation of the 4\dd\ orbital is better due to the smaller hybridization with the ligands. 
 
\begin{figure}
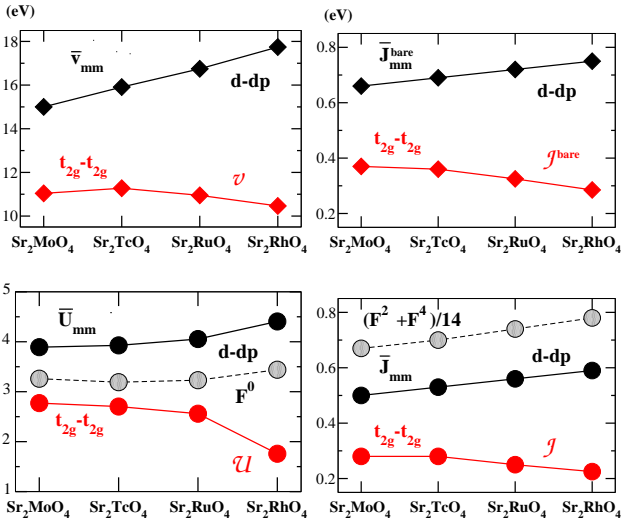

\begin{center}
\includegraphics[width=1.6in,clip]{Fig7a.eps}
\includegraphics[width=1.6in,clip]{Fig7b.eps}
\end{center}
\begin{center}
\includegraphics[width=1.6in,clip]{Fig7c.eps}
\includegraphics[width=1.6in,clip]{Fig7d.eps}
\end{center}
\begin{center}
\caption{\footnotesize{(color online) Bottom panel: On-site interaction $U_{mm}$ (left) and exchange interaction $J_{m}$ (right) between \t2g\ orbitals within the \dd-\dd\pp\ (black curve) model for \srmol\ and on-site interactions $\mathcal{U}$ (left) and $\mathcal{J}$ (right) but within \t2g-\t2g\ (red curve). 
In dashed line with open circles, the Hubbard parameter $U=\textrm{F}^{0}$ and Hund's exchange $J=(\textrm{F}^{2}+\textrm{F}^{4})/14$ are shown. 
Top panel: Same but for the bare interactions between the \t2g\ orbitals.
}}
\label{fig:u4d214plot}
\end{center}
\end{figure}

The bare on-site and exchange interactions increase with the number of the 4\dd\ electrons (Fig.~\ref{fig:u4d214plot}), suggesting a rather atomic-like behavior of the 4\dd\ Wannier orbitals within the \dd-\dd\pp\ Hamiltonian. This agrees with the atomic expectations based on the Slater rules, \emph{ie} a higher localization due to the contraction of the atomic 4\dd\ wavefunction from the left to the right of the periodic classification.

In the \dd-\dd\pp\ model, the \dd\dd\ transitions are removed from the screening in order to obtain the Hubbard interaction matrices (Tab.~\ref{tab:slater4d214} and Appendix D). As the bare interaction, the on-site Hubbard interaction gets larger with the 4\dd\ electron number (Fig.~\ref{fig:u4d214plot}). This is rationalized by the fact that the screening only slightly increases from \srmool\ to \srrhol\ as indicated by the small decrease of the ratio $U_{mm}/v_{mm}$ (Tab.~\ref{tab:Ut2g214}), from $25.8/100$ to $24.8/100$. The increase of the screening is thus not able to counterbalance the effects due to the stronger orbital localization. 

The Hund's exchange interaction (Fig.~\ref{fig:u4d214plot}) also reproduces the atomic trend, increasing with the atomic number. 

\begin{table}
\caption{\footnotesize{Hubbard-Kanamori parameters
    $\mathcal{U},\mathcal{U'},\mathcal{J}$ (in eV) for the \t2g-\t2g\ model and average interactions between the \t2g\ orbitals within \dd-\dd\pp\ in \srmol. The ratios $\mathcal{U}/\mathcal{V}$ in \t2g-\t2g\ and $U_{mm}/v_{mm} $ in \dd-\dd\pp\ have been multiplied by a factor 100.}}
\begin{tabular*}{0.48\textwidth}{@{\extracolsep{\fill}}lcccc|ccc}
\hline \hline 
       &    \t2g-\t2g  &       &   &                         &    \dd-\dd\pp  &  &                               \\ 
(eV)       & $\mathcal{U}$ & $\mathcal{U'}$ & $\mathcal{J}$ & $\mathcal{U}/\mathcal{V}$ & $U_{mm}$ & $J_{m}$ & $U_{mm}/v_{mm}$ \\ 
\hline
Mo      & 2.77  & 2.15   &  0.28  & $24.8$           & 3.8   & 0.48   & $25.3$                  \\ 
Tc       & 2.70  & 2.07   &  0.28  & $24.3$           & 3.9   & 0.52   & $24.5$                      \\ 
Ru       & 2.56  & 1.94   &  0.26  & $23.2$           & 4.0   & 0.55   & $23.9$                    \\  
Rh       & 1.76  & 1.18   &  0.23  & $16.6$           & 4.4   & 0.61   & $24.8$                     \\ 
\hline \hline
\end{tabular*} 
\label{tab:Ut2g214}
\end{table}

\subsection{Hubbard parameters within the \t2g-\t2g\ Hamiltonian}  
Alternatively, one can construct the low-energy Hamiltonian that only includes the (\dxy,\dxz,\dyz) degrees of freedom. The energy windows $\WW_{\t2g}$ are given in Tab.~\ref{tab:windows214}. 

Since the \pp\dd\ charge transfer energy decreases throughout the series, the tail on the oxygen atomic sites of the downfolded local orbitals gets larger from \srmool\ to \srrhol. Consequently, the orbital localization decreases with the 4\dd\ electron number, in contrast to the atomic \dd\ wavefunctions. The trends for the bare interactions 
thus deviate from the atomic ones, which were previously reported for the \dd-\dd\pp\ model (Fig.~\ref{fig:u4d214plot} and Tab.~\ref{tab:Ut2g214}). 

The orbitally-resolved interactions (see Appendix D) calculated within the \t2g-\t2g\ Hamiltonian are more anisotropic than their analogues within the \dd-\dd\pp\ Hamiltonian. This is another signature of the less spherical character of the downfolded orbitals within \t2g-\t2g. The largest interactions are interestingly obtained on the \dxy\ local orbital, for both the screened and bare cases. 

The on-site \t2g\ interaction $\mathcal{U}$ and Hund's exchange $\mathcal{J}$ exhibit trends that are similar to $\mathcal{V}$ and $\mathcal{J}_{\textrm{bare}}$ (Fig.~\ref{fig:u4d214plot}). The decrease from \srmool\ to \srrhol\ is even more pronounced. This is due to the screening which strongly and non-linearly increases between \srruol\ and \srrhol. Indeed, the ratio $\mathcal{U}/\mathcal{V}$ within the \t2g-\t2g\ model (Tab.~\ref{tab:Ut2g214}) is divided by almost a factor two through the series, whereas the ratio $U_{mm}/v_{mm}$ is slowly decreasing within the \dd-\dd\pp\ model. 

The difference in the screening between the two low-energy Hamiltonians comes from the transitions between \t2g\ and \eg\ Kohn-Sham eigenstates, which are removed from the total polarization in the \dd-\dd\pp\ model. These transitions -- by causing a notable increase of the screening between \srruol\ and \srrhol\ -- are responsible for the smaller effective interactions in the latter. Within DFT-LDA (Fig.~\ref{fig:214KS}), the \eg's come closer to the Fermi level in the late \srmol\ perovskites. The \eg\ bands are even partially filled in \srrhol\ leading to a metallic screening that contributes to lower the effective interactions. This is an artefact of the undistorted crystal structure of \srrhol\ and is not the case for the realistic distorted structure \cite{Sr2IrO4-cyril}.

\section{Perspectives : using interactions calculated within cRPA in many-body calculations}

\subsection{Dynamical screening effects}

Both, calculations using a \t2g\ Hamiltonian and a \dd-\dd\pp\ Hamiltonian
do exist in the literature for one of the compounds considered in this
work, namely
\srvo. Indeed, this compound has been chosen as a benchmark system
for nearly every new implementation of LDA+DMFT \cite{SrVO3-sekiyama,AnisimovSrVO3,AmadonSrVO3,Wannier-lechermann-2006,cRPA-DMFT-LaOFeAs-markus}. 
In Ref.~[\onlinecite{cRPA-DMFT-LaOFeAs-markus}] for example, such a calculation within a \t2g\ model was performed
with a Hubbard-Kanamori $\mathcal{U}=4$ eV, yielding a quasi-particle renormalization 
of $Z = 0.6$. A second calculation used an energy window of -8.10 eV 
to 1.90 eV, spanning both, the bands used in the \t2g\ model and the 
oxygen-\pp\ dominated bands located between -8 and -2 eV, thus 
corresponding to the \dd-\dd\pp\ model in the notation of the present work.
For the latter, a value of the intra-orbital interactions, $U_{mm}=6.0$ eV was used, leading to a similar
quasi-particle renormalization ($Z=0.57$) as in the \t2g-only model.

The need for a larger interaction value in the \dd-\dd\pp\ model than in the \t2g\
model for reproducing the same mass renormalization is expected,
and consistent with our results. The interaction values used in the 
calculations of Ref.~[\onlinecite{cRPA-DMFT-LaOFeAs-markus}] are however larger than what is found within cRPA: 
$\mathcal{U}=3.2$ eV for the \t2g\ model (Tab.~\ref{tab:u3dTM}), $U_{mm}=4$ eV
for the \dd-\dd\pp\ one (Tab.~\ref{tab:slaterTMddp}).
This fact corresponds to the general observation 
of interaction values used in many-body calculations
in the literature being quite often somewhat larger than
the ones calculated from cRPA.

This fact has remained a puzzle in the field for quite a while. 
Very recently a solution was proposed in Ref.~[\onlinecite{udyneff-michele}],
based on considerations including effects from the
dynamical screening of the Coulomb interactions
in correlated materials~\cite{udyn-michele,udyn-werner}.
These works concluded that the inclusion
of dynamical screening effects resolves the apparent mismatch 
between cRPA values for the Hubbard interactions 
and what is needed e.g. in standard LDA+DMFT 
to obtain agreement with experiments (e.g. to reproduce the experimental 
mass enhancement). We call here ``standard LDA+DMFT'' the usual
procedure of supplementing a one-particle Hamiltonian obtained
from LDA with static Hubbard (and Hund) interactions, and solving
the resulting many-body Hamiltonian within DMFT.
The extension of this procedure investigated in 
Refs.~[\onlinecite{udyn-michele,udyn-werner}]
takes into account also the frequency-dependence of the 
Hubbard $U$ that results from the dynamical nature of screening.
To lowest order, the net effect of this frequency-dependence is
an additional renormalization of the {\it one-particle part} of the
Hamiltonian (see an explicit discussion of this effect in
Ref.~[\onlinecite{udyneff-michele}]).
When this 
additional modification is taken into account  -- either implicitly
by an explicit inclusion of dynamical $U$ in the DMFT calculation,
as done for \srvo\ e.g. in Ref.~[\onlinecite{cRPA-DMFT-LaOFeAs-markus}],
or explicitly by determining the additional renormalization factor
from the frequency-dependence of $U$ as in Ref.~[\onlinecite{udyneff-michele}] 
-- the correct mass renormalization is
obtained, even when using the seemingly small ab initio value
obtained within cRPA.
One can thus conclude that standard LDA+DMFT calculations
need artificially increased $U$ values to compensate for the lack 
of dynamical screening effects. 
If these effects are however included (e.g. by an explicit
renormalization of the one-particle part of the Hamiltonian
following Ref.~[\onlinecite{udyneff-michele}]) 
the values calculated within the present work should be used.

\subsection{\t2g-\t2g\ or \dd-\dd\pp\ Models ?}

We have calculated, using our cRPA scheme, both, Hubbard and Hund
interaction parameters suitable for \t2g-only and \dd-\dd\pp\ models.
For materials where both choices yield a valid effective low-energy
description of the physical properties, observables calculated
within a many-body calculation using either model should give the
same results. This has been verified explicitly for the case of
\srvo\ (e.g. in Ref.~[\onlinecite{cRPA-DMFT-LaOFeAs-markus}]), 
albeit with the {\it caveat} about additional renormalizations
stemming from dynamical screening.
Analogous results can be expected for early transition metal
oxides, where hybridization effects between $d$ and $p$ states
are not too strong. This includes all the materials considered
in the present work, except possibly \srmno, where the high-spin
character of the half-filled \t2g\ shell and the close-lying \eg\
states may make a \dd-\dd\pp\ model more suitable for the calculation
of some observables.

In the general case, one has to be aware of a trade-off that has
to be done in the construction of the low-energy model. Indeed,
\dd-\pp\ hybridization is dealt with in a very different manner in
the two model constructions that we have discussed.

In the \t2g model, the Wannier functions are built 
from the threefold degenerate low-energy bands only.
While having majority \t2g\ character, 
these bands can contain substantial admixing of \pp-states, 
through the \pp-\dd\ hybridization.
Downfolding then results in an ``effective'' \t2g\ orbital, 
which becomes more and more extended
the stronger the hybridization, displaying pronounced tails on the
ligand sites. 
This may not be a problem as such, but the then
quite extended ``effective \t2g'' orbitals may make subsequent local
approximations (in the construction of the model, when inter-site
interaction terms are neglected, or in the solution of the model,
e.g. within DMFT) questionable.
If the orbitals are too extended, one expects on the contrary
also intersite interactions to become substantial, and the limitation
to a Hubbard-type model with only on-site terms becomes questionable.

At the same time, in late transition metal oxides, one encounters
however a second, related problem that invalidates the \t2g\ only
model: the energetic position of the
ligand \pp-states moves towards the Fermi level, and eventually
the gap forms
not between Hubbard bands of \t2g\ character, but between ligand
states and the upper Hubbard band (charge transfer insulator)
\cite{TMO-zaanen}.
We have limited ourselves in this work to early transition metal
oxides in order to investigate only cases where the \t2g-only
model makes physical sense.

In the \dd-\dd\pp\ model, Wannier functions 
are built from both, bands with majority-\dd\ and bands with majority-\pp\
character: the resulting Wannier functions are more localized
and thus more truly of \dd-character than the ones in the \t2g\ model.
Applying the local Hubbard interaction to the \dd\ states only is therefore 
a priori a well-defined procedure.
The question that arises in this context is however the one about
neglecting interorbital interactions, e.g. $U_{pd}$ between \dd\ and \pp\ 
orbitals, or $U_{pp}$ between the \pp-states. While neglecting the latter 
is likely a good approximation because in reality the \pp-states are 
rather extended and nearly full, the neglect of the former seems 
questionable when the \pp-\dd\ hybridization becomes strong.
This is the reason for considering only early transition metal oxides
in this work.
We note that 
these questions are more questions about the 
nature of the {\it appropriate low-energy model} for a given compound 
than about calculating the parameters of such a model. Indeed, cRPA
is well able to also yield $U_{pd}$ or non-local interactions, once
the orbitals and subspaces are defined.

\subsection{Four-index $U$ vs. Slater parametrization}

Following Eq.~\ref{Wcrpa4}, a four-index $U$ is calculated 
and could in principle be used in a many-body calculation without any restriction from the side of cRPA. 
Two-index or average interaction quantities are introduced only in
the perspective of a further many-body calculation such as LDA+U or
LDA+DMFT. In LDA+DMFT for example, even two-index spin-flip and pair-hopping
terms (of energy scale $J$) are not routinely included due to the
numerical cost of such calculation \cite{cRPA-DMFT-FeSe-markus}. 
We note, for the case of \srvo, that the largest 
three-index term is about $10^{-3}$ eV, 
which compares to the smallest two-index one of
 2.01 eV (see Appendix C).
This order of magnitude suggests that the more than two-index terms can reasonably be
neglected.

Of course, restricting the interactions to density-density (two-index) 
interactions does not necessarily imply that these have to be averaged.
At this point, a subtle issue related to the construction of the
one-particle part of the Hamiltonian comes into play. Indeed, LDA+U
as well as LDA+DMFT comprise a ``double counting correction'' meant
to substract from the LDA Hamiltonian the mean-field like contribution 
due to interactions between the correlated orbitals. This term amounts,
in the standard formulation, to a global shift of correlated against
uncorrelated orbitals. If the many-body interactions were very different
for different correlated orbitals, one would expect that also the
double counting would have to bear orbital-dependence. This is an
important but largely unexplored issue, and the standard procedure
consists therefore in choosing spherically averaged (atomic-like)
interactions. 

In practice, this question becomes of relevance only if the shape
of the correlated orbitals is very anisotropic, so that the Slater
parametrization becomes a poor approximation. The quality of the
latter is related to the degree of localization of the orbitals,
so that the quality of the parametrization can to some extent be 
controlled by the choice of the energy window (the larger the energy 
window, the purer the atomic-like \dd-character of the orbitals).
As we show in Appendix C, this parametrization is excellent for
simple oxide materials such as \srvo.

\section{CONCLUSIONS AND OUTLOOK}

Viewing the Hubbard and Hund interactions of a correlated
electron material as bare interactions within a reference system
with reduced number of degrees of freedom, gives these parameters
the status of auxiliary quantities, only meant to represent the
physical fully screened interaction $W$.
From a conceptual point of view, they are then defined as
soon as the reference system is specified. From a practical
point of view, also the approximations to $W$ and to the
polarization of the reference system $P^{sub}$ have to be
chosen. The lattice version of cRPA corresponds in this
language to the choice of a low-energy subsystem, and the
choice of the random phase approximation for the evaluation
of all polarizations.

In this work, we have presented an implementation of the 
constrained Random Phase Approximation in a 
density functional theory electronic structure code within the linearized augmented plane wave framework. The method gives access to the matrix elements of the Hubbard interaction matrix in a localized basis set of a downfolded lattice Hamiltonian. The strength of the Coulomb interactions is parametrized by the Hubbard $U$ and Hund's exchange $J$, which can be used for the
description of correlated electron systems within interacting lattice
Hamiltonians.  

We have calculated the Hubbard $U$ and Hund's exchange $J$ for the 3\dd\ and 4\dd\ ternary oxides \srmo\ (M= V, Cr, Mn) and (M= Nb, Mo, Tc).
Within a low-energy Hamiltonian (``\dd-\dd\pp'') that includes both \dd\ and oxygen \pp\ degrees of freedom -- and an occupation energy cost on \dd\ only -- $U$ does not change much with the \dd\ electron number. We have rationalized this trend 
as a competition between two aspects: increasing orbital localization -- as evidenced also by the increasing bare interaction $v$ -- and increasing screening due to the reduction of the charge transfer energy with the oxygen ligands -- as illustrated by a decrease of the screened interaction $W$. 
An alternative low-energy Hamiltonian, based on the \t2g\ degrees of freedom only (``\t2g-\t2g''), has also been constructed. Within this framework, both $U$ and -- to a lesser extent -- $J$ decrease with the \dd\ electron number in 3\dd\ oxides, as a consequence of the extended character of the localized \t2g\ orbitals. This is due to the downfolding of the tails of oxygen-\pp\ character. In 4\dd\ oxides, the trends for $U$ and $J$ are flat again, because of the larger charge transfer energy with the oxygen-\pp\ states. The surprisingly smaller $U$ for \srmno\ than for \srtco\ within \crpa\ for such Hamiltonian can be understood in terms of the weaker screening effects in the latter. 
 
We have considered analogous low-energy Hamiltonians for the materials with the layered perovskite structure \srmol\ (M= Mo, Tc, Ru, Rh). For the same reasons as in the ternary oxides, $U$ and $J$ increase with the atomic number within the \dd-\dd\pp\ Hamiltonian, while they decrease within the \t2g-\t2g\ Hamiltonian.   

We have emphasized the dependence of the effective Coulomb interactions on the choice of the one-particle part of the Hamiltonian.   
Determining Hubbard $U$ and Hund's $J$ entirely from first principles for a given low-energy Hamiltonian opens the way to a \emph{truly ab initio} description of corrrelated materials within many-body calculations.    

\section*{ACKNOWLEDGMENTS}

We acknowledge useful discussions with M. Aichhorn, F. Aryasetiawan,
M. Casula, A. Georges, D. Khomskii, M. Imada, C. Martins, T. Miyake, J. Mravlje, L. Pourovskii, P. Rinke, G. Sawatzky,
M. Scheffler, M. van Schilfgaarde and V. Vildosola.
This work was supported by the French ANR under project SURMOTT and
IDRIS/GENCI under project 1393, and Natural Science Foundation of China (Project No. 20973009).

While this work was being written, an alternative implementation of \crpa\ within a pseudopotential code was proposed in Ref.~[\onlinecite{ShihPRB12}].

\section*{APPENDIX A : Technicalities of the \crpa\ implementation within the \LAPWlo\ framework}

We report below on the technicalities that are specific to the \crpa\ part. Our \crpa\ implementation relies on the recent GW implementation within the \LAPWlo\ framework called \textsc{fhi-gap}~\cite{FHI-GAP-hong}. Details related to the construction of an optimized product mixed basis for expanding polarizations, dielectric functions and Coulomb interactions, can be found in Ref.~[\onlinecite{FHI-GAP-hong}]. We will hence employ the same notations as in this reference. The product mixed basis, for example, will be denoted $\{ \chi_{i}^{\qqq}(\rrr) \}$, where $\qqq$-vectors belong to the first Brillouin zone and $i$-indices run over the size of the complete product mixed basis. 

The total polarization at the RPA level can be expanded into the product mixed basis $\{ \chi_{i}^{\qqq}(\rrr) \}$~\cite{GW_Ferdi,GW-MB-kotani}:
\begin{eqnarray*}
P(\rrr,\rrr';\omega) &=& \frac{1}{\NN} \sum_{\qqq} \sum_{ij} [\chi_{i}^{\qqq}(\rrr)]^{*}
P_{ij}(\qqq,\omega) \chi_{j}^{\qqq}(\rrr'),
\end{eqnarray*} 
where the $\qqq$-summation (of $\NN$ size) is performed over the first Brillouin zone. The polarization matrix elements in the product basis, $P_{ij}$, read as~: 
\begin{eqnarray*}
P_{ij}(\qqq,\omega) &=& \frac{1}{\NN} \sum_{\kkk} \sum_{n,n'}^{\occ,\unocc}M_{nn'}^{i}(\kkk,\qqq) F_{nn'\kkk}(\qqq,\omega) [M_{nn'}^{j}(\kkk,\qqq)]^{*},
\end{eqnarray*}
where the overlap between the Bloch wavefunctions, $\psi_{\kkk n}$, and the product mixed basis functions, denoted $M_{nn'}^{i}(\kkk,\qqq)$, is defined as follows~:
\begin{eqnarray*}
M_{nn'}^{i}(\kkk,\qqq) &=& \int_{\Omega} d\rrr \, \psi_{\kkk n}(\rrr) [\chi_{i}^{\qqq}(\rrr)\psi_{\kkk-\qqq,n'}(\rrr)]^{*}, 
\end{eqnarray*}
and~:
\begin{eqnarray*}
\omega_{n\kkk,n'\kkk-\qqq} &=& \epsilon_{n'\kkk - \qqq} - \epsilon_{n\kkk} \\
F_{nn'\kkk}(\qqq,\omega) &=& \frac{1}{\omega - \omega_{n\kkk,n'\kkk -\qqq}+ i \eta} - \frac{1}{\omega+\omega_{n \kkk,n' \kkk - \qqq}-i\eta}.
\end{eqnarray*}

Within the assumption that the correlated subspace $\CC$ is unambiguously defined since the target correlated bands do not energetically overlap with the itinerant ones, it follows for the \dd-restricted polarization $P^{d}$ (Eq.~\ref{Pcrpa2}) expanded into the product mixed basis~: 
\begin{eqnarray*}
P_{ij}^{d}(\qqq,\omega) &=& \frac{1}{\NN} \sum_{\kkk} \sum_{d,d'}^{\occ,\unocc}M_{dd'}^{i}(\kkk,\qqq) F_{dd'\kkk}(\qqq,\omega) [M_{dd'}^{j}(\kkk,\qqq)]^{*}, 
\end{eqnarray*} 
and hence for the constrained polarization, $P^{r}$~:
\begin{eqnarray*}
P_{ij}^{r}(\qqq,\omega) &=& P_{ij}(\qqq,\omega) - P_{ij}^{d}(\qqq,\omega).
\end{eqnarray*}

The total symmetrized dielectric function, $\varepsilon$, in the product mixed basis is defined as follows~\cite{FHI-GAP-hong}~:
\begin{eqnarray*}
\varepsilon_{ij}(\qqq,\omega) &=& \delta_{ij} - v_{i}^{\frac{1}{2}}(\qqq) P_{ij}(\qqq,\omega) v_{j}^{\frac{1}{2}}(\qqq) \nonumber \\
&=& \delta_{ij} - \frac{1}{\NN} \sum_{\kkk} \sum_{n,n'}^{\occ,\unocc} v_{i}^{\frac{1}{2}}(\qqq) M_{nn'}^{i}(\kkk,\qqq) F_{nn'\kkk}(\qqq,\omega) \nonumber \\
&& \times [v_{j}^{\frac{1}{2}}(\qqq) M_{nn'}^{j}(\kkk,\qqq)]^{*}. 
\end{eqnarray*}

Analogously, the \dd-constrained (partial) dielectric function, $\varepsilon^{r}$, reads as 
\begin{eqnarray*}
\varepsilon_{ij}^{r}(\qqq,\omega) &=& \delta_{ij} - v_{i}^{\frac{1}{2}}(\qqq) P_{ij}^{r}(\qqq,\omega) v_{j}^{\frac{1}{2}}(\qqq).
\end{eqnarray*}

From the Dyson-like equation for the screened interaction (Eq.~\ref{eq:Wr}), it follows for $W, W^{r}$ expanded into the product mixed basis~:
\begin{eqnarray*}
W_{ij}(\qqq,\omega) &=& v_{i}^{\frac{1}{2}}(\qqq) \varepsilon_{ij}^{-1}(\qqq,\omega) v_{j}^{\frac{1}{2}}(\qqq) \\
W_{ij}^{r}(\qqq,\omega) &=& v_{i}^{\frac{1}{2}}(\qqq) [\varepsilon^{r}(\qqq,\omega)]_{ij}^{-1} v_{j}^{\frac{1}{2}}(\qqq).
\end{eqnarray*}

The product Kohn-Sham representation of $W^{r}$ (Eq.~\ref{Wcrpa4}) can be expanded into the product mixed basis \emph{via} the overlap quantities $M_{nn'}^{i}(\kkk,\qqq)$~:
\begin{eqnarray*}
&& \langle \psi_{\kkk_{1}n_{1}} \psi_{\kkk_{2}n_{2}}|W^{r}|\psi_{\kkk_{3}n_{3}} \psi_{\kkk_{4}n_{4}} \rangle = \frac{1}{\NN} \sum_{\qqq} \sum_{ij} [M_{n_{1}n_{3}}^{i}(\kkk_{1},\qqq)]^{*} \\ 
&& \times W^{r}_{ij}(\qqq,\omega) M_{n_{4}n_{2}}^{j}(\kkk_{4},\qqq) \delta_{\kkk_{3},\kkk_{1}-\qqq} \delta_{\kkk_{2},\kkk_{4}-\qqq}. 
\end{eqnarray*}
and therefore, it follows for the Hubbard interactions~: 
\begin{eqnarray*}
&& U_{L_{1}L_{2}L_{3}L_{4}}^{\RRR_{1}\RRR_{2}\RRR_{3}\RRR_{4}}(\omega) = \frac{1}{\NN} \sum_{\qqq} \expo^{i\qqq \cdot (\RRR_{3}-\RRR_{2})} \sum_{ij} [M_{L_{1}\RRR_{1},L_{3}\RRR_{3}}^{i}(\qqq)]^{*} \\
&& ~~~~~~~~~~~~~~ \times W_{ij}^{r}(\qqq,\omega)M_{L_{4}\RRR_{4},L_{2}\RRR_{2}}^{j}(\qqq),
\end{eqnarray*}
where $M_{L\RRR,L\RRR'}^{i}(\qqq)$ are defined as follows~:
\begin{eqnarray*}
&& M_{L\RRR,L'\RRR'}^{i}(\qqq) =  \frac{1}{\NN} \sum_{\kkk} \expo^{-i\kkk \cdot (\RRR -\RRR')} \\
&& ~~~~~~~~~~~~~~~~~~~ \times \sum_{n,n' \in \WW} [P_{Ln}(\kkk,)]^{*} M_{nn'}^{i}(\kkk,\qqq) P_{L'n'}(\kkk-\qqq).
\end{eqnarray*} 

\section*{APPENDIX B : Parametrization of the Hubbard interaction matrix with Slater integrals}
The interaction matrix in atoms can be efficiently parametrized by a finite number of Slater integrals~\cite{judd,Slater_book,Sugano_book} thanks to the atomic sphericity. Such parametrization can be extended to solids if one assumes that within the solid~:
\begin{eqnarray}
U_{m_{1}m_{2}m_{3}m_{4}}^{\textrm{(spheric)}} &=& \sum_{k=0}^{2l} \alpha_{k}(m_{1},m_{2},m_{3},m_{4}) \textrm{F}^{k},
\end{eqnarray}
where the angular part is described with Racah-Wigner coefficients. $\alpha_{k}(m_{1},m_{2},m_{3},m_{4})$ are calculated with spherical harmonics, $Y_{lm}$, as follows~:
\begin{eqnarray}
\alpha_{k}(m_{1},m_{2},m_{3},m_{4}) &=& \frac{4\pi}{2k+1}\sum_{q=-k}^{k} \langle Y_{lm_{1}}| Y_{kq} Y_{lm_{3}} \rangle  \nonumber \\
&& \times \langle Y_{lm_{2}} Y_{kq} | Y_{lm_{4}} \rangle, 
\end{eqnarray}
($\langle Y_{l_{1}m_{1}} | Y_{l_{2}m_{2}}  Y_{l_{3}m_{3}} \rangle$ corresponds to the Gaunt coefficient calculated with spherical harmonics), whereas the radial part is expressed in terms of Slater integrals $\{\textrm{F}^{k}\}$. These Slater integrals are deduced from the interaction matrix elements computed by \crpa\ in the Wannier basis, $\{\phi_{m, -2\leq m \leq 2}\}$ that is said ``spheric'' because of its complex representation \cite{PhD-Vaugier}:
\begin{eqnarray}
\textrm{F}^{k}(\omega) &=& \mathcal{C}_{l,k} \sum_{m_{1},m_{2},m_{3},m_{4}} (-1)^{m_{1}+m_{4}} U_{m_{1}m_{2}m_{3}m_{4}}^{\textrm{(spheric)}}(\omega) \nonumber \\
&& \times \left( {\setlength\arraycolsep{1pt}\begin{array}{ccc} l&k&l \\ -m_{1}&m_{1}-m_{3}&m_{3} \end{array}} \right) \left( {\setlength\arraycolsep{1pt}\begin{array}{ccc} l&k&l \\ -m_{2}&m_{2}-m_{4}&m_{4} \end{array}} \right), \nonumber \\
&&
\label{slaterFU1}
\end{eqnarray}  
where the parentheses correspond to the Wigner 3j-symbols and the coefficients $\mathcal{C}_{l,k}$ are defined as follows~: 
\begin{eqnarray}
\mathcal{C}_{l,k} &=& \frac{2k+1}{(2l+1)^{2}\left( {\setlength\arraycolsep{2pt}\begin{array}{ccc} l&k&l \\ 0&0&0 \end{array}} \right)^{2}}.
\end{eqnarray} 
The frequency dependence of the Slater integrals arises from the frequency dependence of the interaction matrix elements induced by the dynamical screening. 

\section*{APPENDIX C : Reduced interaction matrices for \srmo\ (M = V, Cr, Mn)}
We give below the reduced interaction matrices (Eqs.~\ref{reducedU-1},
\ref{reducedU-2} and~\ref{reducedU-3}) within the \dd-\dd\pp\ model
(see Tab.~\ref{tab:windows113} for the choice of the energy windows
used to construct the \dd\ Wannier orbitals) and calculated with cubic
symmetry. In the following, the ordering of the orbitals in these
matrices is $d_{z^{2}},d_{x^{2}-y^{2}},d_{xy},d_{xz},d_{yz}$. The
values are given in eV. 

In the case of \srvo, we also add the reduced interaction matrices,
$\bar{U}_{mm}$ (Eqs.~\ref{slaterFU2}, \ref{slater4}, \ref{slater5} and
\ref{slater6})
that are deduced from the Slater integrals given in
Tab.~\ref{tab:slaterTMddp}. This allows for an estimation of the
accuracy of the Slater parametrization.  

\subsubsection*{\srvo}
\begin{displaymath}
U_{mm'}^{\sigma \bar{\sigma}}= \left( 
\begin{array}{ccccc}
 4.43 & 2.88 &   2.73 &  3.19 &   3.19 \\
 2.88 & 4.43 &   3.35 &  2.88 &   2.88 \\
 2.73 & 3.35 &   3.97 &  2.75 &   2.75 \\
 3.19 & 2.88 &   2.75 &  3.97 &   2.75 \\
 3.19 & 2.88 &   2.75 &  2.75 &   3.97
\end{array}
\right),
\end{displaymath} 
\begin{displaymath}
U_{mm'}^{\sigma \sigma} = \left( 
\begin{array}{ccccc}
 0      & 2.10 & 2.01  & 2.70  & 2.70   \\ 
 2.10 & 0      & 2.94  & 2.24  & 2.24 \\
 2.01 & 2.94 & 0       & 2.15  & 2.15 \\
 2.70 & 2.24 & 2.15  & 0       & 2.15 \\
 2.70 & 2.24 & 2.15  & 2.15  & 0 
\end{array}
\right).
\end{displaymath}

Employing the Slater integrals $\textrm{F}^{0}=3.2$ eV, $\textrm{F}^{2}=6.6$ and $\textrm{F}^{4}=5.3$
eV, we obtain for the Slater-parametrized reduced interaction matrices
with cubic symmetry~:
\begin{displaymath}
\bar{U}_{mm'}^{\sigma \bar{\sigma}}= \left( 
\begin{array}{ccccc}
 4.10 & 2.73 &   2.73 &  3.18 &   3.18 \\
 2.73 & 4.10 &   3.33 &  2.88 &   2.88 \\
 2.73 & 3.33 &   4.10 &  2.88 &   2.88 \\
 3.18 & 2.88 &   2.88 &  4.10 &   2.88 \\
 3.18 & 2.88 &   2.88 &  2.88 &   4.10
\end{array}
\right),
\end{displaymath} 
\begin{displaymath}
\bar{U}_{mm'}^{\sigma \sigma} = \left( 
\begin{array}{ccccc}
 0      & 2.01 & 2.01  & 2.68  & 2.68   \\ 
 2.01 & 0      & 2.91  & 2.24  & 2.24 \\
 2.01 & 2.91 & 0       & 2.24  & 2.24 \\
 2.68 & 2.24 & 2.24  & 0       & 2.24 \\
 2.68 & 2.24 & 2.24  & 2.24  & 0 
\end{array}
\right).
\end{displaymath}

\subsubsection*{\srcro}
\begin{displaymath}
U_{mm'}^{\sigma \bar{\sigma}}= \left( 
\begin{array}{ccccc}
 3.84 & 2.39 & 2.40 & 2.87 & 2.87   \\ 
 2.39 & 3.84 & 3.02 & 2.56 & 2.56 \\
 2.40 & 3.02 & 3.89 & 2.59 & 2.59 \\
 2.87 & 2.56 & 2.59 & 3.89 & 2.59 \\
 2.87 & 2.56 & 2.59 & 2.59 & 3.89 
\end{array}
\right),
\end{displaymath} 
\begin{displaymath}
U_{mm'}^{\sigma \sigma} = \left( 
\begin{array}{ccccc}
 0      & 1.66 & 1.67  & 2.38  & 2.38   \\ 
 1.66 & 0      & 2.61  & 1.91  & 1.91 \\
 1.67 & 2.61 & 0       & 1.94  & 1.94 \\
 2.38 & 1.91 & 1.94  & 0       & 1.94 \\
 2.38 & 1.91 & 1.94  & 1.94  & 0 
\end{array}
\right).
\end{displaymath}

\subsubsection*{\srmno}
\begin{displaymath}
U_{mm'}^{\sigma \bar{\sigma}}= \left( 
\begin{array}{ccccc}
 3.62 & 2.17 & 2.25 & 2.74 & 2.74   \\ 
 2.17 & 3.62 & 2.90 & 2.41 & 2.41 \\
 2.25 & 2.90 & 3.90 & 2.52 & 2.51 \\
 2.74 & 2.41 & 2.52 & 3.91 & 2.52 \\
 2.74 & 2.41 & 2.51 & 2.52 & 3.90 
\end{array}
\right),
\end{displaymath} 
\begin{displaymath}
U_{mm'}^{\sigma \sigma} = \left( 
\begin{array}{ccccc}
 0      & 1.44 & 1.49  & 2.22  & 2.22   \\ 
 1.44 & 0      & 2.47  & 1.73  & 1.73 \\
 1.49 & 2.47 & 0       & 1.81  & 1.81 \\
 2.22 & 1.73 & 1.81  & 0       & 1.81 \\
 2.22 & 1.73 & 1.81  & 1.81  & 0 
\end{array}
\right).
\end{displaymath}

\section*{APPENDIX D : Reduced interaction matrices for \srmol\ (M = Mo, Tc, Ru, Rh)}

\subsection*{``\dd-\dd\pp\ Hamiltonian''}
The reduced interaction matrices within the \dd-\dd\pp\ model (Tab.~\ref{tab:windows214}) and calculated with cubic symmetry, are given below for the layered perovskites. The ordering of the orbitals in these matrices is $d_{z^{2}},d_{x^{2}-y^{2}},d_{xy},d_{xz},d_{yz}$. The values are given in eV. 

\subsubsection*{\srmool}
\begin{displaymath}
U_{mm'}^{\sigma \bar{\sigma}}= \left( 
\begin{array}{ccccc}
 4.20 & 3.04 & 2.90 & 3.22 & 3.22   \\ 
 3.04 & 4.35 & 3.43 & 3.06 & 3.05 \\
 2.90 & 3.43 & 3.97 & 2.93 & 2.92 \\
 3.22 & 3.06 & 2.93 & 3.86 & 2.89 \\
 3.22 & 3.05 & 2.92 & 2.89 & 3.84 
\end{array}
\right),
\end{displaymath} 
\begin{displaymath}
U_{mm'}^{\sigma \sigma} = \left( 
\begin{array}{ccccc}
 0      & 2.48 & 2.38  & 2.83  & 2.83   \\ 
 2.48 & 0      & 3.10  & 2.56  & 2.55 \\
 2.38 & 3.10 & 0       & 2.44  & 2.44 \\
 2.83 & 2.56 & 2.44  & 0       & 2.41 \\
 2.83 & 2.55 & 2.44  & 2.41  & 0 
\end{array}
\right).
\end{displaymath}

\subsubsection*{\srtcol}
\begin{displaymath}
U_{mm'}^{\sigma \bar{\sigma}}= \left( 
\begin{array}{ccccc}
 4.06 & 2.88 & 2.81 & 3.13 & 3.13   \\ 
 2.88 & 4.23 & 3.38 & 2.96 & 2.96 \\
 2.81 & 3.38 & 4.04 & 2.90 & 2.90 \\
 3.13 & 2.96 & 2.90 & 3.86 & 2.84 \\
 3.13 & 2.96 & 2.90 & 2.84 & 3.86 
\end{array}
\right),
\end{displaymath} 
\begin{displaymath}
U_{mm'}^{\sigma \sigma} = \left( 
\begin{array}{ccccc}
 0      & 2.29 & 2.25  & 2.72  & 2.72   \\ 
 2.29 & 0      & 3.04  & 2.44  & 2.44 \\
 2.25 & 3.04 & 0       & 2.38  & 2.38 \\
 2.72 & 2.44 & 2.38  & 0       & 2.33 \\
 2.72 & 2.44 & 2.38  & 2.33  & 0 
\end{array}
\right).
\end{displaymath}

\subsubsection*{\srruol}
\begin{displaymath}
U_{mm'}^{\sigma \bar{\sigma}}= \left( 
\begin{array}{ccccc}
 4.06 & 2.83 & 2.78 & 3.17 & 3.18   \\ 
 2.83 & 4.22 & 3.37 & 2.98 & 2.99 \\
 2.78 & 3.37 & 4.07 & 2.94 & 2.95 \\
 3.17 & 2.98 & 2.94 & 4.02 & 2.93 \\
 3.18 & 2.99 & 2.95 & 2.93 & 4.05 
\end{array}
\right),
\end{displaymath} 
\begin{displaymath}
U_{mm'}^{\sigma \sigma} = \left( 
\begin{array}{ccccc}
 0      & 2.22 & 2.19  & 2.73  & 2.74   \\ 
 2.22 & 0      & 3.01  & 2.42  & 2.44 \\
 2.19 & 3.01 & 0       & 2.39  & 2.39 \\
 2.73 & 2.42 & 2.39  & 0       & 2.38 \\
 2.74 & 2.44 & 2.39  & 2.38  & 0 
\end{array}
\right).
\end{displaymath}

\subsubsection*{\srrhol}
\begin{displaymath}
U_{mm'}^{\sigma \bar{\sigma}}= \left( 
\begin{array}{ccccc}
 4.18 & 2.93 & 2.95 & 3.37 & 3.38   \\ 
 2.93 & 4.34 & 3.55 & 3.15 & 3.16 \\
 2.95 & 3.55 & 4.37 & 3.18 & 3.19 \\
 3.37 & 3.15 & 3.18 & 4.39 & 3.20 \\
 3.38 & 3.16 & 3.19 & 3.20 & 4.44 
\end{array}
\right),
\end{displaymath} 
\begin{displaymath}
U_{mm'}^{\sigma \sigma} = \left( 
\begin{array}{ccccc}
 0      & 2.29 & 2.31  & 2.92  & 2.94   \\ 
 2.29 & 0      & 3.17  & 2.56  & 2.58 \\
 2.31 & 3.17 & 0       & 2.57  & 2.58 \\
 2.92 & 2.56 & 2.57  & 0       & 2.59 \\
 2.94 & 2.58 & 2.58  & 2.59  & 0 
\end{array}
\right).
\end{displaymath}

\subsection*{``\t2g-\t2g\ Hamiltonian''}
We give below the reduced interaction matrices for the \t2g\ local orbitals within the \t2g-\t2g\ model. The Hubbard-Kanamori parameters, \uk\ and \uk', shown in Table~\ref{tab:Ut2g214} correspond to the orbital average, $U_{mm}^{\sigma \bar{\sigma}}, U_{m \ne m'}^{\sigma \bar{\sigma}}$ (Eqs. \ref{reducedU-1} and \ref{reducedU-3}), respectively. The exchange parameter, \jk,  corresponds to the orbital average of $J_{mm'} = (U_{mm'}^{\sigma \bar{\sigma}}-U_{mm'}^{\sigma \sigma}) (1-\delta_{mm'})$ (Eq. \ref{reducedU-2}). The ordering of the orbitals is $(d_{xy},d_{xz},d_{yz})$.  

\subsubsection*{\srmool}
\begin{displaymath}
U_{mm'}^{\sigma \bar{\sigma}}= \left( 
\begin{array}{ccc}
   2.96 &  2.19 &  2.19 \\
   2.19 &  2.68 &  2.09 \\
   2.19 &  2.09 &  2.68
\end{array}
\right),
\end{displaymath} 
\begin{displaymath}
U_{mm'}^{\sigma \sigma} = \left( 
\begin{array}{ccc}
   0.00 &  1.89 & 1.89 \\
   1.89 &  0.00 & 1.82 \\
   1.89 &  1.82 & 0.00 
\end{array}
\right).
\end{displaymath}

\subsubsection*{\srtcol}
\begin{displaymath}
U_{mm'}^{\sigma \bar{\sigma}}= \left( 
\begin{array}{ccc}
   2.89  & 2.10 & 2.10 \\
   2.10  & 2.61 & 2.01 \\
   2.10  & 2.01 & 2.61
\end{array}
\right),
\end{displaymath} 
\begin{displaymath}
U_{mm'}^{\sigma \sigma} = \left( 
\begin{array}{ccc}
   0.00 & 1.81 & 1.81 \\
   1.81 & 0.00 & 1.74 \\
   1.81 & 1.74 & 0.00
\end{array}
\right).
\end{displaymath}

\subsubsection*{\srruol}
\begin{displaymath}
U_{mm'}^{\sigma \bar{\sigma}}= \left( 
\begin{array}{ccc}
   2.72 &  1.97 &  1.97 \\
   1.97 &  2.48 &  1.89 \\
   1.97 &  1.89 &  2.48
\end{array}
\right),
\end{displaymath} 
\begin{displaymath}
U_{mm'}^{\sigma \sigma} = \left( 
\begin{array}{ccc}
   0.00 &  1.71 &  1.71 \\
   1.71 &  0.00 &  1.65 \\
   1.71 &  1.65 &  0.00
\end{array}
\right).
\end{displaymath}

\subsubsection*{\srrhol}
\begin{displaymath}
U_{mm'}^{\sigma \bar{\sigma}}= \left( 
\begin{array}{ccc}
   1.81 &  1.16 &  1.20 \\
   1.16 &  1.69 &  1.19 \\  
   1.20 &  1.19 &  1.77
\end{array}
\right),
\end{displaymath} 
\begin{displaymath}
U_{mm'}^{\sigma \sigma} = \left( 
\begin{array}{ccc}
   0.00 &   0.93 & 0.96 \\
   0.93 &   0.00 & 0.97 \\   
   0.96 &   0.97 & 0.00
\end{array}
\right).
\end{displaymath}

\addcontentsline{toc}{section}{Bibliography}

\begin{thebibliography}{200}
\expandafter\ifx\csname natexlab\endcsname\relax\def\natexlab#1{#1}\fi
\expandafter\ifx\csname bibnamefont\endcsname\relax
  \def\bibnamefont#1{#1}\fi
\expandafter\ifx\csname bibfnamefont\endcsname\relax
  \def\bibfnamefont#1{#1}\fi
\expandafter\ifx\csname citenamefont\endcsname\relax
  \def\citenamefont#1{#1}\fi
\expandafter\ifx\csname url\endcsname\relax
  \def\url#1{\texttt{#1}}\fi
\expandafter\ifx\csname urlprefix\endcsname\relax\def\urlprefix{URL }\fi
\providecommand{\bibinfo}[2]{#2}
\providecommand{\eprint}[2][]{\url{#2}}

\bibitem[{\citenamefont{Imada et~al.}(1998)\citenamefont{Imada, Fujimori, and
  Tokura}}]{review-imada}
\bibinfo{author}{\bibfnamefont{M.}~\bibnamefont{Imada}},
  \bibinfo{author}{\bibfnamefont{A.}~\bibnamefont{Fujimori}}, \bibnamefont{and}
  \bibinfo{author}{\bibfnamefont{Y.}~\bibnamefont{Tokura}},
  \bibinfo{journal}{Rev. Mod. Phys.} \textbf{\bibinfo{volume}{70}},
  \bibinfo{pages}{1039} (\bibinfo{year}{1998}).

\bibitem[{\citenamefont{Kohn}(1999)}]{Kohn-nobel}
\bibinfo{author}{\bibfnamefont{W.}~\bibnamefont{Kohn}}, \bibinfo{journal}{Rev.
  Mod. Phys.} \textbf{\bibinfo{volume}{71}}, \bibinfo{pages}{1253}
  (\bibinfo{year}{1999}).

\bibitem[{\citenamefont{Hohenberg and Kohn}(1964)}]{hohenberg-kohn-1964}
\bibinfo{author}{\bibfnamefont{P.}~\bibnamefont{Hohenberg}} \bibnamefont{and}
  \bibinfo{author}{\bibfnamefont{W.}~\bibnamefont{Kohn}},
  \bibinfo{journal}{Phys. Rev.} \textbf{\bibinfo{volume}{136}},
  \bibinfo{pages}{B864} (\bibinfo{year}{1964}).

\bibitem[{\citenamefont{Kohn and Sham}(1965)}]{DFT-Kohn-Sham}
\bibinfo{author}{\bibfnamefont{W.}~\bibnamefont{Kohn}} \bibnamefont{and}
  \bibinfo{author}{\bibfnamefont{L.~J.} \bibnamefont{Sham}},
  \bibinfo{journal}{Phys. Rev.} \textbf{\bibinfo{volume}{140}},
  \bibinfo{pages}{A1133} (\bibinfo{year}{1965}).

\bibitem[{\citenamefont{Jones and Gunnarsson}(1989)}]{DFT-review}
\bibinfo{author}{\bibfnamefont{R.~O.} \bibnamefont{Jones}} \bibnamefont{and}
  \bibinfo{author}{\bibfnamefont{O.}~\bibnamefont{Gunnarsson}},
  \bibinfo{journal}{Rev. Mod. Phys.} \textbf{\bibinfo{volume}{61}},
  \bibinfo{pages}{689} (\bibinfo{year}{1989}).


\bibitem[{\citenamefont{Anisimov
  et~al.}(1997{\natexlab{a}})\citenamefont{Anisimov, Aryasetiawan, and
  Lichtenstein}}]{LDA+U-anisimov-1997}
\bibinfo{author}{\bibfnamefont{V.~I.} \bibnamefont{Anisimov}},
  \bibinfo{author}{\bibfnamefont{F.}~\bibnamefont{Aryasetiawan}},
  \bibnamefont{and} \bibinfo{author}{\bibfnamefont{A.~I.}
  \bibnamefont{Lichtenstein}}, \bibinfo{journal}{J. Phys. Condens. Matter}
  \textbf{\bibinfo{volume}{9}}, \bibinfo{pages}{767}
  (\bibinfo{year}{1997}{\natexlab{a}}).

\bibitem[{\citenamefont{Anisimov
  et~al.}(1997{\natexlab{b}})\citenamefont{Anisimov, Poteryaev, Korotin,
  Anokhin, and Kotliar}}]{LDA+DMFT-anisimov-1997}
\bibinfo{author}{\bibfnamefont{V.~I.} \bibnamefont{Anisimov}},
  \bibinfo{author}{\bibfnamefont{A.}~\bibnamefont{Poteryaev}},
  \bibinfo{author}{\bibfnamefont{M.}~\bibnamefont{Korotin}},
  \bibinfo{author}{\bibfnamefont{A.}~\bibnamefont{Anokhin}}, \bibnamefont{and}
  \bibinfo{author}{\bibfnamefont{G.}~\bibnamefont{Kotliar}},
  \bibinfo{journal}{J. Phys. Condens. Matter} \textbf{\bibinfo{volume}{9}},
  \bibinfo{pages}{943} (\bibinfo{year}{1997}{\natexlab{b}}).

\bibitem[{\citenamefont{Lichtenstein and Katsnelson}(1998)}]{LDA+DMFT-licht}
\bibinfo{author}{\bibfnamefont{A.~I.} \bibnamefont{Lichtenstein}}
  \bibnamefont{and} \bibinfo{author}{\bibfnamefont{M.~I.}
  \bibnamefont{Katsnelson}}, \bibinfo{journal}{Phys. Rev. B}
  \textbf{\bibinfo{volume}{57}}, \bibinfo{pages}{6884} (\bibinfo{year}{1998}).

\bibitem[{\citenamefont{Kotliar et~al.}(2006)\citenamefont{Kotliar, Savrasov,
  Haule, Oudovenko, Parcollet, and Marianetti}}]{kotliar-review-DMFT}
\bibinfo{author}{\bibfnamefont{G.}~\bibnamefont{Kotliar}},
  \bibinfo{author}{\bibfnamefont{S.~Y.} \bibnamefont{Savrasov}},
  \bibinfo{author}{\bibfnamefont{K.}~\bibnamefont{Haule}},
  \bibinfo{author}{\bibfnamefont{V.~S.} \bibnamefont{Oudovenko}},
  \bibinfo{author}{\bibfnamefont{O.}~\bibnamefont{Parcollet}},
  \bibnamefont{and} \bibinfo{author}{\bibfnamefont{C.~A.}
  \bibnamefont{Marianetti}}, \bibinfo{journal}{Rev. Mod. Phys.}
  \textbf{\bibinfo{volume}{78}}, \bibinfo{pages}{865} (\bibinfo{year}{2006}).

\bibitem[{\citenamefont{Aryasetiawan et~al.}(2004)\citenamefont{Aryasetiawan,
  Imada, Georges, Kotliar, Biermann, and Lichtenstein}}]{cRPA-ferdi-2004}
\bibinfo{author}{\bibfnamefont{F.}~\bibnamefont{Aryasetiawan}},
  \bibinfo{author}{\bibfnamefont{M.}~\bibnamefont{Imada}},
  \bibinfo{author}{\bibfnamefont{A.}~\bibnamefont{Georges}},
  \bibinfo{author}{\bibfnamefont{G.}~\bibnamefont{Kotliar}},
  \bibinfo{author}{\bibfnamefont{S.}~\bibnamefont{Biermann}}, \bibnamefont{and}
  \bibinfo{author}{\bibfnamefont{A.~I.} \bibnamefont{Lichtenstein}},
  \bibinfo{journal}{Phys. Rev. B} \textbf{\bibinfo{volume}{70}},
  \bibinfo{pages}{195104} (\bibinfo{year}{2004}).

\bibitem[{\citenamefont{L\"owdin}(1951)}]{downfolding-lowdin}
\bibinfo{author}{\bibfnamefont{P.}~\bibnamefont{L\"owdin}},
  \bibinfo{journal}{J. Chem. Phys.} \textbf{\bibinfo{volume}{19}},
  \bibinfo{pages}{1396} (\bibinfo{year}{1951}).

\bibitem[{\citenamefont{Andersen and Saha-Dasgupta}(2000)}]{NMTO_andersen}
\bibinfo{author}{\bibfnamefont{O.~K.} \bibnamefont{Andersen}} \bibnamefont{and}
  \bibinfo{author}{\bibfnamefont{T.}~\bibnamefont{Saha-Dasgupta}},
  \bibinfo{journal}{Phys. Rev. B} \textbf{\bibinfo{volume}{62}},
  \bibinfo{pages}{R16219} (\bibinfo{year}{2000}).

\bibitem[{\citenamefont{Blaha et~al.}(2001)\citenamefont{Blaha, Schwarz,
  Madsen, Kvasnicka, and Luitz}}]{blaha_wien2k}
\bibinfo{author}{\bibfnamefont{P.}~\bibnamefont{Blaha}},
  \bibinfo{author}{\bibfnamefont{K.}~\bibnamefont{Schwarz}},
  \bibinfo{author}{\bibfnamefont{G.}~\bibnamefont{Madsen}},
  \bibinfo{author}{\bibfnamefont{D.}~\bibnamefont{Kvasnicka}},
  \bibnamefont{and} \bibinfo{author}{\bibfnamefont{J.}~\bibnamefont{Luitz}},
  \emph{\bibinfo{title}{\textsf{Wien2k}, {A}n {A}ugmented {P}lane
  {W}ave+{L}ocal {O}rbitals {P}rogram for {C}alculating {C}rystal
  {P}roperties}} (\bibinfo{publisher}{Tech. Universit\"at Wien, Austria},
  \bibinfo{year}{2001}).

\bibitem[{\citenamefont{Aichhorn et~al.}(2009)\citenamefont{Aichhorn,
  Pourovskii, Vildosola, Ferrero, Parcollet, Miyake, Georges, and
  Biermann}}]{cRPA-DMFT-LaOFeAs-markus}
\bibinfo{author}{\bibfnamefont{M.}~\bibnamefont{Aichhorn}},
  \bibinfo{author}{\bibfnamefont{L.}~\bibnamefont{Pourovskii}},
  \bibinfo{author}{\bibfnamefont{V.}~\bibnamefont{Vildosola}},
  \bibinfo{author}{\bibfnamefont{M.}~\bibnamefont{Ferrero}},
  \bibinfo{author}{\bibfnamefont{O.}~\bibnamefont{Parcollet}},
  \bibinfo{author}{\bibfnamefont{T.}~\bibnamefont{Miyake}},
  \bibinfo{author}{\bibfnamefont{A.}~\bibnamefont{Georges}}, \bibnamefont{and}
  \bibinfo{author}{\bibfnamefont{S.}~\bibnamefont{Biermann}},
  \bibinfo{journal}{Phys. Rev. B} \textbf{\bibinfo{volume}{80}},
  \bibinfo{pages}{085101} (\bibinfo{year}{2009}).

\bibitem[{\citenamefont{Antonides et~al.}(1977)\citenamefont{Antonides, Janse,
  and Sawatzky}}]{U-xray-sawatzky-1977}
\bibinfo{author}{\bibfnamefont{E.}~\bibnamefont{Antonides}},
  \bibinfo{author}{\bibfnamefont{E.~C.} \bibnamefont{Janse}}, \bibnamefont{and}
  \bibinfo{author}{\bibfnamefont{G.~A.} \bibnamefont{Sawatzky}},
  \bibinfo{journal}{Phys. Rev. B} \textbf{\bibinfo{volume}{15}},
  \bibinfo{pages}{1669} (\bibinfo{year}{1977}).

\bibitem[{\citenamefont{Sawatzky and Allen}(1984)}]{NiO-sawatzky}
\bibinfo{author}{\bibfnamefont{G.~A.} \bibnamefont{Sawatzky}} \bibnamefont{and}
  \bibinfo{author}{\bibfnamefont{J.~W.} \bibnamefont{Allen}},
  \bibinfo{journal}{Phys. Rev. Lett.} \textbf{\bibinfo{volume}{53}},
  \bibinfo{pages}{2339} (\bibinfo{year}{1984}).

\bibitem[{\citenamefont{Hubbard}(1963)}]{Hubbard_model_1963}
\bibinfo{author}{\bibfnamefont{J.}~\bibnamefont{Hubbard}},
  \bibinfo{journal}{Proc. R. Soc. London A} \textbf{\bibinfo{volume}{276}},
  \bibinfo{pages}{238} (\bibinfo{year}{1963}).

\bibitem[{\citenamefont{Kanamori}(1963)}]{Kanamori_paper}
\bibinfo{author}{\bibfnamefont{J.}~\bibnamefont{Kanamori}},
  \bibinfo{journal}{Prog. Theor. Phys.} \textbf{\bibinfo{volume}{30}},
  \bibinfo{pages}{275} (\bibinfo{year}{1963}).

\bibitem[{\citenamefont{Gutzwiller}(1963)}]{Gutzwiller_1963}
\bibinfo{author}{\bibfnamefont{M.~C.} \bibnamefont{Gutzwiller}},
  \bibinfo{journal}{Phys. Rev. Lett.} \textbf{\bibinfo{volume}{10}},
  \bibinfo{pages}{159} (\bibinfo{year}{1963}).

\bibitem[{\citenamefont{de~Boer et~al.}(1984)\citenamefont{de~Boer, Haas, and
  Sawatzky}}]{U-xray-deboer}
\bibinfo{author}{\bibfnamefont{D.~K.~G.} \bibnamefont{de~Boer}},
  \bibinfo{author}{\bibfnamefont{C.}~\bibnamefont{Haas}}, \bibnamefont{and}
  \bibinfo{author}{\bibfnamefont{G.~A.} \bibnamefont{Sawatzky}},
  \bibinfo{journal}{J. Phys. F} \textbf{\bibinfo{volume}{14}},
  \bibinfo{pages}{2769} (\bibinfo{year}{1984}).

\bibitem[{\citenamefont{Ghijsen et~al.}(1990)\citenamefont{Ghijsen, Tjeng,
  Eskes, Sawatzky, and Johnson}}]{CuO-sawatzky}
\bibinfo{author}{\bibfnamefont{J.}~\bibnamefont{Ghijsen}},
  \bibinfo{author}{\bibfnamefont{L.~H.} \bibnamefont{Tjeng}},
  \bibinfo{author}{\bibfnamefont{H.}~\bibnamefont{Eskes}},
  \bibinfo{author}{\bibfnamefont{G.~A.} \bibnamefont{Sawatzky}},
  \bibnamefont{and} \bibinfo{author}{\bibfnamefont{R.~L.}
  \bibnamefont{Johnson}}, \bibinfo{journal}{Phys. Rev. B}
  \textbf{\bibinfo{volume}{42}}, \bibinfo{pages}{2268} (\bibinfo{year}{1990}).

\bibitem[{\citenamefont{van~der Marel}(1985)}]{PhD-vandermarel}
\bibinfo{author}{\bibfnamefont{D.}~\bibnamefont{van~der Marel}}, Ph.D. thesis,
  \bibinfo{school}{Rijksuniversiteit Groningen} (\bibinfo{year}{1985}).

\bibitem[{\citenamefont{Judd}(1998)}]{judd}
\bibinfo{author}{\bibfnamefont{B.~R.} \bibnamefont{Judd}},
  \emph{\bibinfo{title}{Operator Techniques in Atomic Spectroscopy}}
  (\bibinfo{publisher}{Princeton University Press}, \bibinfo{year}{1998}).

\bibitem[{\citenamefont{Sugano et~al.}(1970)\citenamefont{Sugano, Tanabe, and
  Kamimura}}]{Sugano_book}
\bibinfo{author}{\bibfnamefont{S.}~\bibnamefont{Sugano}},
  \bibinfo{author}{\bibfnamefont{Y.}~\bibnamefont{Tanabe}}, \bibnamefont{and}
  \bibinfo{author}{\bibfnamefont{H.}~\bibnamefont{Kamimura}},
  \emph{\bibinfo{title}{Multiplets of transition-metal ions in crystal}},
  vol.~\bibinfo{volume}{1} (\bibinfo{publisher}{Academic Press, New York
  London}, \bibinfo{year}{1970}).

\bibitem[{\citenamefont{Dederichs et~al.}(1984)\citenamefont{Dederichs,
  Bl\"ugel, Zeller, and Akai}}]{cLDA-Ce-dederichs}
\bibinfo{author}{\bibfnamefont{P.~H.} \bibnamefont{Dederichs}},
  \bibinfo{author}{\bibfnamefont{S.}~\bibnamefont{Bl\"ugel}},
  \bibinfo{author}{\bibfnamefont{R.}~\bibnamefont{Zeller}}, \bibnamefont{and}
  \bibinfo{author}{\bibfnamefont{H.}~\bibnamefont{Akai}},
  \bibinfo{journal}{Phys. Rev. Lett.} \textbf{\bibinfo{volume}{53}},
  \bibinfo{pages}{2512} (\bibinfo{year}{1984}).

\bibitem[{\citenamefont{McMahan et~al.}(1988)\citenamefont{McMahan, Martin, and
  Satpathy}}]{cLDA-mcmahan-La2CuO4}
\bibinfo{author}{\bibfnamefont{A.~K.} \bibnamefont{McMahan}},
  \bibinfo{author}{\bibfnamefont{R.~M.} \bibnamefont{Martin}},
  \bibnamefont{and} \bibinfo{author}{\bibfnamefont{S.}~\bibnamefont{Satpathy}},
  \bibinfo{journal}{Phys. Rev. B} \textbf{\bibinfo{volume}{38}},
  \bibinfo{pages}{6650} (\bibinfo{year}{1988}).

\bibitem[{\citenamefont{Hybertsen et~al.}(1989)\citenamefont{Hybertsen,
  Schl\"uter, and Christensen}}]{cLDA-hybertsen-La2CuO4}
\bibinfo{author}{\bibfnamefont{M.~S.} \bibnamefont{Hybertsen}},
  \bibinfo{author}{\bibfnamefont{M.}~\bibnamefont{Schl\"uter}},
  \bibnamefont{and} \bibinfo{author}{\bibfnamefont{N.~E.}
  \bibnamefont{Christensen}}, \bibinfo{journal}{Phys. Rev. B}
  \textbf{\bibinfo{volume}{39}}, \bibinfo{pages}{9028} (\bibinfo{year}{1989}).

\bibitem[{\citenamefont{Gunnarsson et~al.}(1989)\citenamefont{Gunnarsson,
  Andersen, Jepsen, and Zaanen}}]{cLDA-andersen}
\bibinfo{author}{\bibfnamefont{O.}~\bibnamefont{Gunnarsson}},
  \bibinfo{author}{\bibfnamefont{O.~K.} \bibnamefont{Andersen}},
  \bibinfo{author}{\bibfnamefont{O.}~\bibnamefont{Jepsen}}, \bibnamefont{and}
  \bibinfo{author}{\bibfnamefont{J.}~\bibnamefont{Zaanen}},
  \bibinfo{journal}{Phys. Rev. B} \textbf{\bibinfo{volume}{39}},
  \bibinfo{pages}{1708} (\bibinfo{year}{1989}).

\bibitem[{\citenamefont{Gunnarsson}(1990)}]{cLDA-gunnarsson}
\bibinfo{author}{\bibfnamefont{O.}~\bibnamefont{Gunnarsson}},
  \bibinfo{journal}{Phys. Rev. B} \textbf{\bibinfo{volume}{41}},
  \bibinfo{pages}{514} (\bibinfo{year}{1990}).

\bibitem[{\citenamefont{Anisimov and
  Gunnarsson}(1991)}]{cLDA-anisimov-gunnarsson}
\bibinfo{author}{\bibfnamefont{V.~I.} \bibnamefont{Anisimov}} \bibnamefont{and}
  \bibinfo{author}{\bibfnamefont{O.}~\bibnamefont{Gunnarsson}},
  \bibinfo{journal}{Phys. Rev. B} \textbf{\bibinfo{volume}{43}},
  \bibinfo{pages}{7570} (\bibinfo{year}{1991}).

\bibitem[{\citenamefont{Madsen and Nov\'ak}(2005)}]{cLDA-LAPW-madsen}
\bibinfo{author}{\bibfnamefont{G.~K.~H.} \bibnamefont{Madsen}}
  \bibnamefont{and} \bibinfo{author}{\bibfnamefont{P.}~\bibnamefont{Nov\'ak}},
  \bibinfo{journal}{Eur. Phys. Lett.} \textbf{\bibinfo{volume}{69}},
  \bibinfo{pages}{777} (\bibinfo{year}{2005}).

\bibitem[{\citenamefont{Nakamura et~al.}(2006)\citenamefont{Nakamura, Arita,
  Yoshimoto, and Tsuneyuki}}]{cLDA-MLWF-nakamura}
\bibinfo{author}{\bibfnamefont{K.}~\bibnamefont{Nakamura}},
  \bibinfo{author}{\bibfnamefont{R.}~\bibnamefont{Arita}},
  \bibinfo{author}{\bibfnamefont{Y.}~\bibnamefont{Yoshimoto}},
  \bibnamefont{and}
  \bibinfo{author}{\bibfnamefont{S.}~\bibnamefont{Tsuneyuki}},
  \bibinfo{journal}{Phys. Rev. B} \textbf{\bibinfo{volume}{74}},
  \bibinfo{pages}{235113} (\bibinfo{year}{2006}).

\bibitem[{\citenamefont{Pickett et~al.}(1998)\citenamefont{Pickett, Erwin, and
  Ethridge}}]{linearresponse-pickett}
\bibinfo{author}{\bibfnamefont{W.~E.} \bibnamefont{Pickett}},
  \bibinfo{author}{\bibfnamefont{S.~C.} \bibnamefont{Erwin}}, \bibnamefont{and}
  \bibinfo{author}{\bibfnamefont{E.~C.} \bibnamefont{Ethridge}},
  \bibinfo{journal}{Phys. Rev. B} \textbf{\bibinfo{volume}{58}},
  \bibinfo{pages}{1201} (\bibinfo{year}{1998}).

\bibitem[{\citenamefont{Cococcioni and
  de~Gironcoli}(2005)}]{linearresponse-cococcioni}
\bibinfo{author}{\bibfnamefont{M.}~\bibnamefont{Cococcioni}} \bibnamefont{and}
  \bibinfo{author}{\bibfnamefont{S.}~\bibnamefont{de~Gironcoli}},
  \bibinfo{journal}{Phys. Rev. B} \textbf{\bibinfo{volume}{71}},
  \bibinfo{pages}{035105} (\bibinfo{year}{2005}).

\bibitem[{\citenamefont{Solovyev and Imada}(2005)}]{cRPA-solovyev-2005}
\bibinfo{author}{\bibfnamefont{I.~V.} \bibnamefont{Solovyev}} \bibnamefont{and}
  \bibinfo{author}{\bibfnamefont{M.}~\bibnamefont{Imada}},
  \bibinfo{journal}{Phys. Rev. B} \textbf{\bibinfo{volume}{71}},
  \bibinfo{pages}{045103} (\bibinfo{year}{2005}).

\bibitem[{\citenamefont{Miyake and Aryasetiawan}(2008)}]{cRPA-takashi-2008}
\bibinfo{author}{\bibfnamefont{T.}~\bibnamefont{Miyake}} \bibnamefont{and}
  \bibinfo{author}{\bibfnamefont{F.}~\bibnamefont{Aryasetiawan}},
  \bibinfo{journal}{Phys. Rev. B} \textbf{\bibinfo{volume}{77}},
  \bibinfo{pages}{085122} (\bibinfo{year}{2008}).

\bibitem[{\citenamefont{Sasioglu
  et~al.}(2011)\citenamefont{Sasioglu,
  Friedrich, and
  Bl\"ugel}}]{cRPA-friedrich}
\bibinfo{author}{\bibfnamefont{E.}~\bibnamefont{Sasioglu}},
  \bibinfo{author}{\bibfnamefont{C.}~\bibnamefont{Friedrich}},
  \bibnamefont{and} \bibinfo{author}{\bibfnamefont{S.}~\bibnamefont{Bl\"ugel}},
  \bibinfo{journal}{Phys. Rev. B} \textbf{\bibinfo{volume}{83}},
  \bibinfo{pages}{121101} (\bibinfo{year}{2011}).

\bibitem[{\citenamefont{Miyake et~al.}(2010)\citenamefont{Miyake, Nakamura,
  Arita, and Imada}}]{cRPA-pnictides-takashi}
\bibinfo{author}{\bibfnamefont{T.}~\bibnamefont{Miyake}},
  \bibinfo{author}{\bibfnamefont{K.}~\bibnamefont{Nakamura}},
  \bibinfo{author}{\bibfnamefont{R.}~\bibnamefont{Arita}}, \bibnamefont{and}
  \bibinfo{author}{\bibfnamefont{M.}~\bibnamefont{Imada}}, \bibinfo{journal}{J.
  Phys. Soc. Jpn.} \textbf{\bibinfo{volume}{79}}, \bibinfo{pages}{044705}
  (\bibinfo{year}{2010}).

\bibitem[{\citenamefont{Aryasetiawan et~al.}(2006)\citenamefont{Aryasetiawan,
  Karlsson, Jepsen, and Sch\"onberger}}]{Ufirstprinciples-ferdi-2006}
\bibinfo{author}{\bibfnamefont{F.}~\bibnamefont{Aryasetiawan}},
  \bibinfo{author}{\bibfnamefont{K.}~\bibnamefont{Karlsson}},
  \bibinfo{author}{\bibfnamefont{O.}~\bibnamefont{Jepsen}}, \bibnamefont{and}
  \bibinfo{author}{\bibfnamefont{U.}~\bibnamefont{Sch\"onberger}},
  \bibinfo{journal}{Phys. Rev. B} \textbf{\bibinfo{volume}{74}},
  \bibinfo{pages}{125106} (\bibinfo{year}{2006}).

\bibitem[{\citenamefont{Miyake et~al.}(2009)\citenamefont{Miyake, Aryasetiawan,
  and Imada}}]{cRPA-takashi-2009}
\bibinfo{author}{\bibfnamefont{T.}~\bibnamefont{Miyake}},
  \bibinfo{author}{\bibfnamefont{F.}~\bibnamefont{Aryasetiawan}},
  \bibnamefont{and} \bibinfo{author}{\bibfnamefont{M.}~\bibnamefont{Imada}},
  \bibinfo{journal}{Phys. Rev. B} \textbf{\bibinfo{volume}{80}},
  \bibinfo{pages}{155134} (\bibinfo{year}{2009}).

\bibitem[{\citenamefont{Tomczak et~al.}(2009)\citenamefont{Tomczak, Miyake,
  Sakuma, and Aryasetiawan}}]{cRPA-pressure-jan}
\bibinfo{author}{\bibfnamefont{J.~M.} \bibnamefont{Tomczak}},
  \bibinfo{author}{\bibfnamefont{T.}~\bibnamefont{Miyake}},
  \bibinfo{author}{\bibfnamefont{R.}~\bibnamefont{Sakuma}}, \bibnamefont{and}
  \bibinfo{author}{\bibfnamefont{F.}~\bibnamefont{Aryasetiawan}},
  \bibinfo{journal}{Phys. Rev. B} \textbf{\bibinfo{volume}{79}},
  \bibinfo{pages}{235133} (\bibinfo{year}{2009}).

\bibitem[{\citenamefont{Karlsson et~al.}(2010)\citenamefont{Karlsson,
  Aryasetiawan, and Jepsen}}]{cRPA-LDA+U-ferdi}
\bibinfo{author}{\bibfnamefont{K.}~\bibnamefont{Karlsson}},
  \bibinfo{author}{\bibfnamefont{F.}~\bibnamefont{Aryasetiawan}},
  \bibnamefont{and} \bibinfo{author}{\bibfnamefont{O.}~\bibnamefont{Jepsen}},
  \bibinfo{journal}{Phys. Rev. B} \textbf{\bibinfo{volume}{81}},
  \bibinfo{pages}{245113} (\bibinfo{year}{2010}).

\bibitem[{\citenamefont{Aichhorn et~al.}(2010)\citenamefont{Aichhorn, Biermann,
  Miyake, Georges, and Imada}}]{cRPA-DMFT-FeSe-markus}
\bibinfo{author}{\bibfnamefont{M.}~\bibnamefont{Aichhorn}},
  \bibinfo{author}{\bibfnamefont{S.}~\bibnamefont{Biermann}},
  \bibinfo{author}{\bibfnamefont{T.}~\bibnamefont{Miyake}},
  \bibinfo{author}{\bibfnamefont{A.}~\bibnamefont{Georges}}, \bibnamefont{and}
  \bibinfo{author}{\bibfnamefont{M.}~\bibnamefont{Imada}},
  \bibinfo{journal}{Phys. Rev. B} \textbf{\bibinfo{volume}{82}},
  \bibinfo{pages}{064504} (\bibinfo{year}{2010}).

\bibitem[{\citenamefont{Mravlje et~al.}(2011)\citenamefont{Mravlje, Aichhorn,
  Miyake, Haule, Kotliar, and Georges}}]{Sr2RuO4-jernej}
\bibinfo{author}{\bibfnamefont{J.}~\bibnamefont{Mravlje}},
  \bibinfo{author}{\bibfnamefont{M.}~\bibnamefont{Aichhorn}},
  \bibinfo{author}{\bibfnamefont{T.}~\bibnamefont{Miyake}},
  \bibinfo{author}{\bibfnamefont{K.}~\bibnamefont{Haule}},
  \bibinfo{author}{\bibfnamefont{G.}~\bibnamefont{Kotliar}}, \bibnamefont{and}
  \bibinfo{author}{\bibfnamefont{A.}~\bibnamefont{Georges}},
  \bibinfo{journal}{Phys. Rev. Lett.} \textbf{\bibinfo{volume}{106}},
  \bibinfo{pages}{096401} (\bibinfo{year}{2011}).

\bibitem[{\citenamefont{Martins et~al.}(2011)\citenamefont{Martins, Aichhorn,
  Vaugier, and Biermann}}]{Sr2IrO4-cyril}
\bibinfo{author}{\bibfnamefont{C.}~\bibnamefont{Martins}},
  \bibinfo{author}{\bibfnamefont{M.}~\bibnamefont{Aichhorn}},
  \bibinfo{author}{\bibfnamefont{L.}~\bibnamefont{Vaugier}}, \bibnamefont{and}
  \bibinfo{author}{\bibfnamefont{S.}~\bibnamefont{Biermann}},
  \bibinfo{journal}{Phys. Rev. Lett.} \textbf{\bibinfo{volume}{107}},
  \bibinfo{pages}{266404} (\bibinfo{year}{2011}).

\bibitem[{\citenamefont{Anisimov et~al.}(2005)\citenamefont{Anisimov,
      Kondakov, Kozhevnikov, Nekrasov, Pchelkina, Allen, Mo, Kim,
      Metcalf, Suga, Sekiyama, Keller, Leonov, Ren and Vollhardt}}]{AnisimovSrVO3}
\bibinfo{author}{\bibfnamefont{V. I.}~\bibnamefont{Anisimov}},
  \bibinfo{author}{\bibfnamefont{D. E.}~\bibnamefont{Kondakov}},
  \bibinfo{author}{\bibfnamefont{A. V.}~\bibnamefont{Kozhevnikov}},
  \bibinfo{author}{\bibfnamefont{I. A.}~\bibnamefont{Nekrasov}},
  \bibinfo{author}{\bibfnamefont{Z. V.}~\bibnamefont{Pchelkina}}, 
 \bibinfo{author}{\bibfnamefont{J. W.}~\bibnamefont{Allen}},
 \bibinfo{author}{\bibfnamefont{S.-K.}~\bibnamefont{Mo}},   
 \bibinfo{author}{\bibfnamefont{H.-D.}~\bibnamefont{Kim}}, 
 \bibinfo{author}{\bibfnamefont{P.}~\bibnamefont{Metcalf}}, 
 \bibinfo{author}{\bibfnamefont{S.}~\bibnamefont{Suga}}, 
 \bibinfo{author}{\bibfnamefont{A.}~\bibnamefont{Sekiyama}},
 \bibinfo{author}{\bibfnamefont{G.}~\bibnamefont{Keller}},
 \bibinfo{author}{\bibfnamefont{I.}~\bibnamefont{Leonov}},
 \bibinfo{author}{\bibfnamefont{X.}~\bibnamefont{Ren}}, \bibnamefont{and}
  \bibinfo{author}{\bibfnamefont{D.}~\bibnamefont{Vollhardt}},
  \bibinfo{journal}{Phys. Rev. B} \textbf{\bibinfo{volume}{71}},
  \bibinfo{pages}{125119} (\bibinfo{year}{2005}).

\bibitem[{\citenamefont{Amadon et~al.}(2008)\citenamefont{Amadon,
      Lechermann, Georges, Jollet, Wehling and Lichtenstein}}]{AmadonSrVO3}
\bibinfo{author}{\bibfnamefont{B.}~\bibnamefont{Amadon}},
  \bibinfo{author}{\bibfnamefont{F.}~\bibnamefont{Lechermann}},
  \bibinfo{author}{\bibfnamefont{A.}~\bibnamefont{Georges}},
  \bibinfo{author}{\bibfnamefont{F.}~\bibnamefont{Jollet}},
  \bibinfo{author}{\bibfnamefont{T. O.}~\bibnamefont{Wehling}}, \bibnamefont{and}
  \bibinfo{author}{\bibfnamefont{A. I.}~\bibnamefont{Lichtenstein}},
  \bibinfo{journal}{Phys. Rev. B} \textbf{\bibinfo{volume}{77}},
  \bibinfo{pages}{205112} (\bibinfo{year}{2008}).

\bibitem[{\citenamefont{Biermann et~al.}(2003)\citenamefont{Biermann,
  Aryasetiawan, and Georges}}]{GW+DMFT-biermann}
\bibinfo{author}{\bibfnamefont{S.}~\bibnamefont{Biermann}},
  \bibinfo{author}{\bibfnamefont{F.}~\bibnamefont{Aryasetiawan}},
  \bibnamefont{and} \bibinfo{author}{\bibfnamefont{A.}~\bibnamefont{Georges}},
  \bibinfo{journal}{Phys. Rev. Lett.} \textbf{\bibinfo{volume}{90}},
  \bibinfo{pages}{086402} (\bibinfo{year}{2003}).

\bibitem[{\citenamefont{Sun and Kotliar}(2002)}]{GW+DMFT-kotliar}
\bibinfo{author}{\bibfnamefont{P.}~\bibnamefont{Sun}} \bibnamefont{and}
  \bibinfo{author}{\bibfnamefont{G.}~\bibnamefont{Kotliar}},
  \bibinfo{journal}{Phys. Rev. B} \textbf{\bibinfo{volume}{66}},
  \bibinfo{pages}{085120} (\bibinfo{year}{2002}).

\bibitem[{\citenamefont{Aryasetiawan et~al.}(2009)\citenamefont{Aryasetiawan,
  Tomczak, Miyake, and Sakuma}}]{DownfoldedSelfEnergy_Ferdi}
\bibinfo{author}{\bibfnamefont{F.}~\bibnamefont{Aryasetiawan}},
  \bibinfo{author}{\bibfnamefont{J.~M.} \bibnamefont{Tomczak}},
  \bibinfo{author}{\bibfnamefont{T.}~\bibnamefont{Miyake}}, \bibnamefont{and}
  \bibinfo{author}{\bibfnamefont{R.}~\bibnamefont{Sakuma}},
  \bibinfo{journal}{Phys. Rev. Lett.} \textbf{\bibinfo{volume}{102}},
  \bibinfo{pages}{176402} (\bibinfo{year}{2009}).

\bibitem[{\citenamefont{Kutepov et~al.}(2010)\citenamefont{Kutepov, Haule,
  Savrasov, and Kotliar}}]{scGW-kutepov}
\bibinfo{author}{\bibfnamefont{A.}~\bibnamefont{Kutepov}},
  \bibinfo{author}{\bibfnamefont{K.}~\bibnamefont{Haule}},
  \bibinfo{author}{\bibfnamefont{S.~Y.} \bibnamefont{Savrasov}},
  \bibnamefont{and} \bibinfo{author}{\bibfnamefont{G.}~\bibnamefont{Kotliar}},
  \bibinfo{journal}{Phys. Rev. B} \textbf{\bibinfo{volume}{82}},
  \bibinfo{pages}{045105} (\bibinfo{year}{2010}).

\bibitem[{\citenamefont{Imada and Miyake}(2010)}]{review-imada-takashi}
\bibinfo{author}{\bibfnamefont{M.}~\bibnamefont{Imada}} \bibnamefont{and}
  \bibinfo{author}{\bibfnamefont{T.}~\bibnamefont{Miyake}},
  \bibinfo{journal}{J. Phys. Soc. Jpn.} \textbf{\bibinfo{volume}{79}},
  \bibinfo{pages}{112001} (\bibinfo{year}{2010}).

\bibitem[{\citenamefont{Vaugier}(2011)}]{PhD-Vaugier}
\bibinfo{author}{\bibfnamefont{L.}~\bibnamefont{Vaugier}}, Ph.D. thesis,
  \bibinfo{school}{Ecole Polytechnique, France} (\bibinfo{year}{2011}).

\bibitem[{\citenamefont{Nomura et~al.}(2012)\citenamefont{Nomura, Kaltak,
  Nakamura, Taranto, Sakai, Toschi, Arita, Held, Kresse, and
  Imada}}]{nomura-imada-12}
\bibinfo{author}{\bibfnamefont{Y.}~\bibnamefont{Nomura}},
  \bibinfo{author}{\bibfnamefont{M.}~\bibnamefont{Kaltak}},
  \bibinfo{author}{\bibfnamefont{K.}~\bibnamefont{Nakamura}},
  \bibinfo{author}{\bibfnamefont{C.}~\bibnamefont{Taranto}},
  \bibinfo{author}{\bibfnamefont{S.}~\bibnamefont{Sakai}},
  \bibinfo{author}{\bibfnamefont{A.}~\bibnamefont{Toschi}},
  \bibinfo{author}{\bibfnamefont{R.}~\bibnamefont{Arita}},
  \bibinfo{author}{\bibfnamefont{K.}~\bibnamefont{Held}},
  \bibinfo{author}{\bibfnamefont{G.}~\bibnamefont{Kresse}}, \bibnamefont{and}
  \bibinfo{author}{\bibfnamefont{M.}~\bibnamefont{Imada}},
  \bibinfo{journal}{Phys. Rev. B} \textbf{\bibinfo{volume}{86}},
  \bibinfo{pages}{085117} (\bibinfo{year}{2012}).

\bibitem[{\citenamefont{Casula et~al.}(2012{\natexlab{a}})\citenamefont{Casula,
  Rubtsov, and Biermann}}]{udyn-michele}
\bibinfo{author}{\bibfnamefont{M.}~\bibnamefont{Casula}},
  \bibinfo{author}{\bibfnamefont{A.}~\bibnamefont{Rubtsov}}, \bibnamefont{and}
  \bibinfo{author}{\bibfnamefont{S.}~\bibnamefont{Biermann}},
  \bibinfo{journal}{Phys. Rev. B} \textbf{\bibinfo{volume}{85}},
  \bibinfo{pages}{035115} (\bibinfo{year}{2012}{\natexlab{a}}).

\bibitem[{\citenamefont{Werner et~al.}(2012)\citenamefont{Werner, Casula,
  Miyake, Aryasetiawan, Millis, and Biermann}}]{udyn-werner}
\bibinfo{author}{\bibfnamefont{P.}~\bibnamefont{Werner}},
  \bibinfo{author}{\bibfnamefont{M.}~\bibnamefont{Casula}},
  \bibinfo{author}{\bibfnamefont{T.}~\bibnamefont{Miyake}},
  \bibinfo{author}{\bibfnamefont{F.}~\bibnamefont{Aryasetiawan}},
  \bibinfo{author}{\bibfnamefont{A.~J.} \bibnamefont{Millis}},
  \bibnamefont{and} \bibinfo{author}{\bibfnamefont{S.}~\bibnamefont{Biermann}},
  \bibinfo{journal}{Nature Physics} \textbf{\bibinfo{volume}{8}},
  \bibinfo{pages}{331} (\bibinfo{year}{2012}).

\bibitem[{\citenamefont{Casula et~al.}(2012{\natexlab{b}})\citenamefont{Casula,
  Werner, Vaugier, Aryasetiawan, Miyake, Millis, and Biermann}}]{udyneff-michele}
\bibinfo{author}{\bibfnamefont{M.}~\bibnamefont{Casula}},
  \bibinfo{author}{\bibfnamefont{P.}~\bibnamefont{Werner}},
  \bibinfo{author}{\bibfnamefont{L.}~\bibnamefont{Vaugier}},
  \bibinfo{author}{\bibfnamefont{F.}~\bibnamefont{Aryasetiawan}},
  \bibinfo{author}{\bibfnamefont{T.}~\bibnamefont{Miyake}},
  \bibinfo{author}{\bibfnamefont{A.~J.} \bibnamefont{Millis}},
  \bibnamefont{and} \bibinfo{author}{\bibfnamefont{S.}~\bibnamefont{Biermann}},
  \bibinfo{journal}{Phys. Rev. Lett.} \textbf{\bibinfo{volume}{109}}, \bibinfo{pages}{126408}
  (\bibinfo{year}{2012}{\natexlab{b}}).

\bibitem[{\citenamefont{Miyake et~al.}(2008)\citenamefont{Miyake, Pourovskii,
  Vildosola, Biermann, and Georges}}]{cRPA-LaOFeAs-miyake}
\bibinfo{author}{\bibfnamefont{T.}~\bibnamefont{Miyake}},
  \bibinfo{author}{\bibfnamefont{L.}~\bibnamefont{Pourovskii}},
  \bibinfo{author}{\bibfnamefont{V.}~\bibnamefont{Vildosola}},
  \bibinfo{author}{\bibfnamefont{S.}~\bibnamefont{Biermann}}, \bibnamefont{and}
  \bibinfo{author}{\bibfnamefont{A.}~\bibnamefont{Georges}},
  \bibinfo{journal}{J. Phys. Soc. Jpn. : Supplement C}
  \textbf{\bibinfo{volume}{77}}, \bibinfo{pages}{99} (\bibinfo{year}{2008}).

\bibitem[{\citenamefont{Slater}(1960)}]{Slater_book}
\bibinfo{author}{\bibfnamefont{J.~C.} \bibnamefont{Slater}},
  \emph{\bibinfo{title}{Quantum Theory of Atomic Structure}},
  vol.~\bibinfo{volume}{1} (\bibinfo{publisher}{McGraw-Hill, New York},
  \bibinfo{year}{1960}).

\bibitem[{\citenamefont{Anisimov et~al.}(1993)\citenamefont{Anisimov, Solovyev,
  Korotin, Czyzyk, and
  Sawatzky}}]{LDA+U-anisimov-NiO}
\bibinfo{author}{\bibfnamefont{V.~I.} \bibnamefont{Anisimov}},
  \bibinfo{author}{\bibfnamefont{I.~V.} \bibnamefont{Solovyev}},
  \bibinfo{author}{\bibfnamefont{M.~A.} \bibnamefont{Korotin}},
  \bibinfo{author}{\bibfnamefont{M.~T.} \bibnamefont{Czyzyk}}, \bibnamefont{and} \bibinfo{author}{\bibfnamefont{G.~A.}
  \bibnamefont{Sawatzky}}, \bibinfo{journal}{Phys. Rev. B}
  \textbf{\bibinfo{volume}{48}}, \bibinfo{pages}{16929} (\bibinfo{year}{1993}).

\bibitem[{\citenamefont{Lee et~al.}(2003)\citenamefont{Lee, Lee, Noh, Byun,
  Yoo, Yamaura, and Takayama-Muromachi}}]{Lee-SrMO3}
\bibinfo{author}{\bibfnamefont{Y.~S.} \bibnamefont{Lee}},
  \bibinfo{author}{\bibfnamefont{J.~S.} \bibnamefont{Lee}},
  \bibinfo{author}{\bibfnamefont{T.~W.} \bibnamefont{Noh}},
  \bibinfo{author}{\bibfnamefont{D.~Y.} \bibnamefont{Byun}},
  \bibinfo{author}{\bibfnamefont{K.~S.} \bibnamefont{Yoo}},
  \bibinfo{author}{\bibfnamefont{K.}~\bibnamefont{Yamaura}}, \bibnamefont{and}
  \bibinfo{author}{\bibfnamefont{E.}~\bibnamefont{Takayama-Muromachi}},
  \bibinfo{journal}{Phys. Rev. B} \textbf{\bibinfo{volume}{67}},
  \bibinfo{pages}{113101} (\bibinfo{year}{2003}).

\bibitem[{\citenamefont{Torrance et~al.}(1991)\citenamefont{Torrance, Lacorre, Asavaroengchai,
  and Metzger}}]{torrance-deltapd}
\bibinfo{author}{\bibfnamefont{J.}~\bibnamefont{Torrance}},
  \bibinfo{author}{\bibfnamefont{P.}~\bibnamefont{Lacorre.}},
  \bibinfo{author}{\bibfnamefont{C.}~\bibnamefont{Asavaroengchai}}, \bibnamefont{and}
  \bibinfo{author}{\bibfnamefont{R.}~\bibnamefont{Metzger}},
  \bibinfo{journal}{Physica C} \textbf{\bibinfo{volume}{182}},
  \bibinfo{pages}{351} (\bibinfo{year}{1991}).

\bibitem[{\citenamefont{Georges}(2004)}]{georges-DMFT-2004}
\bibinfo{author}{\bibfnamefont{A.}~\bibnamefont{Georges}},
  \bibinfo{journal}{Lectures on the physics of highly correlated electron
  systems VI} \textbf{\bibinfo{volume}{715}}, \bibinfo{pages}{3}
  (\bibinfo{year}{2004}).

\bibitem[{\citenamefont{Zaanen et~al.}(1985)\citenamefont{Zaanen, Sawatzky, and
  Allen}}]{TMO-zaanen}
\bibinfo{author}{\bibfnamefont{J.}~\bibnamefont{Zaanen}},
  \bibinfo{author}{\bibfnamefont{G.~A.} \bibnamefont{Sawatzky}},
  \bibnamefont{and} \bibinfo{author}{\bibfnamefont{J.~W.} \bibnamefont{Allen}},
  \bibinfo{journal}{Phys. Rev. Lett.} \textbf{\bibinfo{volume}{55}},
  \bibinfo{pages}{418} (\bibinfo{year}{1985}).

\bibitem[{\citenamefont{Sekiyama et~al.}(2004)\citenamefont{Sekiyama, Fujiwara,
  Imada, Suga, Eisaki, Uchida, Takegahara, Harima, Saitoh, Nekrasov
  et~al.}}]{SrVO3-sekiyama}
\bibinfo{author}{\bibfnamefont{A.}~\bibnamefont{Sekiyama}},
  \bibinfo{author}{\bibfnamefont{H.}~\bibnamefont{Fujiwara}},
  \bibinfo{author}{\bibfnamefont{S.}~\bibnamefont{Imada}},
  \bibinfo{author}{\bibfnamefont{S.}~\bibnamefont{Suga}},
  \bibinfo{author}{\bibfnamefont{H.}~\bibnamefont{Eisaki}},
  \bibinfo{author}{\bibfnamefont{S.~I.} \bibnamefont{Uchida}},
  \bibinfo{author}{\bibfnamefont{K.}~\bibnamefont{Takegahara}},
  \bibinfo{author}{\bibfnamefont{H.}~\bibnamefont{Harima}},
  \bibinfo{author}{\bibfnamefont{Y.}~\bibnamefont{Saitoh}},
  \bibinfo{author}{\bibfnamefont{I.~A.} \bibnamefont{Nekrasov}},
  \bibnamefont{et~al.}, \bibinfo{journal}{Phys. Rev. Lett.}
  \textbf{\bibinfo{volume}{93}}, \bibinfo{pages}{156402}
  (\bibinfo{year}{2004}).

\bibitem[{\citenamefont{Eguchi et~al.}(2006)\citenamefont{Eguchi, Kiss, Tsuda,
  Shimojima, Mizokami, Yokoya, Chainani, Shin, Inoue, Togashi
  et~al.}}]{SrVO3-eguchi}
\bibinfo{author}{\bibfnamefont{R.}~\bibnamefont{Eguchi}},
  \bibinfo{author}{\bibfnamefont{T.}~\bibnamefont{Kiss}},
  \bibinfo{author}{\bibfnamefont{S.}~\bibnamefont{Tsuda}},
  \bibinfo{author}{\bibfnamefont{T.}~\bibnamefont{Shimojima}},
  \bibinfo{author}{\bibfnamefont{T.}~\bibnamefont{Mizokami}},
  \bibinfo{author}{\bibfnamefont{T.}~\bibnamefont{Yokoya}},
  \bibinfo{author}{\bibfnamefont{A.}~\bibnamefont{Chainani}},
  \bibinfo{author}{\bibfnamefont{S.}~\bibnamefont{Shin}},
  \bibinfo{author}{\bibfnamefont{I.~H.} \bibnamefont{Inoue}},
  \bibinfo{author}{\bibfnamefont{T.}~\bibnamefont{Togashi}},
  \bibnamefont{et~al.}, \bibinfo{journal}{Phys. Rev. Lett.}
  \textbf{\bibinfo{volume}{96}}, \bibinfo{pages}{076402}
  (\bibinfo{year}{2006}).

\bibitem[{\citenamefont{Fujimori et~al.}(1992)\citenamefont{Fujimori, Hase,
  Namatame, Fujishima, Tokura, Eisaki, Uchida, Takegahara, and
  de~Groot}}]{Mott-Hubbard-fujimori}
\bibinfo{author}{\bibfnamefont{A.}~\bibnamefont{Fujimori}},
  \bibinfo{author}{\bibfnamefont{I.}~\bibnamefont{Hase}},
  \bibinfo{author}{\bibfnamefont{H.}~\bibnamefont{Namatame}},
  \bibinfo{author}{\bibfnamefont{Y.}~\bibnamefont{Fujishima}},
  \bibinfo{author}{\bibfnamefont{Y.}~\bibnamefont{Tokura}},
  \bibinfo{author}{\bibfnamefont{H.}~\bibnamefont{Eisaki}},
  \bibinfo{author}{\bibfnamefont{S.}~\bibnamefont{Uchida}},
  \bibinfo{author}{\bibfnamefont{K.}~\bibnamefont{Takegahara}},
  \bibnamefont{and} \bibinfo{author}{\bibfnamefont{F.~M.~F.}
  \bibnamefont{de~Groot}}, \bibinfo{journal}{Phys. Rev. Lett.}
  \textbf{\bibinfo{volume}{69}}, \bibinfo{pages}{1796} (\bibinfo{year}{1992}).

\bibitem[{\citenamefont{Yoshida et~al.}(2005)\citenamefont{Yoshida, Tanaka,
  Yagi, Ino, Eisaki, Fujimori, and Shen}}]{SrVO3-yoshida}
\bibinfo{author}{\bibfnamefont{T.}~\bibnamefont{Yoshida}},
  \bibinfo{author}{\bibfnamefont{K.}~\bibnamefont{Tanaka}},
  \bibinfo{author}{\bibfnamefont{H.}~\bibnamefont{Yagi}},
  \bibinfo{author}{\bibfnamefont{A.}~\bibnamefont{Ino}},
  \bibinfo{author}{\bibfnamefont{H.}~\bibnamefont{Eisaki}},
  \bibinfo{author}{\bibfnamefont{A.}~\bibnamefont{Fujimori}}, \bibnamefont{and}
  \bibinfo{author}{\bibfnamefont{Z.-X.} \bibnamefont{Shen}},
  \bibinfo{journal}{Phys. Rev. Lett.} \textbf{\bibinfo{volume}{95}},
  \bibinfo{pages}{146404} (\bibinfo{year}{2005}).

\bibitem[{\citenamefont{Morikawa et~al.}(1995)\citenamefont{Morikawa, Mizokawa,
  Kobayashi, Fujimori, Eisaki, Uchida, Iga, and Nishihara}}]{SrVO3-morikawa}
\bibinfo{author}{\bibfnamefont{K.}~\bibnamefont{Morikawa}},
  \bibinfo{author}{\bibfnamefont{T.}~\bibnamefont{Mizokawa}},
  \bibinfo{author}{\bibfnamefont{K.}~\bibnamefont{Kobayashi}},
  \bibinfo{author}{\bibfnamefont{A.}~\bibnamefont{Fujimori}},
  \bibinfo{author}{\bibfnamefont{H.}~\bibnamefont{Eisaki}},
  \bibinfo{author}{\bibfnamefont{S.}~\bibnamefont{Uchida}},
  \bibinfo{author}{\bibfnamefont{F.}~\bibnamefont{Iga}}, \bibnamefont{and}
  \bibinfo{author}{\bibfnamefont{Y.}~\bibnamefont{Nishihara}},
  \bibinfo{journal}{Phys. Rev. B} \textbf{\bibinfo{volume}{52}},
  \bibinfo{pages}{13711} (\bibinfo{year}{1995}).

\bibitem[{\citenamefont{Lechermann et~al.}(2006)\citenamefont{Lechermann,
  Georges, Poteryaev, Biermann, Posternak, Yamasaki, and
  Andersen}}]{Wannier-lechermann-2006}
\bibinfo{author}{\bibfnamefont{F.}~\bibnamefont{Lechermann}},
  \bibinfo{author}{\bibfnamefont{A.}~\bibnamefont{Georges}},
  \bibinfo{author}{\bibfnamefont{A.}~\bibnamefont{Poteryaev}},
  \bibinfo{author}{\bibfnamefont{S.}~\bibnamefont{Biermann}},
  \bibinfo{author}{\bibfnamefont{M.}~\bibnamefont{Posternak}},
  \bibinfo{author}{\bibfnamefont{A.}~\bibnamefont{Yamasaki}}, \bibnamefont{and}
  \bibinfo{author}{\bibfnamefont{O.~K.} \bibnamefont{Andersen}},
  \bibinfo{journal}{Phys. Rev. B} \textbf{\bibinfo{volume}{74}},
  \bibinfo{pages}{125120} (\bibinfo{year}{2006}).

\bibitem[{\citenamefont{Chamberland}(1967)}]{SrCrO3-chamberland}
\bibinfo{author}{\bibfnamefont{B.}~\bibnamefont{Chamberland}},
  \bibinfo{journal}{Solid State Commun.} \textbf{\bibinfo{volume}{5}},
  \bibinfo{pages}{663} (\bibinfo{year}{1967}).

\bibitem[{\citenamefont{Zhou et~al.}(2006)\citenamefont{Zhou, Jin, Long, Yang,
  and Goodenough}}]{SrCrO3-zhou}
\bibinfo{author}{\bibfnamefont{J.-S.} \bibnamefont{Zhou}},
  \bibinfo{author}{\bibfnamefont{C.-Q.} \bibnamefont{Jin}},
  \bibinfo{author}{\bibfnamefont{Y.-W.} \bibnamefont{Long}},
  \bibinfo{author}{\bibfnamefont{L.-X.} \bibnamefont{Yang}}, \bibnamefont{and}
  \bibinfo{author}{\bibfnamefont{J.~B.} \bibnamefont{Goodenough}},
  \bibinfo{journal}{Phys. Rev. Lett.} \textbf{\bibinfo{volume}{96}},
  \bibinfo{pages}{046408} (\bibinfo{year}{2006}).

\bibitem[{\citenamefont{SanMartin et~al.}(2007)\citenamefont{SanMartin, Williams, Rodgers, Attfield,
  Heymann, and Huppertz}}]{ortegaPRL07}
\bibinfo{author}{\bibfnamefont{L.~O.} \bibnamefont{SanMartin}},
  \bibinfo{author}{\bibfnamefont{A.~J.} \bibnamefont{Williams}},
  \bibinfo{author}{\bibfnamefont{J.}~\bibnamefont{Rodgers}},
  \bibinfo{author}{\bibfnamefont{J.~P.} \bibnamefont{Attfield}},
  \bibinfo{author}{\bibfnamefont{G.}~\bibnamefont{Heymann}}, \bibnamefont{and}
  \bibinfo{author}{\bibfnamefont{H.}~\bibnamefont{Huppertz}},
  \bibinfo{journal}{Phys. Rev. Lett.} \textbf{\bibinfo{volume}{99}},
  \bibinfo{pages}{255701} (\bibinfo{year}{2007}).

\bibitem[{\citenamefont{Lee and Pickett}(2009)}]{SrCrO3-pickett}
\bibinfo{author}{\bibfnamefont{K.-W.} \bibnamefont{Lee}} \bibnamefont{and}
  \bibinfo{author}{\bibfnamefont{W.~E.} \bibnamefont{Pickett}},
  \bibinfo{journal}{Phys. Rev. B} \textbf{\bibinfo{volume}{80}},
  \bibinfo{pages}{125133} (\bibinfo{year}{2009}).

\bibitem[{\citenamefont{Qian et~al.}(2011)\citenamefont{Qian, Wang, Li, Jin,
  and Fang}}]{SrCrO3-qian}
\bibinfo{author}{\bibfnamefont{Y.}~\bibnamefont{Qian}},
  \bibinfo{author}{\bibfnamefont{G.}~\bibnamefont{Wang}},
  \bibinfo{author}{\bibfnamefont{Z.}~\bibnamefont{Li}},
  \bibinfo{author}{\bibfnamefont{C.}~\bibnamefont{Jin}}, \bibnamefont{and}
  \bibinfo{author}{\bibfnamefont{Z.}~\bibnamefont{Fang}}, \bibinfo{journal}{New
  Journal of Physics} \textbf{\bibinfo{volume}{13}}, \bibinfo{pages}{053002}
  (\bibinfo{year}{2011}).

\bibitem[{\citenamefont{Daoud-Aladine et~al.}(2007)\citenamefont{Daoud-Aladine,
  Martin, Chapon, Hervieu, Knight, Brunelli, and Radaelli}}]{SrMnO3-daoud}
\bibinfo{author}{\bibfnamefont{A.}~\bibnamefont{Daoud-Aladine}},
  \bibinfo{author}{\bibfnamefont{C.}~\bibnamefont{Martin}},
  \bibinfo{author}{\bibfnamefont{L.~C.} \bibnamefont{Chapon}},
  \bibinfo{author}{\bibfnamefont{M.}~\bibnamefont{Hervieu}},
  \bibinfo{author}{\bibfnamefont{K.~S.} \bibnamefont{Knight}},
  \bibinfo{author}{\bibfnamefont{M.}~\bibnamefont{Brunelli}}, \bibnamefont{and}
  \bibinfo{author}{\bibfnamefont{P.~G.} \bibnamefont{Radaelli}},
  \bibinfo{journal}{Phys. Rev. B} \textbf{\bibinfo{volume}{75}},
  \bibinfo{pages}{104417} (\bibinfo{year}{2007}).

\bibitem[{\citenamefont{Takeda and Ohara}(1974)}]{SrMnO3-takeda}
\bibinfo{author}{\bibfnamefont{T.}~\bibnamefont{Takeda}} \bibnamefont{and}
  \bibinfo{author}{\bibfnamefont{S.}~\bibnamefont{Ohara}}, \bibinfo{journal}{J.
  Phys. Soc. Jpn.} \textbf{\bibinfo{volume}{37}}, \bibinfo{pages}{275}
  (\bibinfo{year}{1974}).

\bibitem[{\citenamefont{S\o{}nden\aa{}
  et~al.}(2006)\citenamefont{S\o{}nden\aa{}, Ravindran, St\o{}len, Grande, and
  Hanfland}}]{SrMnO3-sondena}
\bibinfo{author}{\bibfnamefont{R.}~\bibnamefont{S\o{}nden\aa{}}},
  \bibinfo{author}{\bibfnamefont{P.}~\bibnamefont{Ravindran}},
  \bibinfo{author}{\bibfnamefont{S.}~\bibnamefont{St\o{}len}},
  \bibinfo{author}{\bibfnamefont{T.}~\bibnamefont{Grande}}, \bibnamefont{and}
  \bibinfo{author}{\bibfnamefont{M.}~\bibnamefont{Hanfland}},
  \bibinfo{journal}{Phys. Rev. B} \textbf{\bibinfo{volume}{74}},
  \bibinfo{pages}{144102} (\bibinfo{year}{2006}).

\bibitem[{\citenamefont{Chmaissem et~al.}(2001)\citenamefont{Chmaissem,
  Dabrowski, Kolesnik, Mais, Brown, Kruk, Prior, Pyles, and
  Jorgensen}}]{SrMnO3-Neel}
\bibinfo{author}{\bibfnamefont{O.}~\bibnamefont{Chmaissem}},
  \bibinfo{author}{\bibfnamefont{B.}~\bibnamefont{Dabrowski}},
  \bibinfo{author}{\bibfnamefont{S.}~\bibnamefont{Kolesnik}},
  \bibinfo{author}{\bibfnamefont{J.}~\bibnamefont{Mais}},
  \bibinfo{author}{\bibfnamefont{D.~E.} \bibnamefont{Brown}},
  \bibinfo{author}{\bibfnamefont{R.}~\bibnamefont{Kruk}},
  \bibinfo{author}{\bibfnamefont{P.}~\bibnamefont{Prior}},
  \bibinfo{author}{\bibfnamefont{B.}~\bibnamefont{Pyles}}, \bibnamefont{and}
  \bibinfo{author}{\bibfnamefont{J.~D.} \bibnamefont{Jorgensen}},
  \bibinfo{journal}{Phys. Rev. B} \textbf{\bibinfo{volume}{64}},
  \bibinfo{pages}{134412} (\bibinfo{year}{2001}).

\bibitem[{\citenamefont{Saitoh et~al.}(1995)\citenamefont{Saitoh, Bocquet,
  Mizokawa, and Fujimori}}]{saitoh-earlyTM}
\bibinfo{author}{\bibfnamefont{T.}~\bibnamefont{Saitoh}},
  \bibinfo{author}{\bibfnamefont{A.~E.} \bibnamefont{Bocquet}},
  \bibinfo{author}{\bibfnamefont{T.}~\bibnamefont{Mizokawa}}, \bibnamefont{and}
  \bibinfo{author}{\bibfnamefont{A.}~\bibnamefont{Fujimori}},
  \bibinfo{journal}{Phys. Rev. B} \textbf{\bibinfo{volume}{52}},
  \bibinfo{pages}{7934} (\bibinfo{year}{1995}).

\bibitem[{\citenamefont{Bocquet et~al.}(1996)\citenamefont{Bocquet, Mizokawa,
  Morikawa, Fujimori, Barman, Maiti, Sarma, Tokura, and
  Onoda}}]{bocquet-earlyTMO}
\bibinfo{author}{\bibfnamefont{A.~E.} \bibnamefont{Bocquet}},
  \bibinfo{author}{\bibfnamefont{T.}~\bibnamefont{Mizokawa}},
  \bibinfo{author}{\bibfnamefont{K.}~\bibnamefont{Morikawa}},
  \bibinfo{author}{\bibfnamefont{A.}~\bibnamefont{Fujimori}},
  \bibinfo{author}{\bibfnamefont{S.~R.} \bibnamefont{Barman}},
  \bibinfo{author}{\bibfnamefont{K.}~\bibnamefont{Maiti}},
  \bibinfo{author}{\bibfnamefont{D.~D.} \bibnamefont{Sarma}},
  \bibinfo{author}{\bibfnamefont{Y.}~\bibnamefont{Tokura}}, \bibnamefont{and}
  \bibinfo{author}{\bibfnamefont{M.}~\bibnamefont{Onoda}},
  \bibinfo{journal}{Phys. Rev. B} \textbf{\bibinfo{volume}{53}},
  \bibinfo{pages}{1161} (\bibinfo{year}{1996}).

\bibitem[{\citenamefont{Kang et~al.}(2008)\citenamefont{Kang, Lee, Kim, Kim,
  Dabrowski, Kolesnik, Lee, Kim, and Min}}]{SrMnO3-kang}
\bibinfo{author}{\bibfnamefont{J.-S.} \bibnamefont{Kang}},
  \bibinfo{author}{\bibfnamefont{H.~J.} \bibnamefont{Lee}},
  \bibinfo{author}{\bibfnamefont{G.}~\bibnamefont{Kim}},
  \bibinfo{author}{\bibfnamefont{D.~H.} \bibnamefont{Kim}},
  \bibinfo{author}{\bibfnamefont{B.}~\bibnamefont{Dabrowski}},
  \bibinfo{author}{\bibfnamefont{S.}~\bibnamefont{Kolesnik}},
  \bibinfo{author}{\bibfnamefont{H.}~\bibnamefont{Lee}},
  \bibinfo{author}{\bibfnamefont{J.-Y.} \bibnamefont{Kim}}, \bibnamefont{and}
  \bibinfo{author}{\bibfnamefont{B.~I.} \bibnamefont{Min}},
  \bibinfo{journal}{Phys. Rev. B} \textbf{\bibinfo{volume}{78}},
  \bibinfo{pages}{054434} (\bibinfo{year}{2008}).

\bibitem[{\citenamefont{van~der Marel et~al.}(1984)\citenamefont{van~der Marel,
  Sawatzky, and Hillebrecht}}]{U-xray-sawatzky-1984}
\bibinfo{author}{\bibfnamefont{D.}~\bibnamefont{van~der Marel}},
  \bibinfo{author}{\bibfnamefont{G.~A.} \bibnamefont{Sawatzky}},
  \bibnamefont{and} \bibinfo{author}{\bibfnamefont{F.~U.}
  \bibnamefont{Hillebrecht}}, \bibinfo{journal}{Phys. Rev. Lett.}
  \textbf{\bibinfo{volume}{53}}, \bibinfo{pages}{206} (\bibinfo{year}{1984}).

\bibitem[{\citenamefont{Hannerz et~al.}(1999)\citenamefont{Hannerz, Svensson,
  Istomin, and D'yachenko}}]{Hannerz-SrNbO3}
\bibinfo{author}{\bibfnamefont{H.}~\bibnamefont{Hannerz}},
  \bibinfo{author}{\bibfnamefont{G.}~\bibnamefont{Svensson}},
  \bibinfo{author}{\bibfnamefont{S.~Y.} \bibnamefont{Istomin}},
  \bibnamefont{and} \bibinfo{author}{\bibfnamefont{O.~G.}
  \bibnamefont{D'yachenko}}, \bibinfo{journal}{Journal of Solid State
  Chemistry} \textbf{\bibinfo{volume}{147}}, \bibinfo{pages}{421}
  (\bibinfo{year}{1999}).

\bibitem[{\citenamefont{Isawa et~al.}(1993)\citenamefont{Isawa, Sugiyama,
  Matsuura, Nozaki, and Yamauchi}}]{SrNbO3-isawa}
\bibinfo{author}{\bibfnamefont{K.}~\bibnamefont{Isawa}},
  \bibinfo{author}{\bibfnamefont{J.}~\bibnamefont{Sugiyama}},
  \bibinfo{author}{\bibfnamefont{K.}~\bibnamefont{Matsuura}},
  \bibinfo{author}{\bibfnamefont{A.}~\bibnamefont{Nozaki}}, \bibnamefont{and}
  \bibinfo{author}{\bibfnamefont{H.}~\bibnamefont{Yamauchi}},
  \bibinfo{journal}{Phys. Rev. B} \textbf{\bibinfo{volume}{47}},
  \bibinfo{pages}{2849} (\bibinfo{year}{1993}).

\bibitem[{\citenamefont{Nagai et~al.}(2005)\citenamefont{Nagai, Shirakawa,
  Ikeda, Iwasaki, Nishimura, and Kosaka}}]{SrMoO3-nagai}
\bibinfo{author}{\bibfnamefont{I.}~\bibnamefont{Nagai}},
  \bibinfo{author}{\bibfnamefont{N.}~\bibnamefont{Shirakawa}},
  \bibinfo{author}{\bibfnamefont{S.}~\bibnamefont{Ikeda}},
  \bibinfo{author}{\bibfnamefont{R.}~\bibnamefont{Iwasaki}},
  \bibinfo{author}{\bibfnamefont{H.}~\bibnamefont{Nishimura}},
  \bibnamefont{and} \bibinfo{author}{\bibfnamefont{M.}~\bibnamefont{Kosaka}},
  \bibinfo{journal}{Appl. Phys. Lett.} \textbf{\bibinfo{volume}{87}},
  \bibinfo{pages}{024105} (\bibinfo{year}{2005}).

\bibitem[{\citenamefont{Rodriguez et~al.}(2011)\citenamefont{Rodriguez,
  Poineau, Llobet, Kennedy, Avdeev, Thorogood, Carter, Seshadri, Singh, and
  Cheetham}}]{SrTcO3-rodriguez}
\bibinfo{author}{\bibfnamefont{E.~E.} \bibnamefont{Rodriguez}},
  \bibinfo{author}{\bibfnamefont{F.}~\bibnamefont{Poineau}},
  \bibinfo{author}{\bibfnamefont{A.}~\bibnamefont{Llobet}},
  \bibinfo{author}{\bibfnamefont{B.~J.} \bibnamefont{Kennedy}},
  \bibinfo{author}{\bibfnamefont{M.}~\bibnamefont{Avdeev}},
  \bibinfo{author}{\bibfnamefont{G.~J.} \bibnamefont{Thorogood}},
  \bibinfo{author}{\bibfnamefont{M.~L.} \bibnamefont{Carter}},
  \bibinfo{author}{\bibfnamefont{R.}~\bibnamefont{Seshadri}},
  \bibinfo{author}{\bibfnamefont{D.~J.} \bibnamefont{Singh}}, \bibnamefont{and}
  \bibinfo{author}{\bibfnamefont{A.~K.} \bibnamefont{Cheetham}},
  \bibinfo{journal}{Phys. Rev. Lett.} \textbf{\bibinfo{volume}{106}},
  \bibinfo{pages}{067201} (\bibinfo{year}{2011}).

\bibitem[{\citenamefont{Franchini et~al.}(2011)\citenamefont{Franchini, Archer,
  He, Chen, Filippetti, and Sanvito}}]{FranchiniPRB11}
\bibinfo{author}{\bibfnamefont{C.}~\bibnamefont{Franchini}},
  \bibinfo{author}{\bibfnamefont{T.}~\bibnamefont{Archer}},
  \bibinfo{author}{\bibfnamefont{J.}~\bibnamefont{He}},
  \bibinfo{author}{\bibfnamefont{X.-Q.} \bibnamefont{Chen}},
  \bibinfo{author}{\bibfnamefont{A.}~\bibnamefont{Filippetti}},
  \bibnamefont{and} \bibinfo{author}{\bibfnamefont{S.}~\bibnamefont{Sanvito}},
  \bibinfo{journal}{Phys. Rev. B} \textbf{\bibinfo{volume}{83}},
  \bibinfo{pages}{220402} (\bibinfo{year}{2011}).

\bibitem[{\citenamefont{Middey et~al.}(2011)\citenamefont{Middey, Kumar~Nandy,
  Mahadevan, and Sarma}}]{Middey11}
\bibinfo{author}{\bibfnamefont{S.}~\bibnamefont{Middey}},
  \bibinfo{author}{\bibfnamefont{A.}~\bibnamefont{Kumar~Nandy}},
  \bibinfo{author}{\bibfnamefont{P.}~\bibnamefont{Mahadevan}},
  \bibnamefont{and} \bibinfo{author}{\bibfnamefont{D.~D.} \bibnamefont{Sarma}},
  \bibinfo{journal}{arXiv} \textbf{\bibinfo{volume}{1112.5587v1}}
  (\bibinfo{year}{2011}).

\bibitem[{\citenamefont{Mravlje et~al.}(2012)\citenamefont{Mravlje, Aichhorn,
  and Georges}}]{SrTcO3-jernej}
\bibinfo{author}{\bibfnamefont{J.}~\bibnamefont{Mravlje}},
  \bibinfo{author}{\bibfnamefont{M.}~\bibnamefont{Aichhorn}}, \bibnamefont{and}
  \bibinfo{author}{\bibfnamefont{A.}~\bibnamefont{Georges}},
  \bibinfo{journal}{Phys. Rev. Lett.} \textbf{\bibinfo{volume}{108}},
  \bibinfo{pages}{197202} (\bibinfo{year}{2012}).

\bibitem[{\citenamefont{Ikeda et~al.}(2000)\citenamefont{Ikeda, Shirakawa,
  Bando, and Ootuka}}]{Sr2MoO4-ikeda}
\bibinfo{author}{\bibfnamefont{S.-I.} \bibnamefont{Ikeda}},
  \bibinfo{author}{\bibfnamefont{N.}~\bibnamefont{Shirakawa}},
  \bibinfo{author}{\bibfnamefont{H.}~\bibnamefont{Bando}}, \bibnamefont{and}
  \bibinfo{author}{\bibfnamefont{Y.}~\bibnamefont{Ootuka}},
  \bibinfo{journal}{J. Phys. Soc. Jpn.} \textbf{\bibinfo{volume}{69}},
  \bibinfo{pages}{3162} (\bibinfo{year}{2000}).

\bibitem[{\citenamefont{Hussey et~al.}(1998)\citenamefont{Hussey, Mackenzie,
  Cooper, Maeno, Nishizaki, and Fujita}}]{husseyPRB98}
\bibinfo{author}{\bibfnamefont{N.~E.} \bibnamefont{Hussey}},
  \bibinfo{author}{\bibfnamefont{A.~P.} \bibnamefont{Mackenzie}},
  \bibinfo{author}{\bibfnamefont{J.~R.} \bibnamefont{Cooper}},
  \bibinfo{author}{\bibfnamefont{Y.}~\bibnamefont{Maeno}},
  \bibinfo{author}{\bibfnamefont{S.}~\bibnamefont{Nishizaki}},
  \bibnamefont{and} \bibinfo{author}{\bibfnamefont{T.}~\bibnamefont{Fujita}},
  \bibinfo{journal}{Phys. Rev. B} \textbf{\bibinfo{volume}{57}},
  \bibinfo{pages}{5505} (\bibinfo{year}{1998}).

\bibitem[{\citenamefont{Mackenzie and Maeno}(2003)}]{Sr2RuO4-review}
\bibinfo{author}{\bibfnamefont{A.~P.} \bibnamefont{Mackenzie}}
  \bibnamefont{and} \bibinfo{author}{\bibfnamefont{Y.}~\bibnamefont{Maeno}},
  \bibinfo{journal}{Rev. Mod. Phys.} \textbf{\bibinfo{volume}{75}},
  \bibinfo{pages}{657} (\bibinfo{year}{2003}).

\bibitem[{\citenamefont{Liebsch and Lichtenstein}(2000)}]{liebschPRL00}
\bibinfo{author}{\bibfnamefont{A.}~\bibnamefont{Liebsch}} \bibnamefont{and}
  \bibinfo{author}{\bibfnamefont{A.}~\bibnamefont{Lichtenstein}},
  \bibinfo{journal}{Phys. Rev. Lett.} \textbf{\bibinfo{volume}{84}},
  \bibinfo{pages}{1591} (\bibinfo{year}{2000}).

\bibitem[{\citenamefont{Anisimov et~al.}(2002)\citenamefont{Anisimov, Nekrasov,
  Kondakov, Rice, and Sigrist}}]{anisimovEPL02}
\bibinfo{author}{\bibfnamefont{V.~I.} \bibnamefont{Anisimov}},
  \bibinfo{author}{\bibfnamefont{I.~A.} \bibnamefont{Nekrasov}},
  \bibinfo{author}{\bibfnamefont{D.~E.} \bibnamefont{Kondakov}},
  \bibinfo{author}{\bibfnamefont{T.~M.} \bibnamefont{Rice}}, \bibnamefont{and}
  \bibinfo{author}{\bibfnamefont{M.}~\bibnamefont{Sigrist}},
  \bibinfo{journal}{Eur. Phys. Lett.} \textbf{\bibinfo{volume}{25}},
  \bibinfo{pages}{191} (\bibinfo{year}{2002}).

\bibitem[{\citenamefont{Pchelkina et~al.}(2007)\citenamefont{Pchelkina,
  Nekrasov, Pruschke, Sekiyama, Suga, Anisimov, and
  Vollhardt}}]{pchelkinaPRB07}
\bibinfo{author}{\bibfnamefont{Z.~V.} \bibnamefont{Pchelkina}},
  \bibinfo{author}{\bibfnamefont{I.~A.} \bibnamefont{Nekrasov}},
  \bibinfo{author}{\bibfnamefont{T.}~\bibnamefont{Pruschke}},
  \bibinfo{author}{\bibfnamefont{A.}~\bibnamefont{Sekiyama}},
  \bibinfo{author}{\bibfnamefont{S.}~\bibnamefont{Suga}},
  \bibinfo{author}{\bibfnamefont{V.~I.} \bibnamefont{Anisimov}},
  \bibnamefont{and}
  \bibinfo{author}{\bibfnamefont{D.}~\bibnamefont{Vollhardt}},
  \bibinfo{journal}{Phys. Rev. B} \textbf{\bibinfo{volume}{75}},
  \bibinfo{pages}{035122} (\bibinfo{year}{2007}).

\bibitem[{\citenamefont{Huang et~al.}(1994)\citenamefont{Huang, Soubeyroux,
  Chmaissem, Natali~Sora, Santoro, Cava, Krajewski, and
  Peck}}]{HuangJSSC112-1994}
\bibinfo{author}{\bibfnamefont{Q.}~\bibnamefont{Huang}},
  \bibinfo{author}{\bibfnamefont{J.~L.} \bibnamefont{Soubeyroux}},
  \bibinfo{author}{\bibfnamefont{O.}~\bibnamefont{Chmaissem}},
  \bibinfo{author}{\bibfnamefont{I.}~\bibnamefont{Natali~Sora}},
  \bibinfo{author}{\bibfnamefont{A.}~\bibnamefont{Santoro}},
  \bibinfo{author}{\bibfnamefont{R.~J.} \bibnamefont{Cava}},
  \bibinfo{author}{\bibfnamefont{J.~J.} \bibnamefont{Krajewski}},
  \bibnamefont{and} \bibinfo{author}{\bibfnamefont{W.~F.} \bibnamefont{Peck}},
  \bibinfo{journal}{Journal of Solid State Chemistry}
  \textbf{\bibinfo{volume}{112}}, \bibinfo{pages}{355} (\bibinfo{year}{1994}).

\bibitem[{\citenamefont{Moon et~al.}(2006)\citenamefont{Moon, Kim, Kim, Lee,
  Kim, Park, Kim, Oh, Nakatsuji, Maeno et~al.}}]{Sr2RhO4-moon}
\bibinfo{author}{\bibfnamefont{S.~J.} \bibnamefont{Moon}},
  \bibinfo{author}{\bibfnamefont{M.~W.} \bibnamefont{Kim}},
  \bibinfo{author}{\bibfnamefont{K.~W.} \bibnamefont{Kim}},
  \bibinfo{author}{\bibfnamefont{Y.~S.} \bibnamefont{Lee}},
  \bibinfo{author}{\bibfnamefont{J.-Y.} \bibnamefont{Kim}},
  \bibinfo{author}{\bibfnamefont{J.-H.} \bibnamefont{Park}},
  \bibinfo{author}{\bibfnamefont{B.~J.} \bibnamefont{Kim}},
  \bibinfo{author}{\bibfnamefont{S.-J.} \bibnamefont{Oh}},
  \bibinfo{author}{\bibfnamefont{S.}~\bibnamefont{Nakatsuji}},
  \bibinfo{author}{\bibfnamefont{Y.}~\bibnamefont{Maeno}},
  \bibnamefont{et~al.}, \bibinfo{journal}{Phys. Rev. B}
  \textbf{\bibinfo{volume}{74}}, \bibinfo{pages}{113104}
  (\bibinfo{year}{2006}).

\bibitem[{\citenamefont{Tamai et~al.}(2008)\citenamefont{Tamai, Allan, Mercure,
  Meevasana, Dunkel, Lu, Perry, Mackenzie, Singh, Shen et~al.}}]{Sr2RhO4-tamai}
\bibinfo{author}{\bibfnamefont{A.}~\bibnamefont{Tamai}},
  \bibinfo{author}{\bibfnamefont{M.~P.} \bibnamefont{Allan}},
  \bibinfo{author}{\bibfnamefont{J.~F.} \bibnamefont{Mercure}},
  \bibinfo{author}{\bibfnamefont{W.}~\bibnamefont{Meevasana}},
  \bibinfo{author}{\bibfnamefont{R.}~\bibnamefont{Dunkel}},
  \bibinfo{author}{\bibfnamefont{D.~H.} \bibnamefont{Lu}},
  \bibinfo{author}{\bibfnamefont{R.~S.} \bibnamefont{Perry}},
  \bibinfo{author}{\bibfnamefont{A.~P.} \bibnamefont{Mackenzie}},
  \bibinfo{author}{\bibfnamefont{D.~J.} \bibnamefont{Singh}},
  \bibinfo{author}{\bibfnamefont{Z.-X.} \bibnamefont{Shen}},
  \bibnamefont{et~al.}, \bibinfo{journal}{Phys. Rev. Lett.}
  \textbf{\bibinfo{volume}{101}}, \bibinfo{pages}{026407}
  (\bibinfo{year}{2008}).

\bibitem[{\citenamefont{Haverkort et~al.}(2008)\citenamefont{Haverkort,
  Elfimov, Tjeng, Sawatzky, and Damascelli}}]{Sr2RhO4-haverkort}
\bibinfo{author}{\bibfnamefont{M.~W.} \bibnamefont{Haverkort}},
  \bibinfo{author}{\bibfnamefont{I.~S.} \bibnamefont{Elfimov}},
  \bibinfo{author}{\bibfnamefont{L.~H.} \bibnamefont{Tjeng}},
  \bibinfo{author}{\bibfnamefont{G.~A.} \bibnamefont{Sawatzky}},
  \bibnamefont{and}
  \bibinfo{author}{\bibfnamefont{A.}~\bibnamefont{Damascelli}},
  \bibinfo{journal}{Phys. Rev. Lett.} \textbf{\bibinfo{volume}{101}},
  \bibinfo{pages}{026406} (\bibinfo{year}{2008}).

\bibitem[{\citenamefont{Liu et~al.}(2008)\citenamefont{Liu, Antonov, Jepsen,
  and Andersen.}}]{Sr2RhO4-Liu}
\bibinfo{author}{\bibfnamefont{G.-Q.} \bibnamefont{Liu}},
  \bibinfo{author}{\bibfnamefont{V.~N.} \bibnamefont{Antonov}},
  \bibinfo{author}{\bibfnamefont{O.}~\bibnamefont{Jepsen}}, \bibnamefont{and}
  \bibinfo{author}{\bibfnamefont{O.~K.} \bibnamefont{Andersen.}},
  \bibinfo{journal}{Phys. Rev. Lett.} \textbf{\bibinfo{volume}{101}},
  \bibinfo{pages}{026408} (\bibinfo{year}{2008}).

\bibitem[{\citenamefont{Haverkort}(2005)}]{haverkort-tesis}
\bibinfo{author}{\bibfnamefont{M.~W.} \bibnamefont{Haverkort}}, Ph.D. thesis,
  \bibinfo{school}{Universit\"at K\"oln} (\bibinfo{year}{2005}).

\bibitem[{\citenamefont{Shih et~al.}(2012)\citenamefont{Shih, Zhang, Zhang, and
  Zhang}}]{ShihPRB12}
\bibinfo{author}{\bibfnamefont{B.-C.} \bibnamefont{Shih}},
  \bibinfo{author}{\bibfnamefont{Y.}~\bibnamefont{Zhang}},
  \bibinfo{author}{\bibfnamefont{W.}~\bibnamefont{Zhang}}, \bibnamefont{and}
  \bibinfo{author}{\bibfnamefont{P.}~\bibnamefont{Zhang}},
  \bibinfo{journal}{Phys. Rev. B} \textbf{\bibinfo{volume}{85}},
  \bibinfo{pages}{045132} (\bibinfo{year}{2012}).

\bibitem[{\citenamefont{Jiang et~al.}(2012)\citenamefont{Jiang, G\'omez-Abal,
  Li, Meisenbichler, Ambrosch-Draxl, and Scheffler}}]{FHI-GAP-hong}
\bibinfo{author}{\bibfnamefont{H.}~\bibnamefont{Jiang}},
  \bibinfo{author}{\bibfnamefont{R.}~\bibnamefont{G\'omez-Abal}},
  \bibinfo{author}{\bibfnamefont{X.}~\bibnamefont{Li}},
  \bibinfo{author}{\bibfnamefont{C.}~\bibnamefont{Meisenbichler}},
  \bibinfo{author}{\bibfnamefont{C.}~\bibnamefont{Ambrosch-Draxl}},
  \bibnamefont{and}
  \bibinfo{author}{\bibfnamefont{M.}~\bibnamefont{Scheffler}},
  \bibinfo{journal}{(unpublished)}  (\bibinfo{year}{2012}).

\bibitem[{\citenamefont{Aryasetiawan and Gunnarsson}(1998)}]{GW_Ferdi}
\bibinfo{author}{\bibfnamefont{F.}~\bibnamefont{Aryasetiawan}}
  \bibnamefont{and}
  \bibinfo{author}{\bibfnamefont{O.}~\bibnamefont{Gunnarsson}},
  \bibinfo{journal}{Rep. Prog. Phys.} \textbf{\bibinfo{volume}{61}},
  \bibinfo{pages}{237} (\bibinfo{year}{1998}).

\bibitem[{\citenamefont{Kotani and van Schilfgaarde}(2002)}]{GW-MB-kotani}
\bibinfo{author}{\bibfnamefont{T.}~\bibnamefont{Kotani}} \bibnamefont{and}
  \bibinfo{author}{\bibfnamefont{M.}~\bibnamefont{van Schilfgaarde}},
  \bibinfo{journal}{Solid State Commun.} \textbf{\bibinfo{volume}{121}},
  \bibinfo{pages}{461} (\bibinfo{year}{2002}).

\end{thebibliography}
\bibliographystyle{apsrev}

\end{document}